\documentclass{IEEEoj}

\usepackage[T1]{fontenc}
\usepackage[utf8]{inputenc}
\usepackage[mathscr]{euscript}
 \let\mathscr\relax
\usepackage[scr]{rsfso}
\usepackage{calrsfs}
\usepackage[cmex10]{amsmath}
\usepackage{amsfonts,amsbsy,amssymb,amsthm}
\usepackage{physics}
\usepackage[final]{graphicx}
\usepackage{url}
\usepackage{enumerate}
\usepackage{cite}
\usepackage{color}
\usepackage{xcolor}
\usepackage{soul}
\usepackage{algorithmic}
\usepackage[ruled,vlined]{algorithm2e}
\usepackage{xfrac,bigints}
\usepackage{float}
\usepackage{textcomp}
\usepackage{multirow}
\usepackage{anyfontsize}

\def\BibTeX{{\rm B\kern-.05em{\sc i\kern-.025em b}\kern-.08em
    T\kern-.1667em\lower.7ex\hbox{E}\kern-.125emX}}
\AtBeginDocument{\definecolor{ojcolor}{cmyk}{0.93,0.59,0.15,0.02}}
\def\OJlogo{\vspace{-4pt}\hskip-4pt\includegraphics[height=18pt]{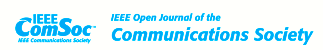}}

\newcommand{\Ep}{{u_\text{d}}}

\newcommand{\E}{{E}}
\newcommand{\Ai}{{\text{Ai}}}
\newcommand{\dAi}{\dot{\text{A}}\text{i}}
\newcommand{\ddAi}{\ddot{\text{A}}\text{i}}

\newcommand{\eqdef}{\triangleq}
\newcommand{\bm}[1]{{\mathbf{#1}}}
\newcommand{\en}{\EuScript{E}}
\newcommand{\banda}{\EuScript{B}}
\DeclareMathAlphabet{\pazocal}{OMS}{zplm}{m}{n}

\def\bdm#1\edm{\begin{displaymath}#1\end{displaymath}}
\def\be#1\ee{\begin{equation}#1\end{equation}}
\def\barr#1\earr{\begin{align}#1\end{align}}
\newtheorem{proposition}{Proposition}

\begin{document}

\receiveddate{16 June, 2026}
\reviseddate{20 July, 2026}
\accepteddate{14 August, 2026}
\publisheddate{XX Month, 2026}
\doiinfo{OJCOMS.2026.xxxxxx}

\title{Airy beams for radiative near-field communications: 
Fundamentals, potentials, and limitations}

\author{DONATELLA~DARSENA\authorrefmark{1} (SENIOR MEMBER, IEEE),
FRANCESCO~VERDE\authorrefmark{2} (SENIOR MEMBER, IEEE),
MARCO~DI RENZO\authorrefmark{3,4} (FELLOW, IEEE), 
\\ AND VINCENZO~GALDI\authorrefmark{5} (FELLOW, IEEE)
}
\affil{Department of Electrical Engineering and
Information Technology, University Federico II, I-80125 Naples, Italy}
\affil{Department of Engineering, University of Campania Luigi Vanvitelli, I-81031 Aversa, Italy}
\affil{CNRS and CentraleSup\'elec, Institute of Electronics and Digital Technologies (IETR), Avenue de la Boulaie, 35576 Cesson-S\'evign\'e, France}
\affil{King's College London, Department of Engineering - Centre for Telecommunications Research, WC2R 2LS London, United Kingdom}
\affil{Fields \& Waves Lab, Department of Engineering, University of Sannio,  
I-82100 Benevento, Italy}
\corresp{CORRESPONDING AUTHOR: F.~VERDE (e-mail: francesco.verde@unicampania.it).}
\authornote{The work of D.~Darsena and V.~Galdi was partially supported by 
the European Union-Next Generation EU under the Italian
National Recovery and Resilience Plan (NRRP), Mission 4,
Component 2, Investment 1.3, CUP E63C22002040007, partnership
on ``Telecommunications of the Future" (PE00000001
- program ``RESTART"). 
The work of M. Di Renzo was supported in part by the European Research Council (ERC) under the European Union’s Horizon Europe Programme WePhICom (agreement number 101225119), as well as by the European Union through the Horizon Europe project COVER under grant agreement number 101086228, the Horizon Europe project UNITE under grant agreement number 101129618, the Horizon Europe project INSTINCT under grant agreement number 101139161, and the Horizon Europe project TWIN6G under grant agreement number 101182794, as well as by the Agence Nationale de la Recherche (ANR) through the France 2030 project ANR-PEPR Networks of the Future under grant agreement NF-SYSTERA 22-PEFT-0006, and by the CHIST-ERA project PASSIONATE under grant agreements CHIST-ERA-22-WAI-04 and ANR-23-CHR4-0003-01. Also, the work of M. Di Renzo was supported in part by the Engineering and Physical Sciences Research Council (EPSRC), part of UK Research and Innovation, and the UK Department of Science, Innovation and Technology through the CHEDDAR Telecom Hub under grant EP/Y037421/1, through the HASC Telecom Hub under grant EP/Y037197/1, and through the TITAN Telecom Hub under grant EP/Y037243/1.}

\markboth{Airy Beams for Radiative Near-Field Communications: Fundamentals, Potentials, and Limitations}%
{Darsena \textit{et al.}}

\begin{abstract}

In next-generation wireless networks, 
electromagnetic signals are envisioned to be radiated from 
large-scale radio-frequency transmitters operating in the millimeter-wave and 
sub-terahertz bands. 
The combination of electrically large radiating apertures 
and high-frequency transmission extends the radiative 
near-field region around the transmitter. 
In this region, 
unlike in the far field,
the wavefront is nonplanar,  which provides additional degrees of freedom to shape and steer the transmitted beam as desired.
In this paper, we focus on Airy beams, which may exhibit several highly desirable
properties in the radiative near-field region. In their ideal form, these beams can follow {\em self-accelerating} (curved) 
trajectories, exhibit resilience to perturbations through self-healing, and maintain
a shape-preserving intensity profile in a co-moving transverse frame, making them effectively {\em diffraction-resistant}. 
Specifically, starting from  diffraction theory as
the foundational propagation model for radiative near-field  free-space beam manipulation, 
we first present the underlying principles of self-accelerating 
beams radiated by continuous aperture field distributions.
We then address several challenges regarding the generation of Airy beams, including
their exponential decay due to finite energy constraints 
and spatial truncation of the aperture. Moreover, we examine 
their free-space propagation characteristics, focusing on a
generalized link budget formulation and a polychromatic representation. 
The second part of the paper focuses on the propagation behavior of 
Airy beams in non-line-of-sight (NLoS) scenarios, which are particularly relevant for radiative near-field wireless communication applications. 
We also present a comparison between Airy beams and Gaussian beams, which represent the most common solution for focused beam transmission in the near field, evaluating their performance in terms of received energy.
Our theoretical and numerical results show that Airy beams may offer a performance advantage over Gaussian beams in certain NLoS channels, provided that their key properties are largely preserved, specifically, self-acceleration along a parabolic 
trajectory and diffraction-resistant propagation. In the presence of an
obstacle, this requires that the portion of the transmit 
aperture with a clear line-of-sight to the receiver is 
sufficiently large.
These findings underscore the intriguing potential of Airy beams in radiative near-field wireless links, while also highlighting the importance of a concurrent electromagnetic and telecommunication design to fully harness their advantages in practical systems.
\end{abstract}


\maketitle

\section{Introduction}
\IEEEPARstart{I}{n} the 1860s, Maxwell identified light as an 
electromagnetic (EM) wave, a discovery of fundamental importance. 
Despite this, the study of how EM waves behave and interact 
with matter has historically progressed along two separate yet 
complementary paths: One pursued by wireless communication 
engineers and the other by optical physicists.
This divergence arises mainly because traditional
wireless networks were designed to operate 
in the sub-6 GHz frequency band, which is well below the optical 
range. However, the growing demand for mobile data 
has recently pushed next-generation wireless networks, such as 
beyond-fifth-generation (B5G) and sixth-generation (6G) systems,  
toward the millimeter-wave (mmWave) and terahertz (THz) frequency 
bands \cite{Petrov-2024,Cui-2023,Singh-2025}.
Moreover, in addition to expanding the available bandwidth, B5G/6G 
wireless systems have achieved increased Shannon capacity by 
leveraging the spatial dimension through the use of multiple 
antennas. This approach enables both power gains and increased 
degrees of freedom for communication \cite{Tse-book}.
Nowadays, these technological trends are providing an opportunity 
to bridge and unify the fields of wireless communication 
engineering and optics.

The propagation of an EM wave radiated by a source aperture 
is typically described by tracking surfaces 
of constant phase, i.e., {\em wavefronts}, as they evolve 
during propagation.
By virtue of the Huygens-Fresnel principle \cite{Goodman}, every 
wavefront can be viewed as a collection of infinitesimally small 
secondary point sources, each radiating wavelets within a 
forward-facing hemisphere. The wavefront at a subsequent instant 
in time is then constructed by coherently superimposing the 
radiation contributions from all these wavelets.
This superposition process is known as {\em diffraction} 
\cite{Goodman}, and the EM wave at any given location in space 
can be mathematically described using Rayleigh-Sommerfeld 
diffraction theory \cite{Orfanidis-2002}. According to this 
theory, the radiated field from a source aperture is determined 
by the amplitude and phase distribution of the complex EM field 
across the aperture.
From a communications perspective, controlling this aperture distribution is precisely 
 {\em transmit beamforming}, while from a wave-optics perspective it is {\em wavefront synthesis}. In this context, a {\em beam} 
is, more precisely, a spatial pattern resulting from constructive 
and destructive interference within the wavefront.
Diffraction is typically categorized into two regions: The 
{\em near field} and the {\em far field} (also known as the 
{\em Fraunhofer diffraction region}).
The boundary between the two domains is defined by the 
Fraunhofer distance \cite{Goodman}, which scales quadratically 
with the physical size of the aperture and linearly with the 
frequency of the EM field.

In many conventional sub-6 GHz systems, the electrically moderate aperture size may lead to Fraunhofer distances of only  
a few meters. As a consequence, the design of conventional 
wireless systems has largely focused on the far-field region, 
where the wavefront is approximately planar, and the EM field 
observed can be determined, up to a multiplicative phase factor, 
by applying the Fourier transform to the aperture field 
distribution \cite{Goodman}.
In the far field, a linear phase ramp across the aperture steers the beam toward a prescribed angle. Because the wavefront is approximately planar, the transmitter can control the angular distribution of radiated energy but has limited ability to perform range-dependent field shaping.

\begin{table*}[!t]
    \small
    \centering
    \caption{Comparison of representative structured beam families 
    relevant to radiative near-field communications.}
    \renewcommand{\arraystretch}{1.2}
    \begin{tabular}{|p{2.2cm}|p{2.7cm}|p{3.2cm}|p{2.8cm}|p{3.2cm}|}
        \hline
        \textbf{Beam Family} 
        & \textbf{Governing Equation / Coordinate System} 
        & \textbf{Transverse Evolution} 
        & \textbf{Propagation Invariance (Ideal Case)} 
        & \textbf{Practical Considerations} \\ 
        \hline
        Gaussian 
        & Paraxial wave equation (Cartesian coordinates) 
        & Straight-axis propagation with diffractive spreading 
        & Not propagation-invariant 
        & Finite energy; diffraction length set by beam waist; 
        standard aperture illumination \\ 
        \hline
        Bessel 
        & Helmholtz equation (cylindrical coordinates) 
        & Straight central core formed by conical wave interference 
        & Yes (invariant transverse profile under ideal infinite 
        aperture) 
        & Ideal solution non-square-integrable; finite aperture 
        limits invariant range; typically generated via axicons 
        or annular apertures \\ 
        \hline
        Mathieu 
        & Helmholtz equation (elliptic coordinates) 
        & Elliptic transverse structure; invariant in separable form 
        & Yes (ideal separable solution) 
        & Non-square-integrable in ideal form; invariant behavior 
        aperture-limited; requires elliptic phase modulation \\ 
        \hline
        Airy (paraxial) 
        & Paraxial wave equation 
        & Parabolic caustic; intensity maximum follows curved 
        trajectory 
        & Not invariant in fixed frame; shape-preserving in 
        co-moving accelerating frame 
        & Ideal solution non-square-integrable; finite-energy 
        realizations require apodization or truncation; diffraction 
        resistance persists over finite range \\ 
        \hline
        Pearcey 
        & Paraxial wave equation (Cartesian coordinates) 
        & Cusp caustic; auto-focusing followed by intensity inversion 
        and self-similar re-expansion; 2D counterpart of Airy 
        function 
        & Not invariant in fixed frame; auto-focusing and 
        form-reconstruction in co-propagating frame 
        & Ideal solution non-square-integrable; finite-energy 
        realizations require quartic phase mask or metasurface; 
        self-healing demonstrated under partial obstruction; 
        suitable for multi-user NLoS near-field 
        links~\cite{Qin_Pearcey_2026} \\ 
        \hline
        Weber (nonparaxial) 
        & Helmholtz equation (parabolic coordinates) 
        & Nonparaxial curved caustic 
        & Not invariant in fixed frame; nonparaxial accelerating 
        solution 
        & Ideal separable solutions generally non-square-integrable; 
        finite aperture limits accelerating range; angular spectrum 
        engineering required; experimentally demonstrated via 
        transparent metasurface for obstacle-avoiding 
        communications~\cite{Luo_Weber_2026} \\ 
        \hline
    \end{tabular}
    \label{Table1}
\end{table*}

In the radiative near-field region \cite{Goodman}, reactive coupling to the aperture is negligible, but the wavefront has not yet reached its far-field  planar form.
The presence of nonplanar wavefronts provides additional spatial 
degrees of freedom that can be exploited through suitable shaping 
of the aperture field distribution, enabling the synthesis of 
structured beam patterns that differ substantially from 
conventional Gaussian beams, which provide a canonical finite-energy model for focused beam transmission and are widely used as a benchmark in optics and near-field wireless 
communications.
Certain structured beams exhibit propagation-invariant or 
diffraction-resistant behavior \cite{Bandres-2009}, meaning that 
their transverse profile remains unchanged, or approximately 
unchanged, over a finite propagation distance.
As summarized in Table~\ref{Table1}, beam families relevant to 
near-field communications can be broadly classified into three 
categories. Gaussian beams correspond to finite-energy paraxial 
solutions whose energy spreads monotonically along a straight 
propagation axis \cite{Siegman}. Bessel beams \cite{Durnin-1987} 
and Mathieu beams \cite{Vega} arise as separable solutions of 
the Helmholtz equation in cylindrical and elliptic coordinates, 
respectively, and are strictly propagation-invariant under ideal 
infinite-aperture conditions. In contrast, accelerating beams 
preserve their functional form in a co-moving frame while their 
intensity maxima evolve along curved caustics. Airy beams in 
the paraxial regime \cite{Siviloglou-2007,Siviloglou-2007-2} and 
Weber beams in the nonparaxial case \cite{Bandres} constitute 
representative examples of this class. All ideal nondiffracting 
or accelerating solutions are non-square-integrable and, thus,  
require truncation or apodization in practice, so that structured 
behavior persists only over a finite, aperture-dependent range. 
Differences in self-reconstruction capability and implementation 
strategies, such as amplitude-and-phase versus phase-only 
modulation, reflect important trade-offs for near-field wireless 
platforms.

Within the class of accelerating beams, Airy beams play a 
central role. Their origin can be traced to a non-spreading 
solution of the free-particle Schr\"{o}dinger equation 
identified in 1979 \cite{Berry-1979,Messiah-2014,Unn-1996}.
Through the mathematical analogy between the 
Schr\"{o}dinger equation and the paraxial wave equation of 
diffraction, this solution was translated into optics, leading 
to the formulation of paraxial Airy beams 
\cite{Siviloglou-2007,Siviloglou-2007-2,Sivi-2008,
Zamboni-2012,Kaganovsky-2010,Efremidis-2019,Vallee-2004}.
Under ideal infinite-aperture conditions in one transverse 
dimension, the Airy beam preserves its functional form while 
its intensity maximum follows a parabolic trajectory. The 
apparent transverse acceleration results from structured 
interference within the wavefront. Consistent with Ehrenfest's 
theorem \cite{Messiah-2014}, the beam centroid propagates along 
a straight line in homogeneous free space, reflecting 
conservation of transverse momentum \cite{Jackson-2007}. Airy 
beams are also known to exhibit self-reconstruction after 
partial obstruction \cite{Broky-2008,Chu-2012,Aiello}, provided 
that a sufficient portion of the aperture remains unobstructed.

Self-accelerating beams, including Airy beams, have attracted 
increasing interest for their potential to improve B5G and 6G 
wireless networks \cite{Petrov-2024,Guerb2024,Lee2025,Drou2025}.
Their unique properties, such as diffraction 
resistance and self-healing, may enable complementary strategies that mitigate some limitations of conventional far-field beam steering in radiative near-field links. 
By shaping the wavefront at the transmitter, these beams may offer precise spatial control, potentially improved spatial selectivity and 
interference management, and the possibility of redirecting energy around obstacles.
These features make them potentially relevant to 
high-throughput links and spatial multiplexing, especially in 
the dense and dynamic environments expected in 6G systems.
The rapidly growing body of literature on this topic can be 
organized along four main research directions, as summarized 
below.

\subsection{Foundational experimental and theoretical studies}

In~\cite{Guerb2024}, the authors demonstrated a communication 
link at THz frequencies using a self-accelerating beam, and 
experimentally validated its ability to avoid obstacles. A model 
is developed to discuss bandwidth limits and to highlight that 
ray optics is insufficient in the near field. However, the 
findings are limited to specific blockage scenarios and do not 
generalize to broader system-level design considerations. A 
related framework in~\cite{GueEUCAP2024} uses caustic-beam 
design to engineer a desired trajectory and to compute link 
budgets, with comparisons against conventional Gaussian beams. 
The experimental study in~\cite{Lee2025} validates sub-THz 
Airy-beam generation, trajectory control, multi-stream 
transmission at very high rates, and self-healing under partial 
blockage, but it is centered on experimental feasibility and 
system demonstrations. As in~\cite{Lee2025}, the main 
emphasis of~\cite{Drou2025} lies in applying self-accelerating 
beams to multiuser scenarios, with a particular focus on 
interference management, based largely on numerical simulations.

\subsection{Beam synthesis, optimization, and management}

In~\cite{Liu_ArXiV2025}, bending-beam synthesis is formulated 
as an optimization problem of 
the received power along a trajectory, and solved via successive 
convex approximation, showing improved numerical performance and 
robustness. Reference~\cite{Canals_ArXiV2025} shows that, compared to flat apertures, curved surfaces 
impose less stringent constraints on the phase of the source EM 
wave over the aperture. From a beam-management 
perspective,~\cite{Ye_ArXiV2025} proposes trajectory-adaptive 
beam shaping to eliminate real-time beam management by shaping 
the wavefront to follow a user's predefined trajectory, 
exploiting the self-accelerating nature of structured beams. A 
unified curved beam framework for multi-user communications in 
blockage-prone environments is developed in~\cite{Yao_2026}, 
where a continuous-aperture model with steering, focusing, and 
curving functions is introduced, the performance gap between 
continuous and discrete aperture implementations is analytically 
characterized, and an iterative optimization algorithm based on 
fractional programming is proposed to maximize the weighted 
sum-rate.

\subsection{Structured beam families and hardware implementations}

Reference \cite{Zhao_DataCenter_2026} presents a system-level comparison between Airy and Gaussian beams for 
THz wireless data centers, where quasi  line-of-sight (LoS) propagation due to 
obstacles is common,
showing regimes in which  Airy beams can outperform Gaussian beams for blockage mitigation 
via hybrid beamforming. Closed-form analytical expressions for 
Airy beam propagation in the THz band, complementing the 
wave-optics framework developed here, are derived 
in~\cite{Zhao_ClosedForm_2026}. The self-healing and 
obstacle-avoiding capabilities of Weber beams, a nonparaxial 
counterpart of Airy beams, have been experimentally demonstrated 
using a transparent metasurface in~\cite{Luo_Weber_2026}, 
providing a compelling hardware validation of the concepts 
discussed in Section~III-B. Pearcey beams, a higher-order 
caustic solution arising from quartic wavefront shaping, have 
been recently proposed for multi-user non-line-of-sight (NLoS) near-field 
communications in~\cite{Qin_Pearcey_2026}. Reconfigurable accelerated 
beams at THz frequencies have been demonstrated using 
inverse-designed bilayer diffractive structures 
in~\cite{Jia_Reconfigurable_2026}, while 4-D Bessel beam steering 
and near-field communications using cascaded metasurfaces have 
been demonstrated in~\cite{Xu_Bessel_2025}.
More recently, a multilayer dielectric metasurface operating over
15-25~GHz and capable of modulating amplitude and phase
simultaneously and independently has been experimentally shown to
generate self-healing non-diffracting beams, which maintain up to
$34\%$ signal integrity at a propagation distance of $120$
wavelengths despite an obstruction as large as $16$ wavelengths, and
achieve a $20.5$~dB path-loss reduction over conventional Gaussian
beams under LoS blockage \cite{Xia2026}.

\subsection{System-level, security, and application-oriented studies}

The potential of accelerated caustic beams for physical-layer 
security has been explored in~\cite{Liu_Secure_2026}, where a 
robust beamforming strategy based on the caustic effect is 
proposed to simultaneously bypass eavesdropping regions and 
illuminate legitimate users. From a broader networking 
viewpoint,~\cite{Petrov-2024} proposes wavefront hopping, 
consisting in dynamically switching among different wavefronts, 
to improve reliability, interference control, and security in 
radiative near-field THz systems. A quantitative roadmap for 
non-diffracting beams in near-field mmWave systems, including 
advantage-regime maps under aperture and blockage constraints, 
is provided in~\cite{Qin_Roadmap_2025}. A geometric model for 
the reflections of accelerating beams from arbitrary surfaces, 
validated experimentally via the Legendre transform framework, 
is introduced in~\cite{Spindel_2026}. Learning-based approaches 
for blockage-resilient beam training using Airy beams have been 
investigated in~\cite{Weng_2025}, while a broad system-level 
perspective on beam manipulation for THz communications is 
discussed in~\cite{Li_Survey_2025}.

On the experimental side, sub-THz measurement campaigns employing
structured (Bessel) beams have demonstrated both their benefits over
conventional Gaussian beamforming and the limited range over which
such benefits persist \cite{Reddy2023,Bodet2024}. Notably, the recent
experimental study in \cite{Kludze2026} has shown that the favorable
properties of near-field beams, such as focusing and self-healing,
are observed only within an \emph{effective near-field range} lying
well within the nominal Fresnel region, and has proposed distributed
(multi-aperture) wavefront shaping as a practical strategy to extend
this range without resorting to impractically large monolithic
apertures. In principle, this strategy is directly applicable
to the Airy beams considered herein.

\subsection{Open challenges and motivation for the present work}

While these works provide important insights, the practical use of
self-accelerating beams in wireless communications remains uncertain and widely debated \cite{Inserra_2022}. Many existing models rely on idealized assumptions, including infinite-energy beam profiles, continuous aperture distributions, and ideal amplitude-and-phase control, whereas practical systems involve finite apertures, spatial discretization, quantized excitations, energy constraints, and hardware losses. In addition, sub-THz implementations may require sophisticated aperture designs or metasurfaces whose scalability and efficiency remain
open challenges. This motivates a wave-optics diffraction framework, grounded in EM propagation, that characterizes Airy-beam generation and propagation under realistic constraints and highlights the key design trade-offs for radiative near-field wireless communications.

\subsection{Contributions}

The contributions of this article are summarized as follows.

\begin{enumerate}

\item 
We present the fundamentals of scalar diffraction theory 
from a wireless communication perspective, emphasizing that the 
radiative near-field channel can be modeled as a linear 
space-invariant (LSI) system. Moreover, even though 
self-accelerating beams have been extensively studied in optics, 
we introduce a structured and self-contained framework that 
highlights the key properties relevant to wireless 
communications, such as diffraction-resistant propagation and 
caustic formation, grounded in well-established optical 
principles.

\item 
We conduct a theoretical study of Airy beams in 
free space, starting from their ideal formulation and 
progressing toward physically realizable configurations 
accounting for the finite-energy constraint and limited transmit-aperture size. This analysis reveals the interplay 
among system parameters, including carrier frequency, aperture 
dimension, tapering of the aperture field, and the extent of 
the diffraction-resisting region. We derive expressions 
for the near-field path loss and provide a spectral characterization 
of Airy beams.

\item 
We investigate the performance of Airy beams in 
NLoS scenarios. Using knife-edge diffraction 
theory and an analytical model for an infinite soft obstacle, 
we investigate their self-healing capabilities. Our findings 
uncover a critical limitation that has not been systematically quantified in prior communication-oriented studies, 
related to the dependence of beam recovery on the portion of 
the transmit aperture that remains in LoS with 
the receiver, relative to the size of the obstruction.

\end{enumerate}

We emphasize the differences between the present 
work and prior studies such as \cite{Inserra_2022}, which have 
also investigated the behavior of Airy beams in obstructed 
scenarios. In \cite{Inserra_2022}, the analysis is mainly 
conducted from an optimization perspective, focusing on discrete 
antenna arrays and on the maximization of power transfer 
efficiency through optimal excitation strategies. In that 
context, it is shown that near-field focusing based on the LoS 
portion of the aperture provides superior performance, leading 
to the conclusion that residual-aperture near-field focusing often provides higher-power transfer efficiency.
Conversely, the present work adopts a complementary viewpoint 
based on a continuous-aperture, wave-optics framework grounded 
in scalar diffraction theory. Within this setting, we provide a 
unified analytical and numerical characterization of Airy beams 
under realistic constraints, including finite energy, finite 
aperture size, and aperture truncation. Furthermore, we 
introduce a generalized link-budget formulation and explicitly 
quantify the trade-offs among aperture configuration, 
obstruction conditions, and beam properties.
This approach confirms the key limitation identified in
\cite{Inserra_2022}, namely that the self-healing capability of Airy beams deteriorates when a substantial portion of the transmit aperture is obstructed. It also identifies operating regimes in which Airy beams can retain performance advantages over conventional Gaussian beams. In particular, when the main lobe of the aperture field remains in LoS with the receiver and only the secondary lobes are partially obstructed, Airy beams can deliver higher received energy than a collimated Gaussian beam. These results refine the conditions under which Airy beams are beneficial and complement the optimization-based
assessment in \cite{Inserra_2022} with a wave-optics characterization of the underlying physical mechanisms.

\subsection{Organization}

This article is organized as follows.
Section~\ref{sec:Diff-theory} outlines the basic principles of 
scalar theory of diffraction in free space.
Section~\ref{sec:self-acc-beam} introduces the concept of 
caustics in the context of diffraction-resistant self-accelerating 
beams, considering both paraxial and nonparaxial regimes.
Section~\ref{sec:airy-free} addresses key aspects of the 
generation, characterization, and propagation of paraxial Airy 
beams in free space.
Section~\ref{sec:self-healing} presents an analysis of the 
behavior of these beams in communication scenarios with partial 
obstructions.
Finally, Section~\ref{sec:concl} summarizes the main findings 
and outlines potential directions for future research.

\begin{figure}[t]
\centering
\includegraphics[width=0.8\linewidth]{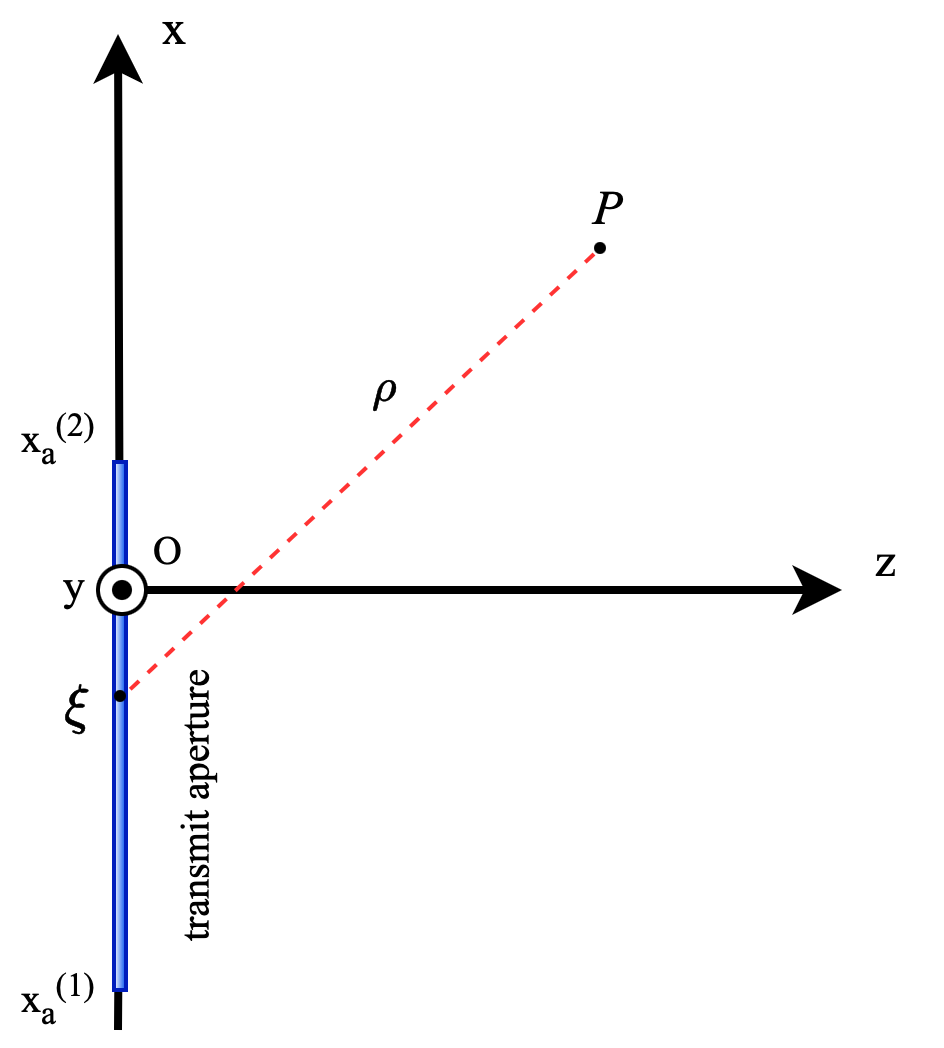} 
\caption{Geometry of the considered diffraction problem.}
\label{fig:fig_1_rev}
\end{figure}

\section{Scalar theory of diffraction in free space}
\label{sec:Diff-theory}

We consider the problem of describing the propagation
of an EM wave radiated from a source aperture in free space.
This propagation 
is governed by wave functions 
that describe the spatial and temporal variation of the electric field 
vector $\boldsymbol{\mathcal E}(\bm{P};t)$, with 
Cartesian components ${\mathcal E}_x(\bm{P};t), {\mathcal E}_y(\bm{P};t),
{\mathcal E}_z(\bm{P};t)$, and 
the magnetic field  vector 
$\boldsymbol{\mathcal H}(\bm{P};t)$, with 
Cartesian components ${\mathcal H}_x(\bm{P};t), {\mathcal H}_y(\bm{P};t),
{\mathcal H}_z(\bm{P};t)$, where $\bm{P}=(x,y,z)$ denotes the spatial coordinates and $t$ denotes time.
In most treatments of diffraction theory, the propagation medium is assumed to be nonmagnetic, implying that 
the magnetic permeability is equal to that of  free space, $\mu_0$.

In a dielectric medium that is linear, isotropic, homogeneous,
and nondispersive, both $\boldsymbol{\mathcal E}(\bm{P};t)$ and
$\boldsymbol{\mathcal H}(\bm{P};t)$ satisfy the same 
vector wave equation \cite{Goodman}. As a result, each Cartesian component of these vectors  obeys an identical scalar wave equation. Under these conditions, the EM  propagation  
can be completely described by a single scalar wave equation.
To formalize this, let the EM
radiation be represented by a real-valued scalar function $\widetilde{\E}(\bm{P};t)$.
Starting from Maxwell's equations, it can be shown \cite{Goodman} that, under the aforementioned 
assumptions, the field $\widetilde{\E}(\bm{P};t)$ satisfies the scalar wave equation
\be
\nabla^2 \, \widetilde{\E}(\bm{P};t) - \frac{n^2}{c^2} \, \frac{\partial^2 \, \widetilde{\E}(\bm{P};t)}{\partial t^2} =0, 
\label{eq:max}
\ee
where $\nabla^2= \tfrac{\partial^2}{\partial x^2}+\tfrac{\partial^2}{\partial y^2}+\tfrac{\partial^2}{\partial z^2}$
is the Laplacian operator, with $\tfrac{\partial^k (\cdot) }{\partial u^k}$ denoting the $k$-th order partial derivative with respect to 
the real-valued variable $u$,
$n$ is the refractive index of the medium,\footnote{
Note that $n = 1$ in vacuum, whereas $n > 1$ in a 
homogeneous dielectric medium.}  
and $c\approx 3 \cdot 10^8$ m/s is the speed of light in vacuum.

To highlight the key aspects of the problem,
as schematically illustrated in Fig.~\ref{fig:fig_1_rev}, 
we assume throughout the paper that the transmitter is 
modeled as a one-dimensional (1-D) radiating flat aperture extending from $x_\text{a}^{(1)}$ to $x_\text{a}^{(2)}$
along the $x$ {\em transverse} direction, with $x_\text{a}^{(2)} > x_\text{a}^{(1)}$,
lying in the plane $z=0$, and extending infinitely along the $y$-axis.
This model serves as a reasonable abstraction of a practical  
transmitter shaped as a rectangular strip, where the finite length $\Delta y_\text{a}$ 
along the $y$-axis is much larger than the width $\Delta x_\text{a} \eqdef x_\text{a}^{(2)} - x_\text{a}^{(1)}$, i.e., 
$\Delta y_\text{a} \gg \Delta x_\text{a}$.
Specifically, we consider a scenario featuring a $y$-polarized electric field, where
the propagation is perpendicular to $x$ towards the $z>0$ {\em longitudinal} direction, 
and there is no dependence on the $y$-coordinate, i.e., 
$\widetilde{\E}(\bm{P};t) \equiv \widetilde{\E}(z,x;t)$.
Moreover, we focus on a spatially continuous radiating surface. 
In practice, this can be reasonably
approximated by lens antennas \cite{Lee2025} or metasurfaces composed of densely spaced subwavelength elements
 \cite{Malevich.2025}. 

\begin{figure}[t]
\centering
\includegraphics[width=\linewidth]{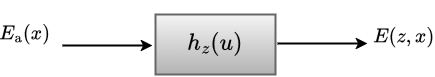} 
\caption{Interpretation of the Rayleigh-Sommerfeld diffraction integral
as an input-output relationship of a linear space-invariant system with 
impulse response $h_z(u)$.}
\label{fig:fig_2_rev}
\end{figure}

For now, we focus on the case of a purely monochromatic wave, postponing 
the extension to polychromatic waves to Subsection~IV-\ref{sec:polyf}.
In the monochromatic case, the scalar field can be expressed in complex form as 
(see, e.g.,  \cite{Madhow})
\be
\widetilde{\E}(z,x;t) = \Re 
\left\{\E(z,x) \, e^{ j 2 \pi f_0 t} \right\},
\label{eq:rf}
\ee 
where $\Re\{\cdot\}$ denotes the real part, $f_0 >0$ is the carrier frequency,  
$\lambda_0=\tfrac{c}{f_0}$ is the corresponding wavelength in vacuum, 
and the phasor function $\E(z,x)$ represents the complex spatial envelope of the wave.
By substituting \eqref{eq:rf} in \eqref{eq:max}, it follows that $\E(z,x)$ satisfies the 
time-independent, source-free \textit{Helmholtz equation}
\be
\frac{\partial^2 \,\E(z,x)}{\partial x^2} + \frac{\partial^2 \, \E(z,x)}{\partial z^2} + k_0^2 \, \E(z,x) = 0,
\label{eq:Helm}
\ee
where $k_0 = \tfrac{2 \pi}{\lambda_0}$ denotes the wave number.
The field $\E(z,x)$ at an observation point $\bm{P}=(z,x)$  can be computed
using Green's theorem \cite{Goodman}, which
forms the foundation of scalar diffraction theory.
Let $\E_\text a(\xi) \eqdef \E(0,\xi)$ denote the {\em aperture} field distribution.  
Assuming Sommerfeld's outgoing radiation condition,
the solution $\E(z,x)$ to the Helmholtz equation \eqref{eq:Helm} for $z>0$ can be expressed in terms of the aperture field distribution $\E_\text{a}(x)$ using the
{\em Rayleigh-Sommerfeld diffraction integral} \cite{Orfanidis-2002}, which reads as 
\be
\E(z,x) = -2 \, \frac{\partial}{\partial z} \int_{x_\text{a}^{(1)}}^{x_\text{a}^{(2)}}
\E_\text a(\xi) \, \text{G}_2(z,x-\xi) \, \mathrm{d}\xi,
\label{eq:Ray}
\ee
where $\text{G}_2(z,x-\xi)$ represents the 2-D Green's function  
\be
\text{G}_2(z,x-\xi) = -\frac{j}{4} \, \text{H}^{(2)}_0\left(k_0 \sqrt{(x-\xi)^2+z^2}\right).
\label{eq:Green}
\ee 
Here, $\text{H}^{(2)}_0(\cdot)$ denotes the zeroth-order second-kind Hankel function \cite{Abram},
and $\xi$ is a transverse integration variable representing a coordinate on the aperture plane.
Substituting \eqref{eq:Green} into \eqref{eq:Ray}, and applying the differentiation identity for the Hankel function of the second kind, i.e., 
\be
\frac{\mathrm{d}}{\mathrm{d}u} \text{H}^{(2)}_0(u) = - \text{H}^{(2)}_1(u), 
\ee
where $\text{H}^{(2)}_1(\cdot)$ is the first-order second-kind Hankel function,
we obtain for $z>0$
\be
\E(z,x) =  \frac{1}{2 j} \int_{x_\text{a}^{(1)}}^{x_\text{a}^{(2)}} 
\frac{k_0 \, z}{\rho(\xi)} \, \E_\text a(\xi) \, \text{H}^{(2)}_1(k_0 \, \rho(\xi)) \, \mathrm{d}\xi,
\label{eq:Ray-2}
\ee
where $\rho(\xi) \eqdef \sqrt{(x-\xi)^2+z^2}$ is the distance between the observation point $\bm{P}=(z,x)$ and the 
infinitesimally small point source located at $(0,\xi)$ on the aperture
(see Fig.~\ref{fig:fig_1_rev}).
Equation \eqref{eq:Ray-2} can be interpreted as a convolution integral. 
To make this perspective explicit, we rewrite it as
\be
\E(z,x) = \int_{x_\text{a}^{(1)}}^{x_\text{a}^{(2)}} \E_\text a(\xi) \, h_z(x-\xi) \, \mathrm{d}\xi,
\label{eq:Ray-2-conv}
\ee
where the impulse response $h_z(u)$ is defined as
\be
h_z(u) = \frac{k_0 \, z}{2 j} \frac{\text{H}^{(2)}_1\left(k_0 \sqrt{u^2+z^2}\right)}{\sqrt{u^2+z^2}},
\ee
for a given $z>0$. This formulation reveals that  the propagation channel from the transmit 
aperture to any observation point $\bm{P}$ can be modeled by an LSI 
system, fully characterized by the impulse response $h_z(u)$. This
interpretation is illustrated in Fig.~\ref{fig:fig_2_rev}.
The linearity of the propagation channel arises from the assumed linearity of the medium, 
as stated earlier, whereas its space invariance follows from the assumption that the medium is homogeneous. 

Additional approximations can be applied to simplify \eqref{eq:Ray-2}
into a more tractable form. Specifically, 
when the observation point $\bm{P}$ is located far enough from the radiating aperture, 
such that the transverse coordinate difference satisfies the 
{\em Fresnel approximation}:
\be
|x-\xi | \ll z, 
\label{eq:fre}
\ee 
the distance $\rho(\xi)$ can be approximated by
\be
\rho(\xi) = \sqrt{(x-\xi)^2 + z^2} \approx z + \frac{(x-\xi)^2}{2 \, z},
\label{eq:dist}
\ee
where the first-order Maclaurin expansion $\sqrt{1+x} \approx 1+\tfrac{x}{2}$, for $x \ll 1$, has been applied.
The Fresnel approximation is valid when the observation point lies close to the propagation axis $z$, a condition
also referred to as {\em paraxial approximation} in geometrical optics.
Under this assumption, the error introduced by approximating $\rho(\xi)$ with $z$ in the denominator of \eqref{eq:Ray-2} is typically small and can be neglected. 

In contrast, for the argument of the Hankel function in \eqref{eq:Ray-2}, 
the approximation $\rho(\xi) \approx z$ is generally not acceptable, because the wave number
$k_0$ can be very large, and even small changes in $\rho(\xi)$ can lead to significant variations.
Hence, the more accurate approximation \eqref{eq:dist} should be retained. Accordingly, in the paraxial regime, the expression for the field simplifies to
\be
\E(z,x) =  \frac{k_0}{2 j} \int_{x_\text{a}^{(1)}}^{x_\text{a}^{(2)}} \E_\text a(\xi) \, \text{H}^{(2)}_1\left(k_0 \, z + \frac{k_0 \, (x-\xi)^2}{2 \, z} \right) \, \mathrm{d}\xi,
\label{eq:Ray-2-par}
\ee
for $z>0$. A further simplification of \eqref{eq:Ray-2-par} can be achieved  
by using the asymptotic approximation of the Hankel function \cite{Orfanidis-2002} for large arguments
\be
H^{(2)}_1(u) \approx -\sqrt{\frac{2}{j \pi u}} \, e^{-j u} \:.
\label{eq:approx-hank}
\ee
This  approximation is valid and accurate for $u\gtrsim2$, which is generally satisfied in practical scenarios where $k_0 \, \rho(\xi)\gg 1$.
Using \eqref{eq:approx-hank} in \eqref{eq:Ray-2-par}, 
the Rayleigh-Sommerfeld diffraction integral
simplifies to the well-known {\em Huygens-Fresnel diffraction formula}
\begin{multline}
\E(z,x) = \sqrt{\frac{j}{\lambda_0 z}} \, e^{-j k_0 z} \int_{x_\text{a}^{(1)}}^{x_\text{a}^{(2)}} \E_\text a(\xi) \, e^{-j \frac{k_0}{2 z} (x-\xi)^2} \, \mathrm{d}\xi  \\
= \sqrt{\frac{j}{\lambda_0 z}} \, e^{-j k_0 z} \, e^{-j \frac{k_0}{2 z}x^2} \int_{x_\text{a}^{(1)}}^{x_\text{a}^{(2)}} 
\E_\text a(\xi) \, e^{-j \frac{k_0}{2 z} \xi^2}  \, e^{j\frac{k_0}{ z}  x \xi} \, \mathrm{d}\xi, 
\label{eq:int}
\end{multline}
where the ``obliquity''
factor ${z}/{\rho(\xi)}$ has again been approximated with $1$. 
When the approximation \eqref{eq:int} holds, the observation point $\bm P$ is said to lie in the {\em region of
Fresnel diffraction}, also referred to as the {\em radiative near field} of the transmitting aperture \cite{Goodman}.
More precisely, by comparing \eqref{eq:Ray-2} and \eqref{eq:int}, it is evident that 
the Huygens-Fresnel approximation effectively  replaces the Hankel function 
with parabolic wavelets, represented by quadratic-phase exponential terms.
The first line of \eqref{eq:int} shows that propagation within 
the radiative near-field region can be still modeled as an LSI system, with the corresponding impulse
response given by
\be
h_z(u) = \sqrt{\frac{j}{\lambda_0 z}} \, e^{-j k_0 z} \, \, e^{-j \frac{k_0}{2 z} u^2} \:.
\ee
The second line of \eqref{eq:int} offers an alternative interpretation. Up 
to multiplicative constants, the Huygens-Fresnel integral can be regarded as 
the Fourier transform of the product between the 
aperture field distribution and 
a quadratic phase factor.
This highlights the key role of spatial frequency modulation in radiative near-field diffraction and beam shaping.

Beyond the Huygens-Fresnel approximation, a stronger simplification can be made 
by noting that when
\be
\frac{k_0}{2 \, z} \, \xi^2 \ll 1 \quad \forall \xi \in [x_\text{a}^{(1)},x_\text{a}^{(2)}] 
\, \Leftrightarrow \, 
z \gg z_\text{F} \eqdef \frac{\pi}{\lambda_0} \left(x_\text{a}^\text{max}\right)^2,
\label{eq:frau}
\ee
with $x_\text{a}^\text{max} \eqdef \max\{|x_\text{a}^{(1)}|, |x_\text{a}^{(2)}|\}$, 
the quadratic phase term in the integral becomes negligible, i.e., $e^{-j k \frac{\xi^2}{2 z}} \approx 1$.
Under this condition,  the integral in \eqref{eq:int} reduces to the Fourier transform of 
the aperture field $\E_\text{a}(x)$, aside from multiplicative phase terms that appear outside the integral. The condition
in \eqref{eq:frau} defines the {\em Fraunhofer approximation} \cite{Goodman}, which describes the far-field regime of diffraction,
and $z_\text{F}$ is referred to as the Fraunhofer distance.
Thus, when condition \eqref{eq:frau} holds,  the field reduces to
\begin{multline}
\E(z,x) 
\approx \sqrt{\frac{j k_0}{2 \pi z}} \, e^{-jk_0z} \, e^{-j \frac{k_0}{2 z}x^2} 
\int_{x_\text{a}^{(1)}}^{x_\text{a}^{(2)}}  \E_\text a(\xi) \, e^{j\frac{k_0}{ z}  x \xi} \, \mathrm{d}\xi, 
\label{eq:far-field}
\end{multline}
and the observation point $\bm{P}$ is said to be in the {\em far-field} region of the 
transmit aperture.\footnote{A common antenna-design rule of thumb adopts the more stringent $\pi/8$ phase-error threshold, usually expressed as
$z > 2\,D^{2}/\lambda_0$, where $D$ denotes the full aperture extent.}
At first glance, no impulse response 
can be directly associated with Fraunhofer diffraction. This is because the approximation in \eqref{eq:frau} eliminates the quadratic phase term that ensures
shift invariance in the Huygens-Fresnel formulation.
The transition from the near field to the far field is gradual; 
hence, no sharp boundary exists between these regions. Accordingly, 
several metrics have been recently proposed to quantify this 
transition~\cite{Sanguinetti,Schober}. In addition, unified channel models 
have been developed to characterize propagation across the 
far- and near-field regimes~\cite{Wang.2025}.
In our 
work, we adopt the \emph{phase-error perspective}~\cite{Selvan} 
to delineate the near-field and far-field regimes, starting from 
the general phase factor of the field diffracted by a planar aperture.
However, it should be emphasized that the fixed phase-error threshold 
underlying~\eqref{eq:frau} -- which, by construction, corresponds to a maximum residual quadratic phase of one radian across the aperture at 
$z=z_\text{F}$ -- is a conventional criterion inherited from classical antenna theory, rather than a physically fundamental boundary~\cite{Schober}: by construction, it does not quantify how the residual wavefront curvature translates into link-level performance.
For this
reason, recent communication-oriented studies advocate
performance-defined boundaries in place of $z_{\mathrm{F}}$. In
\cite{CuiDai2024}, the \emph{effective Rayleigh distance} is defined
as the distance beyond which far-field (planar-wavefront) beamforming
incurs a gain loss smaller than a prescribed threshold; at boresight,
it equals $0.367\, z_{\mathrm{F}}$ for a $5\%$ gain-loss threshold,
i.e., it is significantly shorter than the Fraunhofer distance. Along
similar lines, it has been shown that the classical Fraunhofer and
Fresnel formulas require nontrivial corrections when applied to
phased arrays \cite{Monemi2024}, and recent sub-THz experiments have
found that the favorable properties of near-field beams, such as
focusing and self-healing, are observed only within a subset of the
nominal Fresnel region, referred to as the effective near-field
range \cite{Kludze2026}. Consistent with this view, in this article
the Fraunhofer distance $z_{\mathrm{F}}$ in \eqref{eq:frau} is retained only as
a convenient reference and normalization for the radiative near-field
region, whereas the operationally meaningful ranges of Airy beams are
characterized directly through performance-defined quantities,
namely, the diffraction-resisting distances derived in Section~\ref{sec:airy-free}
[see \eqref{eq:cond-df-apo-zc} and \eqref{eq:xeff}] and the 
link-budget corner distance of Section~IV-\ref{sec:link} [see \eqref{eq:Zcorner}].

In the following, we assume that the receiver is located within the radiative near-field region,
at a distance of several wavelengths from the transmitting aperture. This ensures that 
surface waves and near-field coupling effects are negligible.

\section{Diffraction-resistant and self-accelerating beams in free space}
\label{sec:self-acc-beam}

In the radiative near-field region, wavefronts are no longer 
constrained to be locally planar and may assume a variety of 
spatial structures. This additional geometric freedom enables 
the synthesis of structured beams through suitable shaping of 
the aperture field distribution. Among these, particular 
interest is devoted to beams whose intensity maxima evolve 
along curved caustics in free space, as summarized in 
Table~\ref{Table1}.
Despite this apparent curvature, it is well established 
\cite{Jackson-2007} that, in a homogeneous medium and in the 
absence of external potentials, the centroid of the beam's 
intensity propagates along a straight-line trajectory. This 
behavior reflects conservation of transverse momentum and is 
consistent with Ehrenfest's theorem \cite{Messiah-2014}. As 
discussed in Section~I, the observed transverse 
``acceleration'' of certain beams does not imply a global 
change in momentum, but rather results from structured 
interference within the wavefront. This phenomenon is 
commonly referred to as {\em self-acceleration}.
Before analyzing the principles underlying self-accelerating 
beams, we introduce the notion of {\em intensity} for a 
scalar monochromatic field. Let $\E(z,x)$ denote the complex 
envelope of the field at the point $\bm{P}=(z,x)$. The 
intensity is defined as
\be
I(z,x) \eqdef \left|\E(z,x)\right|^2 \: .
\label{eq:int-cost}
\ee
Within the scalar, monochromatic approximation adopted here, 
the intensity is proportional to the time-averaged power 
density and characterizes the spatial distribution of EM 
energy.
A scalar monochromatic wave field is said to be 
{\em propagation-invariant} if its transverse complex profile 
remains unchanged upon propagation, up to a longitudinal 
phase factor. More generally, a beam is said to be 
{\em diffraction-resistant} or {\em shape-preserving} if its 
transverse intensity satisfies the condition
\be
I(z,x) = I(0,x-\delta_z), 
\label{eq:diff-free}
\ee
where $\delta_z$ denotes a possible transverse shift of the 
intensity distribution during propagation. When $\delta_z=0$, 
the beam is strictly propagation-invariant in a fixed 
transverse coordinate frame. When $\delta_z\neq 0$, the beam 
preserves its functional form in a co-moving accelerating 
frame, while its intensity maximum follows a curved trajectory.
Diffraction-resistant propagation is particularly attractive 
for wireless communications, as maintaining spatial energy 
concentration over distance can enhance the received 
signal-to-noise ratio (SNR) in radiative near-field links. 
The beam families discussed in Table~\ref{Table1} differ 
precisely in how they achieve, or fail to achieve, this 
property under realistic aperture constraints.
In the following subsections, we present the fundamental 
principles underlying paraxial and nonparaxial 
self-accelerating beams, with particular emphasis on the 
aperture phase profiles that give rise to diffraction-resistant 
caustic solutions.

\subsection{Paraxial regime}
\label{eq:parax}

For paraxial self-accelerating beams, the analysis begins with the 
Huygens-Fresnel diffraction formula \eqref{eq:int}, which is valid under 
the paraxial approximation given in \eqref{eq:fre}-\eqref{eq:dist}.
Let us assume that the aperture field distribution
is composed of a magnitude $A_\text{a}(\xi)$ 
modulated by a phase term $\Phi_\text a(\xi)$, i.e., 
$\E_\text{a}(\xi)= A_\text{a}(\xi) \, e^{-j \Phi_\text a(\xi)}$.
Substituting this into ~\eqref{eq:int}, the field becomes 
\be
\E(z,x) 
= \sqrt{\frac{j}{\lambda_0 z}} \, e^{-j k_0 z} \int_{x_\text{a}^{(1)}}^{x_\text{a}^{(2)}} 
A_\text{a}(\xi) \, e^{-j Q_\xi(z,x)} \, \mathrm{d}\xi,  
\label{eq:int-2}
\ee
where the total phase of the integrand is given by
\be
Q_\xi(z,x) \eqdef \Phi_\text a(\xi)+
\frac{k_0}{2 z} (x-\xi)^2.
\label{eq:Q}
\ee
Under the assumptions of
first- and second-order phase-rate dominance (see Appendix~\ref{app:PSP} for details), 
a simple approximate solution to the integral in \eqref{eq:int-2} can be derived
using the stationary-phase approximation \cite{Wong-2001}. 
According to this method, the integral contribution to the field is negligible across most of the interval 
$[x_\text{a}^{(1)},x_\text{a}^{(2)}]$, except near points 
where the derivative of the phase function with respect to $\xi$, denoted as $\dot{Q}_\xi(z,x)$,
is zero. These points are referred to as {\em stationary phase points}. 
As detailed in Appendix~\ref{app:PSP}, 
the contribution of the integrand near each stationary phase
point can be approximated using the second derivative of
the phase,  $\ddot{Q}_\xi(z,x)$, which governs the local curvature of the phase. 
Explicitly, the condition for phase stationarity 
$\dot{Q}_\xi(z,x)=0$ yields
\be
x= \xi + z \, \frac{\dot{\Phi}_\text a(\xi)}{k_0}\:, 
\quad \text{for $\xi \in [x_\text{a}^{(1)},x_\text{a}^{(2)}]$} \: .
\label{eq:rays}
\ee
This equation describes a family of rays, each parameterized 
by the transverse coordinate $\xi$ 
on the aperture. Specifically, a ray originating from
point $\xi$ on the aperture follows a straight trajectory
in the $(x,z)$ plane with slope $\tfrac{\dot{\Phi}_\text a(\xi)}{k_0}$, 
contributing to the field at the observation point $\bm{P}$.
The envelope of the family of curves described by \eqref{eq:rays}
defines a {\em caustic}, which is the curve 
formed by the envelope
of infinitely close rays originating from different points
on the aperture.
Regions near the caustic exhibit strong constructive interference, and the
maxima of the wave intensity tend to follow its shape.
The caustic can be determined by solving a system of equations, as stated  
in the following proposition (see Appendix~\ref{app:prop-1} for the proof).

\vspace{2mm}
\begin{proposition}
\label{prop:1}
The envelope of the family of rays in \eqref{eq:rays}, known as the caustic, is obtained 
by eliminating the parameter $\xi$ from the following system of equations:
\be
\begin{cases}
x_\text{c}= \xi + z_\text{c} \, \frac{\dot{\Phi}_\text a(\xi)}{k_0} \:,
\vspace{1mm} \\ 
0= 1+ z_\text{c} \, \frac{\ddot{\Phi}_\text a(\xi)}{k_0} \:,
\end{cases}
\label{eq:syst}
\ee
where the subscript ``c'' denotes quantities evaluated along the caustic.
\end{proposition}

\begin{figure}[t]
\centering
\includegraphics[width=0.8\linewidth]{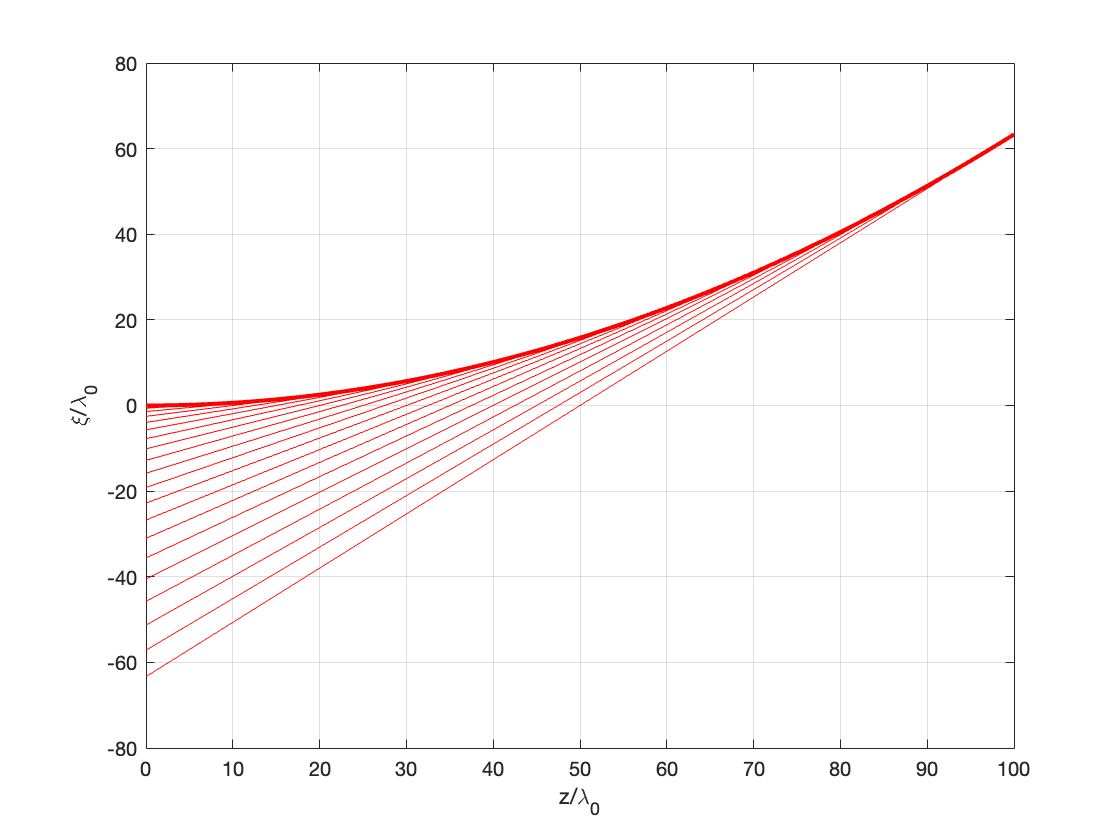} 
\caption{A geometrical construction of a caustic from a family of rays.}
\label{fig:fig_3_rev}
\end{figure}

\vspace{2mm}
We note that, since $z_\text{c}>0$ by assumption, the second equation in \eqref{eq:syst} implies that $\ddot{\Phi}_\text a(\xi_\text{c})=-k_0/z_\text{c}<0$. Hence, local ray convergence leading to caustic formation in the forward propagation direction requires $\ddot{\Phi}_\text a(\xi_\text{c})<0$.
This condition ensures that neighboring rays converge to form an envelope rather than diverge. 
The geometric construction of a caustic is illustrated in Fig.~\ref{fig:fig_3_rev}. 
The family of rays described by \eqref{eq:rays}
provides a systematic framework for synthesizing paraxial self-accelerating beams that follow prescribed caustic trajectories, enabling control over the path of peak intensity during propagation.

Not all caustics are expected to yield diffraction-resistant beams.
To understand this, it is useful to analytically evaluate
the field in the vicinity of the caustic for a general
phase profile of the aperture field distribution. 
This analysis is particularly insightful   
for self-accelerating beams, as the region 
near the caustic typically contains the highest concentration of energy.
To this end, we consider the third-order Taylor  expansion
of the phase function $Q_\xi(z,x)$ around the point $(\xi_\text{c},x_\text{c})$  
\cite{Chest-1957,Chre-2013}, where
$(z_\text{c},x_\text{c})$ is a fixed point on the caustic, and
$\xi_\text{c}$ is the unique solution of 
\be
x_\text{c}= \xi + z_\text{c} \, \frac{\dot{\Phi}_\text a(\xi)}{k_0}\:, 
\quad \text{for $\xi \in [x_\text{a}^{(1)},x_\text{a}^{(2)}]$} \: .
\ee
This yields 
\begin{multline}
Q_{\xi}(z_{\text{c}},x) \approx Q_{\xi_\text{c}}(z_\text{c},x_\text{c}) + \Omega_\text{c}(x-x_\text{c})
\\ + \ddot{\Phi}_\text{a}(\xi_\text{c}) \, 
(\xi-\xi_\text{c}) \, (x-x_\text{c}) + \frac{1}{6} \, \dddot{\Phi}_\text{a}(\xi_\text{c}) \, (\xi-\xi_\text{c})^3,
\label{eq:Q-Taylor}
\end{multline}
where 
\be
\Omega_\text{c}(x-x_\text{c}) \eqdef \frac{k_0}{z_\text{c}} (x-x_\text{c}) \left[ x_\text{c}-\xi_\text{c}  
+ \frac{1}{2} (x-x_\text{c})\right], 
\ee
and the third derivative of
the transmit phase is assumed nonzero at the caustic, i.e., 
$\dddot{\Phi}_\text{a}(\xi_\text{c}) \neq 0$. Moreover, the identity 
$\ddot{Q}_{\xi_\text{c}}(z_\text{c},x_\text{c})=\ddot{\Phi}_\text{a}(\xi_\text{c})+\frac{k_0}{z_\text{c}}=0$ is utilized, which follows directly from the second condition in the caustic-defining system \eqref{eq:syst}.
By substituting the Taylor expansion \eqref{eq:Q-Taylor} into \eqref{eq:int-2}, and assuming 
an unbounded aperture, i.e., $x_\text{a}^{(1)}=-\infty$ and  $x_\text{a}^{(2)}=+\infty$, the analysis simplifies considerably under the
assumption that the  amplitude $A_\text{a}(\xi)$  varies slowly
in a small neighborhood around $\xi_\text{c}$, i.e., 
$A_\text{a}(\xi) \approx A_\text{a}(\xi_\text{c})$ for $|\xi-\xi_\text{c}|\ll 1$. 
With these assumptions, and after lengthy but straightforward
manipulations, we obtain the following closed-form approximation for the field near the caustic:
\begin{multline}
\E(z_c,x) 
\approx \sqrt{2 \, \pi} \, e^{j \frac{\pi}{4}} \, e^{-j k_0 z_\text{c}} \, e^{-j Q_{\xi_\text{c}}(z_\text{c},x_\text{c})} \, e^{-j \Omega_\text{c}(x-x_\text{c})}
\\
\cdot \frac{A_\text{a}(\xi_\text{c}) \, \sqrt{-\ddot{\Phi}_\text a(\xi_\text{c})}}{\left[\frac{\dddot{\Phi}_\text{a}(\xi_\text{c})}{2}\right]^{\frac{1}{3}}}
\, \Ai\left(\frac{\ddot{\Phi}_\text a(\xi_\text{c}) \, (x-x_\text{c})}{\left[\frac{\dddot{\Phi}_\text{a}(\xi_\text{c})}{2}\right]^\frac{1}{3}}\right).
\label{eq:int-2-caustic}
\end{multline}
Here, 
\be
\Ai(u) \eqdef \frac{1}{2 \, \pi} \int_{-\infty}^{+\infty} e^{j \left(\frac{v^3}{3}+u v\right)} \, \mathrm{d}v
\label{eq_fun-airy}
\ee
denotes the  Airy function \cite{Vallee-2004} and, using 
the second equation in \eqref{eq:syst}, we have applied the identity 
$z_\text{c} = -\tfrac{k_0}{\ddot{\Phi}_\text a(\xi_\text{c})}$. 
Under the cubic expansion assumption in \eqref{eq:Q-Taylor} and for a nonvanishing third derivative $\dddot{\Phi}_\text a(\xi_\text{c})$, the field near the caustic locally assumes an Airy-type profile that varies linearly with $x-x_\text{c}$.
However, the 
expansion in \eqref{eq:int-2-caustic} 
also reveals that the beam  intensity depends on
$\xi_\text{c}$, which is itself a function of the propagation distance $z_\text{c}$. 
Consequently, paraxial self-accelerating beams designed to follow prescribed caustic trajectories do not, in general, satisfy the diffraction-resistant condition in \eqref{eq:diff-free}, since the prefactor in \eqref{eq:int-2-caustic} depends on $\xi_\text{c}$ and therefore varies with the propagation distance $z_\text{c}$.
An important exception arises when the aperture amplitude satisfies
\be
A_\text{a}(\xi_\text{c}) \sim \sqrt{-\ddot{\Phi}_\text a(\xi_\text{c})},
\label{eq:A-par}
\ee
where $\sim$ denotes proportionality. 
This condition removes the explicit $\xi_\text{c}$-dependence of the prefactor in \eqref{eq:int-2-caustic}. 
Requiring further that the remaining $\xi_\text{c}$-dependence vanish leads to a scaling relation between the second and third derivatives of the aperture phase, namely
\be
\ddot{\Phi}_\text a(\xi_\text{c}) \sim 
\left[\dddot{\Phi}_\text{a}(\xi_\text{c})\right]^{1/3}.
\label{eq:nldiffeq}
\ee
Up to multiplicative constants, this relation defines a nonlinear differential equation for $\Phi_\text a(\xi)$ whose general solution is
\be
\Phi_\text a(\xi)= c_0+c_1 \xi^{3/2} + c_2 \xi,
\label{eq:phi-nondif}
\ee
where $c_0$, $c_1$, and $c_2$ are real constants. 
Under the assumptions of an infinite transverse aperture and validity of the cubic expansion, these conditions ensure that the resulting beam satisfies the diffraction-resistant condition in \eqref{eq:diff-free}.
It follows from Proposition~\ref{prop:1} that the caustic
associated with the aperture phase \eqref{eq:phi-nondif} traces out a {\em parabolic} trajectory, i.e., 
\be
x_\text{c}= a \, z_\text{c}^2 + b \, z_\text{c} + c,
\ee
where $a$, $b$, and $c$ are real-valued constants.
This result is consistent with the formal proof in \cite{Bandres-2009}, which shows that, under the assumptions of a 1-D transverse coordinate, the paraxial approximation, and an infinite transverse aperture, diffraction-resistant propagation of self-accelerating beams occurs only along parabolic trajectories.
Such parabolic caustics
can be obtained by synthesizing  an Airy-type aperture field distribution, as will be discussed in
detail in Section~\ref{sec:airy-free}.

\subsection{Nonparaxial regime}
\label{eq:nonparax}

In the case of nonparaxial self-accelerating beams, the starting point is the
Rayleigh-Sommerfeld diffraction integral \eqref{eq:Ray-2}. By substituting the aperture field as $\E_\text{a}(\xi)= A_\text{a}(\xi) \, e^{-j \Phi_\text a(\xi)}$
and applying the asymptotic approximation for the Hankel function given in \eqref{eq:approx-hank}, the field can be approximated as
\be
\E(z,x) = \sqrt{\frac{j}{\lambda_0}} \, z\int_{x_\text{a}^{(1)}}^{x_\text{a}^{(2)}} 
\frac{A_\text{a}(\xi)}{\sqrt{\rho^3(\xi)}} \, e^{-j P_\xi(z,x)} \, \mathrm{d}\xi,  
\label{eq:Ray-2-approx}
\ee
where
\be
P_\xi(z,x) \eqdef \Phi_\text a(\xi)+ k_0 \sqrt{(x-\xi)^2+z^2}.
\label{eq:P}
\ee
This approximation is valid in the regime where $\rho(\xi) \gg \lambda_0$. 
Following the same reasoning as in Subsection~III-\ref{eq:parax}, 
we apply again the stationary phase method to
the integral in \eqref{eq:Ray-2-approx}.  
The condition for phase stationarity 
$\dot{P}_\xi(z,x)=0$, where the derivative is taken with respect to $\xi$, yields
\be
x= \xi + \sqrt{(x-\xi)^2+z^2} \, \frac{\dot{\Phi}_\text a(\xi)}{k_0}\:, 
\quad \text{for $\xi \in [x_\text{a}^{(1)},x_\text{a}^{(2)}]$}, 
\ee
which simplifies, after straightforward algebra, to
\be
x= \xi + z \, \frac{\dot{\Phi}_\text a(\xi)}{\sqrt{k_0^2-[\dot{\Phi}_\text a(\xi)]^2}}\:, 
\quad \text{for $\xi \in [x_\text{a}^{(1)},x_\text{a}^{(2)}]$}  \:.
\label{eq:rays-nonparax}
\ee
Equation~\eqref{eq:rays-nonparax} is the nonparaxial counterpart of 
\eqref{eq:rays}. In this case,  the ray originating from point $\xi$ 
on the aperture and contributing to the field at 
$\bm{P}$ follows a linear trajectory with slope 
$\tfrac{\dot{\Phi}_\text a(\xi)}{\sqrt{k_0^2-[\dot{\Phi}_\text a(\xi)]^2}}$.
The caustic corresponding to the family of rays in \eqref{eq:rays-nonparax}
can be derived as follows.

\vspace{2mm}
\begin{proposition}
\label{prop:2}
The envelope of the family of rays described by \eqref{eq:rays-nonparax} is obtained 
by eliminating the parameter $\xi$ from the system of equations
\be
\begin{cases}
x_\text{c}= \xi + z_\text{c} \, \frac{\dot{\Phi}_\text a(\xi)}{\sqrt{k_0^2-[\dot{\Phi}_\text a(\xi)]^2}}\:;
\vspace{1mm} \\
0= 1+ z_\text{c} \, \frac{k_0^2 \, \ddot{\Phi}_\text a(\xi)}{\left\{k_0^2-[\dot{\Phi}_\text a(\xi)]^2\right\}^\frac{3}{2}} \:.
\end{cases}
\label{eq:syst-2}
\ee
\end{proposition}
It follows directly from Proposition~\ref{prop:1}, with the paraxial approximation replaced by its  nonparaxial counterpart.

\vspace{2mm}
Let us consider the problem of synthesizing a nonparaxial
self-accelerating beam that follows a prescribed convex
trajectory of the form $x_\text{c}=g(z_\text{c})$.
For each $\xi \in [x_\text{a}^{(1)},x_\text{a}^{(2)}]$, 
the ray described by \eqref{eq:rays-nonparax} must be 
tangent to the curve $x_\text{c}=g(z_\text{c})$. This condition requires that
the slope of the ray equals the derivative of $g(z_\text{c})$ with respect to $z_\text{c}$, i.e., 
\be
\frac{\dot{\Phi}_\text a(\xi)}{\sqrt{k_0^2-[\dot{\Phi}_\text a(\xi)]^2}} = \dot{g}(z_\text{c}),
\label{eq:phase-g}
\ee
where, according to the first equation in \eqref{eq:syst-2}, the propagation coordinate $z_\text{c}$ is
implicitly defined by
\be
x_\text{c}=g(z_\text{c}) = \xi + z_\text{c} \, \dot{g}(z_\text{c}) \:.
\label{eq:g-z}
\ee
In \eqref{eq:phase-g}--\eqref{eq:g-z},
we have omitted the explicit dependence of 
$z_\text{c}$ on $\xi$ for notational convenience. 
Rearranging \eqref{eq:phase-g}, the required phase gradient at the aperture becomes 
\be
\dot{\Phi}_\text a(\xi) = k_0 \frac{\dot{g}(z_\text{c})}{\sqrt{1+[\dot{g}(z_\text{c})]^2}} \:.
\label{eq:phi}
\ee
This expression establishes the link between the desired beam trajectory and 
the  initial phase profile on the aperture.
It can be used to compute  $\Phi_\text a(\xi)$ either analytically or numerically  for any given function $g(z_\text{c})$.

As shown in \cite{Efremidis.2018}, under the assumptions
$x_\text{a}^{(1)}=-\infty$, $x_\text{a}^{(2)}=+\infty$, and
$A_\text{a}(\xi) \approx A_\text{a}(\xi_\text{c})$ for $|\xi-\xi_\text{c}|\ll 1$,
where $\xi_\text{c}$ is the unique solution of the equation
\be
x_\text{c}= \xi + z_\text{c} \, \frac{\dot{\Phi}_\text a(\xi)}{\sqrt{k_0^2-[\dot{\Phi}_\text a(\xi)]^2}}, 
\label{eq:rays-nonparax-c}
\ee
the local behavior of the field in \eqref{eq:Ray-2-approx} near the caustic depends
on the propagation distance $z_\text{c}$ only through the local curvature of the trajectory,
defined as 
\be
\kappa(z_\text{c}) \eqdef 
\frac{\left|\ddot{g}(z_\text{c})\right|}{\left\{1+\left[\dot{g}(z_\text{c}) \right]^2\right\}^{\tfrac{3}{2}}} = 
\frac{\left|\ddot{g}(z_\text{c})\right|}{\left[1+\frac{(x_\text{c}-\xi_\text{c})^2}{z_\text{c}^2}\right]^{\tfrac{3}{2}}},
\label{eq:curv}
\ee
where the second expression follows directly from \eqref{eq:g-z}.
In the paraxial limit, i.e., for
$|x_\text{c}-\xi_\text{c}| \ll z_\text{c}$, the curvature simplifies to 
$\kappa(z_\text{c}) \approx |\ddot{g}(z_\text{c})|$.

Within the present planar $(x,z)$ formulation, constant 
curvature $\kappa(z_\text{c})=\kappa_0$ corresponds to 
circular caustics in the nonparaxial regime. Nonparaxial 
accelerating beams following circular arcs have been 
demonstrated both theoretically and experimentally 
in~\cite{Kam.2012,Cour.2012,Zhang.2012}, and closed-form 
phase expressions for circular and elliptic trajectories 
have been derived in~\cite{Penciu2015}. More general 
accelerating beams in three spatial dimensions may follow 
nonplanar space curves, such as helical or spiraling 
trajectories, which require a fully three-dimensional 
treatment and lie beyond the scope of the present 
two-dimensional scalar analysis.
The above framework describes the synthesis of nonparaxial 
accelerating beams with prescribed curved caustics. It is 
important to distinguish such accelerating solutions from 
propagation-invariant beams, whose key differences are 
summarized in Table~\ref{Table1}. Exact propagation-invariant 
solutions of the Helmholtz equation arise from separability 
in specific coordinate systems, such as cylindrical 
coordinates for Bessel beams and elliptic coordinates for 
Mathieu beams, and maintain a straight propagation axis. 
In contrast, nonparaxial accelerating beams, such as Weber 
beams~\cite{Bandres}, exhibit curvature of the intensity 
maximum without preserving invariance in a fixed transverse 
coordinate frame.

In the remainder of this work, we restrict attention to 
the paraxial regime, where the analysis simplifies 
considerably and parabolic caustics naturally emerge as 
the unique class of diffraction-resistant accelerating 
solutions, as formally established in~\cite{Bandres-2009}.

\section{Self-accelerating paraxial Airy beams in free space}
\label{sec:airy-free}

A notable example is the case of {\em ideal} Airy beams \cite{Siviloglou-2007, Kaganovsky-2010, Morandotti-2012, Efremidis-2019},
generated by setting 
the aperture field equal to the  Airy function defined in \eqref{eq_fun-airy}, i.e., 
\be
\E_\text a(\xi) =\Ai(\xi) 
\label{eq_Ea-airy},
\ee
for $\xi \in \mathbb{R}$. This function satisfies the Airy equation 
\be
\frac{\mathrm{d}^2}{\mathrm{d} \xi^2}  E(\xi) - \xi \, E(\xi)=0, 
\label{eq:eq-Airy}
\ee
with the boundary condition $E(\xi) \to 0$ as $\xi \to + \infty$.
The term ``ideal'' refers to the fact that
the Airy function is not square integrable over $\mathbb{R}$, implying that the resulting beam 
has infinite energy when the aperture spans the entire real axis, i.e., $x_\text{a}^{(1)}=-\infty$ and 
$x_\text{a}^{(2)}=+\infty$ in \eqref{eq:int-2}.
For $|\xi| \gg 1$,  the Airy function admits the following asymptotic expansions \cite{Vallee-2004}
\be
\Ai(\xi) \sim \begin{cases}
\frac{\cos\left[\frac{\pi}{4}-\frac{2}{3} \, (-\xi)^\frac{3}{2}\right]}{\sqrt{\pi} \, (-\xi)^\frac{1}{4}}\:, & \text{for $\xi<0$}, 
\vspace{1mm} \\ 
\frac{e^{-\frac{2}{3} \xi^\frac{3}{2}}}
{2 \, \sqrt{\pi} \, \xi^\frac{1}{4}}\:, & \text{for $\xi>0$},
\end{cases}
\label{eq:expansions}
\ee
which shows that $\Ai(\xi)$ oscillates for $\xi<0$, with
a discrete set of real negative zeros,  and it decays exponentially for $\xi>0$.
Numerical verification confirms that the approximation in \eqref{eq:expansions}
for negative arguments is already accurate for $\xi \lesssim -2$.
Given this behavior, contributions from $\xi>0$ are often neglected in asymptotic, ray-based analyses because $\Ai(\xi)$ decays exponentially for $\xi>0$; however, $\Ai(\xi)$ is not identically zero on $\xi>0$. For $\xi<0$, the function can be expressed in terms of its modulus $M(\xi)$ and  phase $\Theta(\xi)$:
\barr
\Ai(\xi) & = M(\xi) \, \cos[\Theta(\xi)] 
\nonumber \\ &  = \frac{M(\xi)}{2} \, e^{j \Theta(\xi)} + 
\frac{M(\xi)}{2} \, e^{-j \Theta(\xi)}, 
\label{eq:A-split}
\earr
with the approximations
\barr
M(\xi) & \approx \frac{1}{\sqrt{\pi} (-\xi)^\frac{1}{4}},
\label{eq:mod-esp}
\\
\Theta(\xi) & \approx \frac{\pi}{4} - \frac{2}{3} \, (-\xi)^\frac{3}{2} \:.
\label{eq:phase-esp}
\earr
Notably, the phase of the aperture field approximately follows a power law.

By substituting \eqref{eq:A-split} into \eqref{eq:int-2}, 
the radiated beam $\E(z,x)$ splits into the sum of two contributions.
The first corresponds to an aperture field with 
$A_\text{a}(\xi) = \tfrac{M(\xi)}{2}$ and 
$\Phi_\text a(\xi)=-\Theta(\xi) $; the second has the same amplitude but  phase  $\Phi_\text a(\xi)=\Theta(\xi)$.
From direct differentiation, it follows that $\dot{\Theta}(\xi)=\sqrt{-\xi} >0$ and  
$\ddot{\Theta}(\xi)=-\frac{1}{2 \sqrt{-\xi}} <0$ for $\xi <0$. Consequently, only the second term contributes to the formation of the caustic for
$z_\text{c}>0$, since it satisfies the necessary condition 
$\ddot{\Phi}_\text a(\xi)<0$ in the second equation of \eqref{eq:syst}. In contrast, the first term does not generate a 
forward caustic and thus does not contribute to the maximum of the field strength. 
By substituting the derivatives $\dot{\Phi}_\text a(\xi)=\dot{\Theta}(\xi)$
and $\ddot{\Phi}_\text a(\xi)=\ddot{\Theta}(\xi)$
into the system 
\eqref{eq:syst}, and eliminating the parameter $\xi$, the caustic trajectory
\be
x_\text{c}= \frac{z_\text{c}^2}{4 \, k_0^2}
\label{eq:caustic-airy}
\ee
is obtained, which describes a convex {\em parabola}. 
The parabolic coefficient in \eqref{eq:caustic-airy} scales as $1/k_0^2$, i.e., $x_\text c=(4k_0^2)^{-1}z_\text c^2$. Thus, increasing $k_0$ reduces the transverse displacement $x_\text c$ attained at a given $z_\text c$, yielding a less strongly deflected (``flatter'') trajectory. Note that the geometric curvature of a parabola is not constant and varies with $z_\text c$.
Moreover, inserting $\dot{\Phi}_\text a(\xi)=\sqrt{-\xi} >0$
into \eqref{eq:rays} and solving for $\xi <0$ leads
to a quadratic equation, whose discriminant $\Delta_c= z_\text{c}^2-4 \, k_0^2 \, x_\text{c}$
determines the number and nature of the stationary-phase points contributing to the field at $(z_\text{c},x_\text{c})$.
This discriminant vanishes exactly on the caustic curve in \eqref{eq:caustic-airy}, indicating 
a transition between single- and double-valued ray contributions, corresponding to 
$\Delta_c=0$ (i.e., one stationary point) and $\Delta_c>0$ 
(i.e., two distinct stationary points), respectively.

By applying the Huygens-Fresnel integral \eqref{eq:int} with aperture limits $x_\text{a}^{(1)}=-\infty$ and $x_\text{a}^{(2)}=+\infty$,  the ideal Airy beam
radiated by an infinite-sized aperture is given by \cite{Berry-1979}
\be
\E(z,x)=u(z,x) \, e^{-j k_0 z},
\label{eq:field-decomp}
\ee
for $z>0$, where 
\begin{multline}
u(z,x) = \sqrt{\frac{j}{\lambda_0 z}} \int_{-\infty}^{+\infty} \Ai(\xi) \, e^{-j \frac{k_0}{2 z} (x-\xi)^2} \, \mathrm{d}\xi  \\
= \Ai \left(x-\frac{z^2}{4 \, k_0^2}\right) \, e^{-j \left(\frac{ x \, z}{2 k_0} -\frac{z^3}{12 k_0^3}\right)} \:.
\label{eq:Airy}
\end{multline}

It follows from \eqref{eq:Airy} that the Airy field satisfies the diffraction-resistant (shape-preserving) condition in \eqref{eq:diff-free}, with transverse shift $\delta_z=\tfrac{z^2}{4k_0^2}$. Moreover, evaluating \eqref{eq:Airy} on the caustic $x=x_\text c(z)$ yields a constant on-caustic intensity, namely $I(z_\text c,x_\text c(z_\text c))=\Ai^2(0)\approx 0.1260$.
The fact that the Airy beam maintains its transverse profile during propagation, 
without spreading due to diffraction, is also confirmed
by the observation that the amplitude and phase expressions in \eqref{eq:mod-esp}-\eqref{eq:phase-esp} 
satisfy the diffraction-resistant conditions in \eqref{eq:A-par}-\eqref{eq:nldiffeq}.
Within the 1-D paraxial framework, and under the standard definition of shape-preserving accelerating solutions, the Airy beam plays a canonical role. In particular, \cite{Unn-1996} established a uniqueness result under specific admissibility conditions.

\begin{figure}[t]
\centering
\includegraphics[width=\linewidth]{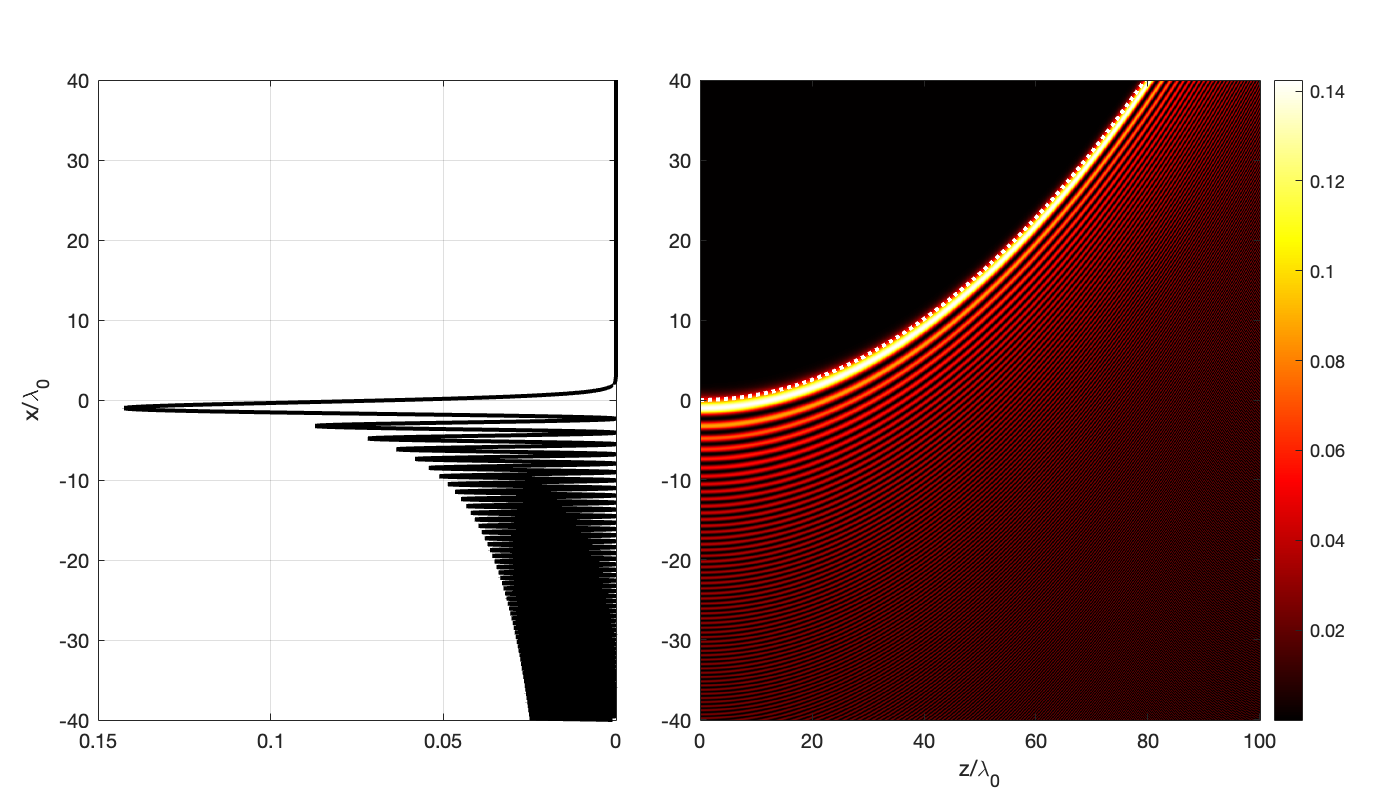} 
\caption{Left: Intensity of the aperture field in \eqref{eq_Ea-airy} as a function of the transverse coordinate. 
Right: Intensity distribution of the corresponding ideal, infinite-energy Airy beam as a function of $z$ and $x$.
The white dotted curve in the right-side panel represents the caustic trajectory in \eqref{eq:caustic-airy}. 
All spatial coordinates are normalized with respect to the wavelength $\lambda_0$.
}
\label{fig:fig_4_new}
\end{figure}

The left-side plot of Fig.~\ref{fig:fig_4_new} shows the
intensity of the  initial field distribution in \eqref{eq_Ea-airy}
as a function of the transverse coordinate $\xi$ on the  aperture.
The corresponding  Airy beam intensity, computed from \eqref{eq:Airy}, is illustrated in the 
right-side plot as a function of $(z,x)$. As expected, the Airy wave propagates indefinitely 
while maintaining constant transverse acceleration and  without exhibiting 
diffraction. This idealized behavior reflects the nondispersive nature of the underlying paraxial solution: the intensity profile is preserved in a co-moving transverse frame while the location of the main lobe follows a parabolic law.

The parabolic caustic of an ideal, infinite-energy Airy beam can be tailored by applying a spatial scaling at the aperture and 
adding a linear phase  \cite{Sivi-2008}.  
To this end, we generalize the aperture field in \eqref{eq_Ea-airy} as
\be
\E_\text a(\xi) =\Ai(\gamma_\text{a} \, \xi) \, e^{-j \nu_\text{a} \xi},
\label{eq_Ea-airy-gen}
\ee
where $\gamma_\text{a} >0$ is the spatial scaling factor and 
$\nu_\text{a}$ sets the initial
launch angle of the beam. The resulting aperture field 
is linearly stretched if $0< \gamma_\text{a} < 1$ and 
 compressed if $\gamma_\text{a} > 1$.\footnote{\label{note:1}In  
principle, a negative value of $\gamma_\text{a}$ is also possible.
In this case, the Airy profile is both scaled and spatially mirrored across the origin.
}
Substituting \eqref{eq_Ea-airy-gen} into 
\eqref{eq:int} and applying the method of 
stationary phase yields
the following family of rays
\be
x= \xi + z \, \frac{\gamma_\text{a} \, \dot{\Phi}_\text a(\gamma_\text{a} \, \xi) +\nu_\text{a}}{k_0}\:, 
\quad \text{for $\xi \in [x_\text{a}^{(1)},x_\text{a}^{(2)}]$}. 
\label{eq:rays-gen}
\ee
This generalizes  \eqref{eq:rays}, which is recovered 
by setting $\gamma_\text{a}=1$ and $\nu_\text{a}=0$.
The  caustic associated with the beam is the envelope of the ray family in \eqref{eq:rays-gen}, and can be determined via the following proposition
(the proof of which is omitted since it follows similar steps as 
Proposition~\ref{prop:1}). 

\vspace{2mm}
\begin{proposition}
\label{prop:3}
The envelope of the family of rays described by \eqref{eq:rays-gen} is obtained 
by eliminating the parameter $\xi$ from the system of equations
\be
\begin{cases}
x_\text{c} = \xi + z_\text{c} \, \frac{\gamma_\text{a} \, \dot{\Phi}_\text a(\gamma_\text{a} \, \xi) +\nu_\text{a}}{k_0} \:;
\vspace{1mm} \\
0= 1+ z_\text{c} \, \frac{\ddot{\Phi}_\text a(\gamma_\text{a} \, \xi)}{k_0} \:.
\end{cases}
\label{eq:syst-gen}
\ee
\end{proposition}
\begin{figure}[t]
\centering
\includegraphics[width=\linewidth]{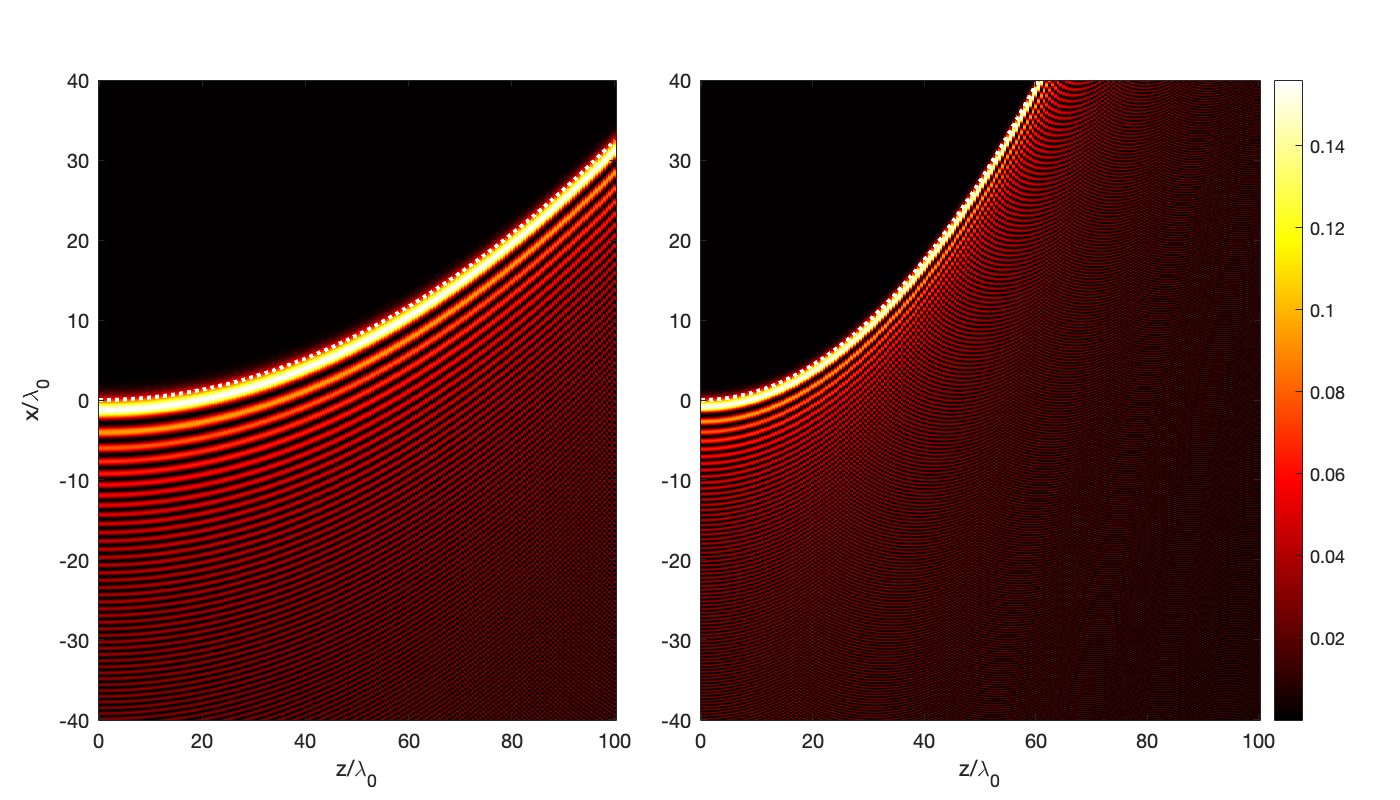} 
\caption{Intensity distribution of the infinite-energy Airy beam corresponding to
the aperture field in \eqref{eq_Ea-airy-gen} as a function of $z$ and $x$, for $\gamma_\text{a}=k_0/7$ (left plot) 
and $\gamma_\text{a}=k_0/5$ (right plot), with $\nu_\text{a}=0$.
The white dotted curve represents the caustic trajectory in \eqref{eq:caustic-airy-gen}. 
All spatial coordinates are normalized with respect to the wavelength $\lambda_0$.
}
\label{fig:fig_5_rev}
\end{figure}

\vspace{2mm}
Let us assume that $\E_\text a(\xi) \approx 0$ for $\xi >0$.
Using the asymptotic expansion in \eqref{eq:phase-esp}, it follows that, 
for $\xi <0$, $\dot{\Phi}_\text a(\gamma_\text{a} \, \xi)=\gamma_\text{a} \sqrt{-\gamma_\text{a} \, \xi} >0$ and  
$\ddot{\Phi}_\text a(\gamma_\text{a} \, \xi)=-\frac{\gamma_\text{a}^2}{2 \sqrt{-\gamma_\text{a} \, \xi}} <0$.
Substituting these expressions into the system in  
\eqref{eq:syst-gen} and eliminating the parameter $\xi$, we obtain 
\be
x_\text{c}= \frac{\gamma_\text{a}^3 \, z_\text{c}^2}{4 \, k_0^2} + \frac{\nu_\text{a} \, z_\text{c}}{k_0},
\label{eq:caustic-airy-gen}
\ee
which generalizes the convex parabolic trajectory \eqref{eq:caustic-airy}.\footnote{A concave 
parabolic caustic may be obtained by choosing negative values of $\gamma_\text{a}$ (see also footnote~\ref{note:1}).}
In particular, the spatial scaling factor $\gamma_\text{a}$ controls 
the width of the parabola: As $\gamma_\text{a}$ increases, the caustic becomes narrower, and vice versa. 
\begin{figure}[t]
\centering
\includegraphics[width=\linewidth]{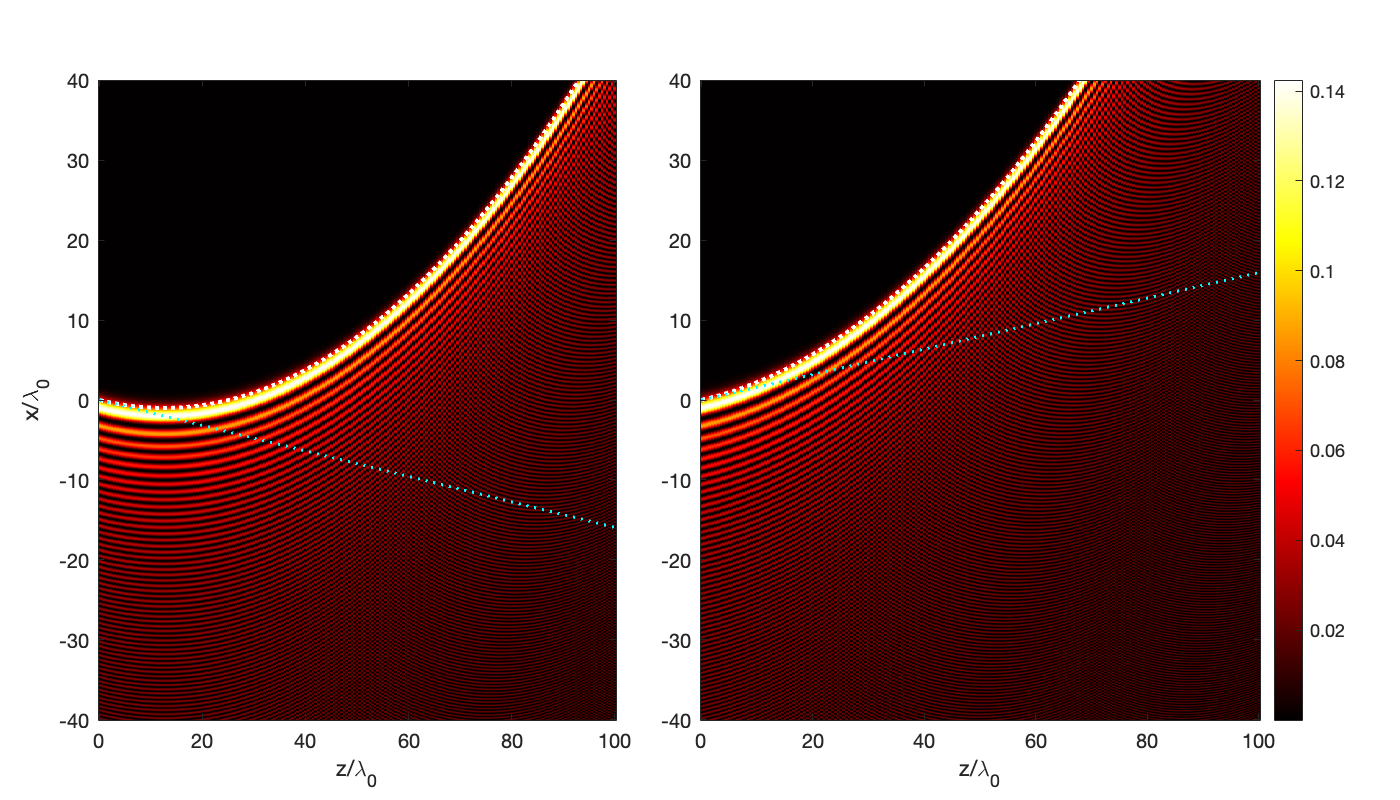} 
\caption{Intensity distribution of the infinite-energy Airy beam corresponding to
the aperture field in \eqref{eq_Ea-airy-gen} as a function of $z$ and $x$, for $\nu_\text{a}=-1/\lambda_0$ (left plot) and 
$\nu_\text{a}=1/\lambda_0$ (right plot), with $\gamma_\text{a}=k_0/6$.
The white dotted curve represents the caustic trajectory in \eqref{eq:caustic-airy-gen},
whereas the blue dotted line indicates the tangent to the caustic at the point $(0,0)$.
All  spatial coordinates are normalized with respect to the wavelength $\lambda_0$.
}
\label{fig:fig_6_rev}
\end{figure}
On the other hand, the linear phase parameter $\nu_\text{a}$ acts as a \emph{beam steering} control: 
from~\eqref{eq:caustic-airy-gen}, it is seen that $\nu_\text{a}$ 
shifts the initial launch angle of the parabolic caustic without 
altering its curvature or the self-acceleration property. 
Specifically, the slope of the tangent to the caustic at the 
aperture plane is $\tfrac{\nu_\text{a}}{k_0}$, so that varying $\nu_\text{a}$ 
steers the beam toward a desired angular direction while 
preserving all other propagation characteristics.

By substituting \eqref{eq_Ea-airy-gen} into \eqref{eq:int}, assuming $x_\text{a}^{(1)}=-\infty$ and $x_\text{a}^{(2)}=+\infty$,
and evaluating the integral, the resulting beam still takes the form \eqref{eq:field-decomp}, where 
\begin{multline}
u(z,x)  =  \Ai \left(\gamma_\text{a} \, x-\frac{z^2 \, \gamma_\text{a}^4}{4 \, k_0^2}- \frac{z \, \gamma_\text{a} \, \nu_\text{a}}{k_0}\right) 
\, e^{-j \phi(z,x)},
\label{eq:Airy-gen}
\end{multline}
with
\be
\phi(z,x) \eqdef \nu_\text{a} \, x +
\frac{z \, \gamma_\text{a}^2}{2 \, k_0} \left(\gamma_\text{a} \, x -\frac{z^2 \, \gamma_\text{a}^4}{6 \, k_0^2} - 
 \frac{z \, \gamma_\text{a} \, \nu_\text{a}}{k_0} - \frac{\nu_\text{a}^2}{\gamma_\text{a}^2}\right)  \:.
\ee
From the definition \eqref{eq:diff-free}, it follows that the spatial scaling and 
the additional linear phase preserve the diffraction-resistant property of infinite-energy Airy beams.

Fig.~\ref{fig:fig_5_rev} shows the intensity of the infinite-energy Airy beam for two different values of the spatial scaling factor 
$\gamma_\text{a}$, with $\nu_\text{a}=0$. As expected,
the width of the caustic depends on $\gamma_\text{a}$: Larger values produce narrower trajectories, while smaller values yield wider ones. Since $\nu_\text{a}=0$, the launch angle is zero, meaning the caustic is tangent to the longitudinal axis at the origin.
Fig.~\ref{fig:fig_6_rev} illustrates the effect of 
varying $\nu_\text{a}$, while keeping
$\gamma_\text{a}=k_0/6$. 
For $\nu_\text{a}<0$ (left plot), the beam is initially launched downwards, with 
the tangent to the caustic at the origin lying in the negative half $x$-plane.
When $\nu_\text{a}>0$ (right plot), the launch is upwards, with the caustic initially directed toward the positive $x$-half-plane. Although both $\gamma_\text{a}$ and $\nu_\text{a}$ are design parameters, their values 
must obey the paraxial condition in \eqref{eq:fre}.

In phased-array or metasurface implementations, 
the parameters $\nu_\text{a}$ and $\gamma_\text{a}$ can in 
principle be made \emph{time-varying}, i.e., $\nu_\text{a}(t)$ 
and $\gamma_\text{a}(t)$, by dynamically updating the phase 
profile of the aperture elements. This enables \emph{adaptive 
beam steering and curvature control}: varying $\nu_\text{a}(t)$ 
redirects the caustic trajectory toward a moving receiver, 
while varying $\gamma_\text{a}(t)$ adapts the parabolic curvature 
and diffraction-resisting range to the instantaneous propagation 
environment. Such dynamic control is directly compatible with 
standard reconfigurable intelligent surface (RIS) control 
mechanisms~\cite{Malevich.2025}, in which the phase of each 
aperture element can be updated on a per-slot basis. 
The trajectory-adaptive beam shaping approaches 
of~\cite{Ye_ArXiV2025} and~\cite{Yao_2026} rely on similar 
principles of dynamic phase profile adjustment, and the 
wave-optics framework developed here provides the 
analytical foundation for quantifying the resulting beam 
performance under varying obstruction geometries, as 
analyzed in Section~\ref{sec:self-healing}.

\subsection{Exponentially modulated finite-energy Airy beams}
\label{sec:expo}

The Airy beam in \eqref{eq:Airy} inherently exhibits infinite power flux across any plane orthogonal to 
its propagation direction \cite{Zamboni-2012}, rendering it physically unrealizable. 
To overcome this issue, a finite-energy Airy beam can be synthesized by modulating the ideal, 
infinite-energy Airy beam with an exponentially decaying envelope at the aperture. This approach remains valid even when the aperture is unbounded, i.e.,
$x_\text{a}^{(1)}=-\infty$ and $x_\text{a}^{(2)}=+\infty$.
The resulting aperture field 
takes the form
\be
\E_\text a(\xi) = U_\text{a} \, \Ai(\gamma_\text{a} \, \xi) \, \, e^{\alpha_\text{a} \xi} \, e^{-j \nu_\text{a} \xi},
\label{eq_Ea}
\ee
where $U_\text{a}>0$ is a normalization constant, and $\alpha_\text{a} > 0$ is an {\em apodization} parameter, which suppresses the divergent Airy tail and ensures finite energy. 
By invoking Parseval's theorem, the total energy of the exponentially modulated Airy field defined in \eqref{eq_Ea} can be computed as
\be
\en_\text{a} \eqdef \int_{-\infty}^{+\infty} |\E_\text a(\xi)|^2 \, \mathrm{d}\xi = \frac{U_\text{a}^2 }{\sqrt{8 \, \pi \, \alpha_\text{a} \, \gamma_\text{a}}}\, 
e^{\frac{2}{3} \left(\frac{\alpha_\text{a}}{\gamma_\text{a}}\right)^3} \:.
\label{eq:Ea-new}
\ee
In what follows, we normalize the energy of the aperture field to unity, i.e., $\en_\text{a} =1$, which yields 
\be
U_\text{a} = (8 \, \pi \, \alpha_\text{a} \gamma_\text{a})^\frac{1}{4} \, e^{-\frac{1}{3} \left(\frac{\alpha_\text{a}}{\gamma_\text{a}}\right)^3}\: .
\label{eq:U0}
\ee

By substituting \eqref{eq_Ea} into \eqref{eq:int} and evaluating the resulting integral along a suitable  path in the complex plane \cite{Vallee-2004}, the propagated 
beam can again be expressed 
in the form \eqref{eq:field-decomp}, where the field envelope now becomes
\begin{multline}
u(z,x) = U_\text{a} \, \Ai \left(\gamma_\text{a} \, x-\frac{z^2 \, \gamma_\text{a}^4}{4 \, k_0^2}- \frac{z \, \gamma_\text{a} \, \nu_\text{a}}{k_0}
- j \frac{z \, \gamma_\text{a} \, \alpha_\text{a}}{k_0} \right)  \\
\cdot e^{\alpha_\text{a} \left( x-\frac{z^2 \, \gamma_\text{a}^3}{2 \, k_0^2}-\frac{z \, \nu_\text{a}}{k_0}\right)}  \, e^{-j \phi(z,x)},
\label{eq:airy-def}
\end{multline}
and the accumulated phase during propagation is given by
\begin{multline}
\phi(z,x) \eqdef \nu_\text{a} \, x +
\frac{z \, \gamma_\text{a}^2}{2 \, k_0} \\ \cdot \left(\gamma_\text{a} \, x -\frac{z^2 \, \gamma_\text{a}^4}{6 \, k_0^2} 
- 
 \frac{z \, \gamma_\text{a} \, \nu_\text{a}}{k_0} 
  + \frac{\alpha_\text{a}^2}{\gamma_\text{a}^2}-\frac{\nu_\text{a}^2}{\gamma_\text{a}^2} \right). 
\end{multline}

The EM fields in \eqref{eq:Airy} and \eqref{eq:Airy-gen} are retrieved from \eqref{eq:airy-def} as special cases.
We note that the argument of the Airy function in \eqref{eq:airy-def}
is complex, and therefore the function itself also takes complex values.

At this point, two important observations are in order. 
First, since the exponential modulation is performed using a real-valued exponential function, it
does not contribute to the phase term
of the Huygens-Fresnel integral \eqref{eq:int}. Consequently, the stationary phase points remain unaffected, and the 
caustic of the exponentially modulated finite-energy Airy beam is still 
described by the parabolic trajectory in \eqref{eq:caustic-airy-gen}.
Second, the exponential modulation in \eqref{eq_Ea}
compromises the diffraction-resistant nature of the beam. In other words, 
condition \eqref{eq:diff-free} is no longer  strictly satisfied.
This affects the 
maximum range over which the beam can maintain its shape and intensity, which is a critical parameter in practical applications involving long-distance propagation. 
To better understand this behavior, let us compute
the intensity of the beam \eqref{eq:airy-def} along its caustic. 
By substituting the expression of the caustic from \eqref{eq:caustic-airy-gen} into \eqref{eq:airy-def}
and evaluating the squared magnitude of the resulting field $u(z_\text{c},x_\text{c})$, we obtain
\be
I(z_\text{c}) = U_\text{a}^2 \, \left|\Ai\left(- j \frac{z_\text{c} \, \gamma_\text{a} \, \alpha_\text{a}}{k_0}\right)\right|^2  
\, e^{-\frac{z_\text{c}^2 \, \gamma_\text{a}^3 \, \alpha_\text{a}}{2 \, k_0^2}}
\: .
\label{eq:airy-def-new}
\ee
This expression shows that  the beam intensity along the caustic decreases as $z_\text{c}$ increases, confirming that finite-energy Airy beams do not preserve a strictly diffraction-resistant profile over arbitrarily large distances. 
However, over a finite range within the radiative near-field region, 
the beam can still display quasi-diffraction-resistant behavior in the sense described in \cite{Siviloglou-2007-2}. This occurs when the intensity $I(z_\text{c})$ remains approximately constant, which requires the
argument of the Airy function in \eqref{eq:airy-def-new}
to remain small and 
the exponential term in 
\eqref{eq:airy-def-new} to vary slowly. These conditions are satisfied when
\be
\frac{z_\text{c} \, \gamma_\text{a} \, \alpha_\text{a}}{k_0} =\epsilon \ll 1
\quad \text{and} \quad \frac{\epsilon^2 \, \gamma_\text{a}}{2 \, \alpha_a} \ll 1,
\label{eq:cond-df-apo}
\ee
which implies that  $I(z_\text{c}) \approx U_\text{a}^2 \, \Ai^2\left(0\right)$.
From a practical perspective, this means that the diffraction-resistant property of the finite-energy Airy beam is preserved up to a certain propagation distance, given approximately by 
\be
z_\text{c} \ll \frac{2 \pi}{\lambda_0 \, \gamma_\text{a}  \, \alpha_\text{a}}.
\label{eq:cond-df-apo-zc}
\ee
This upper bound 
becomes smaller 
as the apodization parameter $\alpha_\text{a}$ increases.
Therefore, the more the field is localized at the aperture, the shorter the range over which it retains its shape.

\begin{figure}[t]
\centering
\includegraphics[width=\linewidth]{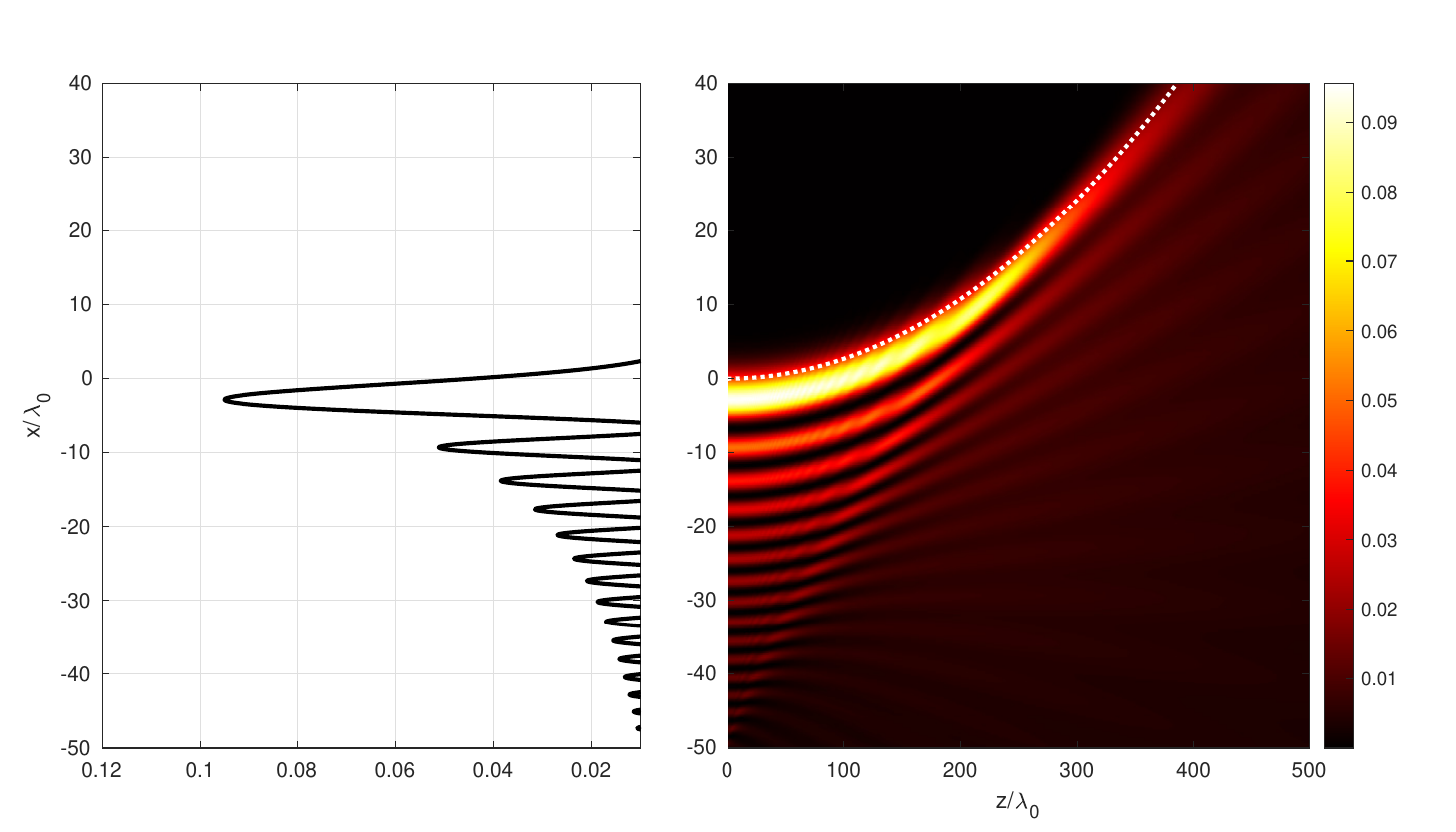} 
\caption{Left: Intensity of the aperture field in \eqref{eq_Ea} as a function of the transverse coordinate,
with $\nu_\text{a}=0$, $\alpha_\text{a}=0.01/\lambda_0$,  and $\gamma_\text{a}=\tfrac{k_0}{18}$. Right: 
Intensity distribution of the corresponding finite-energy Airy beam as a function of $z$ and $x$. 
The white dotted curve in the right-side panel represents the caustic trajectory in \eqref{eq:caustic-airy-gen}. 
All  spatial coordinates are normalized with respect to  the wavelength $\lambda_0$.}
\label{fig:fig_7_rev}
\end{figure}

Fig.~\ref{fig:fig_7_rev}  shows the intensity profile of the
aperture field \eqref{eq_Ea} (left-side plot), where the parameters are set to $\nu_\text{a}=0$, $\alpha_\text{a}=0.01/\lambda_0$, and $\gamma_\text{a}=\tfrac{k_0}{18}$.
The corresponding intensity distribution of the finite-energy Airy beam is displayed in the right-side plot.
Despite the exponential apodization, the beam still retains its characteristic behavior: It undergoes 
self-acceleration along a parabolic caustic and maintains 
a nearly diffraction-resistant profile over a propagation distance of nearly $200$ wavelengths.
The diffraction-resisting region can be extended by reducing the value of the apodization parameter $\alpha_\text{a}$.

\subsection{Effect of a finite-sized aperture}
\label{sec:apert}

An implicit assumption in the previous analysis of Airy waves
is the presence of an infinite transverse aperture located in the plane
$z=0$. This, however, is not physically feasible due to practical limitations in available space.
It is therefore crucial to 
examine how truncating the aperture affects the propagation characteristics of Airy beams. 
To this aim, we note that the aperture Airy field decays exponentially for $\xi>0$ 
[see eq.~\eqref{eq:expansions}], which motivates setting $x_\text{a}^{(2)}=0$. Accordingly, the effective aperture width along the $x$-axis becomes $\Delta x_\text{a} =- x_\text{a}^{(1)}$, which we 
denote by $x_\text{eff}>0$ for simplicity. 
The parameter $x_\text{eff}$ plays a key role in determining the total transmitted energy and the resulting  propagation behavior of the beam. 

To evaluate the energy of the source Airy field, for simplicity, 
we neglect the exponential 
modulation at the aperture by setting $\alpha_\text{a}=0$, 
and we assume a zero initial launch angle, 
i.e., $\nu_\text{a}=0$. Under these assumptions, the total transmitted energy across the finite aperture is given by
\begin{multline}
\en_\text{a}  = \int_{-x_\text{eff}}^{0} \Ai^2(\gamma_\text{a} \, \xi) \, \mathrm{d}\xi
= \frac{1}{\gamma_\text{a}} \int_{-\gamma_\text{a} \, x_\text{eff}}^{0} \Ai^2(v) \, \mathrm{d}v \\
= \frac{1}{\gamma_\text{a}} \left\{ 
\left[ v \, \Ai^2(v)\right]_{v=-\gamma_\text{a} x_\text{eff}}^{v=0} \right.
\\ \left.
- 2 \int_{-\gamma_\text{a} x_\text{eff}}^{0} v \, \Ai(v) \,  \dAi(v)\, \mathrm{d}v
\right\},
\end{multline}
where  we introduced the change of variable $v = \gamma_\text{a} \xi$, and
$\dAi(v)$ denotes the derivative of $\Ai(v)$.
Recalling that the Airy function satisfies the differential equation in \eqref{eq:eq-Airy}, we can express
the transmitted energy analytically  as
\barr
\en_\text{a} & = \frac{1}{\gamma_\text{a}} \Big[
\gamma_\text{a} \, x_\text{eff} \, \Ai^2(-\gamma_\text{a} \, x_\text{eff}) 
- 2 \int_{-\gamma_\text{a} \, x_\text{eff}}^{0} \dAi(v) \, \ddAi(v) \, \mathrm{d}v
\Big] 
\nonumber \\
& = \frac{1}{\gamma_\text{a}} \left\{
\gamma_\text{a} \, x_\text{eff} \, \Ai^2(-\gamma_\text{a} \, x_\text{eff}) - \left[\dAi^2(v) \right]_{v=-\gamma_\text{a} \, x_\text{eff}}^{v=0}
\right\} 
\nonumber \\
& = \frac{1}{\gamma_\text{a}} \left [
\gamma_\text{a} \, x_\text{eff} \, \Ai^2(-\gamma_\text{a} \, x_\text{eff}) - \dAi^2(0) + \dAi^2(-\gamma_\text{a} \, x_\text{eff}) \right],
\label{eq:pow_tx}
\earr
where $\ddAi(v)$ denotes the second derivative of $\Ai(v)$, and it follows that  
$\dAi(0) \approx -0.26$ \cite{Vallee-2004}.
The transmission energy depends on both the spatial scaling factor $\gamma_\text{a}$ and the
effective aperture width $x_\text{eff}$. In particular, for a fixed value of $\gamma_\text{a}$, it can 
be readily verified that $\en_\text{a}$  increases monotonically with $x_\text{eff}$. 
This behavior is consistent with  physical intuition: A larger aperture allows  
a greater portion of the Airy field to be transmitted, thereby capturing more energy.

\begin{figure}[t]
\centering
\includegraphics[width=\linewidth]{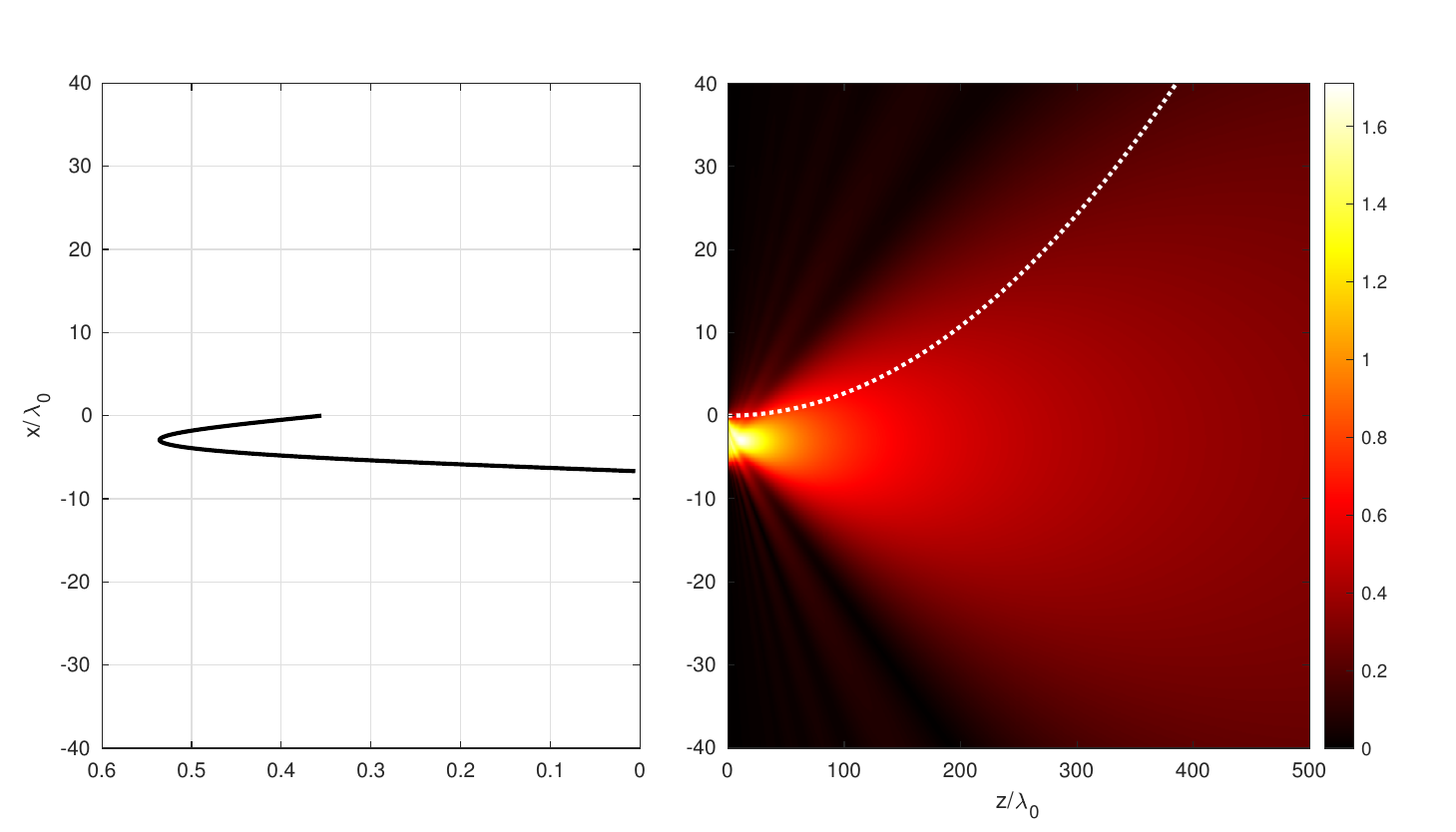} 
\caption{Left: Intensity of the aperture field in \eqref{eq_Ea} as a function of the transverse coordinate
with $\nu_\text{a}=\alpha_\text{a}=0$ and $\gamma_\text{a}=\tfrac{k_0}{18}$. Right:
Intensity distribution of the corresponding radiated beam as a function of $z$ and $x$,
when $x_\text{a}^{(1)}=\xi_1$ and $x_\text{a}^{(2)}=0$. 
The radiated beam is computed numerically by evaluating the Rayleigh-Sommerfeld diffraction integral in \eqref{eq:Ray-2}. 
The white dotted curve in the right-side panel represents the caustic trajectory in \eqref{eq:caustic-airy-gen}. 
All  spatial coordinates are normalized with respect to  the wavelength $\lambda_0$.}
\label{fig:fig_8_rev}
\end{figure}

A central question is how to select the effective aperture size to preserve, as far as possible, 
the main characteristics of the ideal Airy beam.
For $\xi<0$, the Airy function $\Ai(\gamma_\text{a} \, \xi)$ 
exhibits oscillatory behavior and vanishes at a discrete set of real,
negative zeros, whose locations are given, for $n \in \mathbb{N}$, by the
asymptotic expansion~\cite{Vallee-2004}
\be
\xi_n = -\frac{1}{\gamma_\text{a}} \, \left[ \frac{3 \pi}{2} \left(n-\frac{1}{4} \right)\right]^{2/3},
\ee
whose accuracy improves with $n$, yet is already within $0.8\%$ of the exact
value for $n=1$ and within $0.2\%$ for $n \ge 2$. In particular, the first
zero is located at $\xi_1 \simeq -2.338/\gamma_\text{a}$, which defines the
extent of the Airy function's main lobe in the negative $\xi$-domain.
In principle, one could set the aperture width equal to the span of the main lobe by choosing 
$x_\text{eff} = -\xi_1$.
However, as demonstrated in \cite{Kaganovsky-2010}, such a truncated field does not produce a diffraction-resistant or self-accelerating 
Airy beam. 
To corroborate the inadequacy of this choice, we present
in Fig.~\ref{fig:fig_8_rev} the intensity profile of 
the aperture field defined in \eqref{eq_Ea} (left-side plot), where $\nu_\text{a}=\alpha_\text{a}=0$ and $\gamma_\text{a}=\tfrac{k_0}{18}$.
The corresponding intensity of the resulting  beam is shown in the right-side plot, obtained numerically by evaluating the Rayleigh-Sommerfeld integral in 
\eqref{eq:Ray-2} with $x_\text{a}^{(1)}=\xi_1$ and $x_\text{a}^{(2)}=0$.
It is evident that when the aperture width is limited to  the span of the main lobe of the Airy
field, the radiated beam completely loses its hallmark characteristics, namely, self-accelerating 
trajectory and diffraction-resistant propagation.
The result of Fig.~\ref{fig:fig_8_rev} suggests that the secondary lobes
of the input Airy field on the negative $\xi$-axis play a key role in forming 
the curved beam trajectory.

Following a fundamentally different approach, we now employ caustic theory to analyze the propagation dynamics of Airy beams
generated by a finite-size aperture \cite{Dudley.2011}. 
Mathematically, caustic curves represent the envelopes of the family of rays.
In the case of a paraxial Airy beam, the caustic can be described in parametric form by the system
in \eqref{eq:syst-gen}.
According to this formulation, each point $(z_\text{c},x_\text{c})$
on the caustic is associated with the transverse coordinate $\xi$ from which the ray originates
at the aperture.
The relationship between the longitudinal coordinate $z_\text{c}$ and the aperture coordinate $\xi <0$ can be obtained by substituting 
the parabolic trajectory in \eqref{eq:caustic-airy-gen} into \eqref{eq:g-z}, yielding
\be
\xi= - \frac{\gamma_\text{a}^3 \, z_\text{c}^2}{4 \, k_0^2} \:.
\label{eq:xi-zc}
\ee
This expression highlights a fundamental link between aperture size and
propagation distance. Specifically, by assuming $x_\text{a}^{(2)}=0$, the beam's intensity 
characteristics remain essentially invariant up to a maximum propagation 
distance $z_\text{max}$ provided that 
\be
x_\text{eff} \ge  \frac{\gamma_\text{a}^3 \, z_\text{max}^2}{4 \, k_0^2} = 
\frac{\gamma_\text{a}^3 \, z_\text{max}^2 \, \lambda_0^2}{16 \, \pi^2} \:.
\label{eq:xeff}
\ee
This result shows that the effective aperture size must grow quadratically with 
the propagation distance in order to preserve the quasi-diffraction-resistant 
nature of the Airy beam, for given values of the spatial scaling factor 
$\gamma_\text{a}$ and the operating wavelength $\lambda_0$.
Equivalently, for fixed values of $z_\text{max}$ and 
$\gamma_\text{a}$, a shorter wavelength (i.e., a higher 
carrier frequency $f_0$) enables the same propagation range 
with a smaller effective aperture. This behavior follows 
directly from~\eqref{eq:xi-zc}: for a given propagation distance 
$z_c$ and spatial scaling factor $\gamma_\text{a}$, the magnitude 
of the aperture coordinate $\xi$ associated with the corresponding 
point on the caustic decreases as $k_0$ increases. Hence, a smaller 
portion of the transmit aperture is required to sustain the Airy 
trajectory up to $z_\text{max}$.
It is also noteworthy that the minimum required aperture size scales as the cube of the spatial scaling factor $\gamma_\text{a}$. 
Hence, for a fixed aperture size $x_\text{eff}$, decreasing $\gamma_\text{a}$ extends the maximum diffraction-resisting propagation range $z_\text{max}$. 
Physically, as follows from~\eqref{eq:caustic-airy-gen},
a smaller $\gamma_\text{a}$ produces a more weakly curved Airy trajectory, so that the aperture point contributing 
to the caustic moves toward the aperture boundary more slowly as the propagation distance increases. 
By contrast, increasing $\gamma_\text{a}$ results in a more strongly curved trajectory and causes the finite-aperture limitation to be reached at a shorter distance.
The parameter $\gamma_\text{a}$ also affects the spatial scale and, consequently, the spatial bandwidth of the aperture field. 
In particular, decreasing $\gamma_\text{a}$ stretches the Airy profile in the spatial domain, since its characteristic lobe 
spacing scales as $1/\gamma_\text{a}$, and correspondingly compresses its spectrum toward lower spatial frequencies. Therefore, smaller 
values of $\gamma_\text{a}$ lead to slower spatial variations of the aperture field and relax the spatial-sampling 
requirement. Accordingly, larger values of $\gamma_\text{a}$ compresses the Airy profile, broadens its spatial spectrum, and requires a finer 
spatial discretization to avoid aliasing.
A quantitative guideline for the spatial sampling step will be derived in Section~IV-\ref{sec:sampling}, where the 
spectral properties of the aperture field are analyzed in detail.

\begin{figure}[t]
\centering
\includegraphics[width=\linewidth]{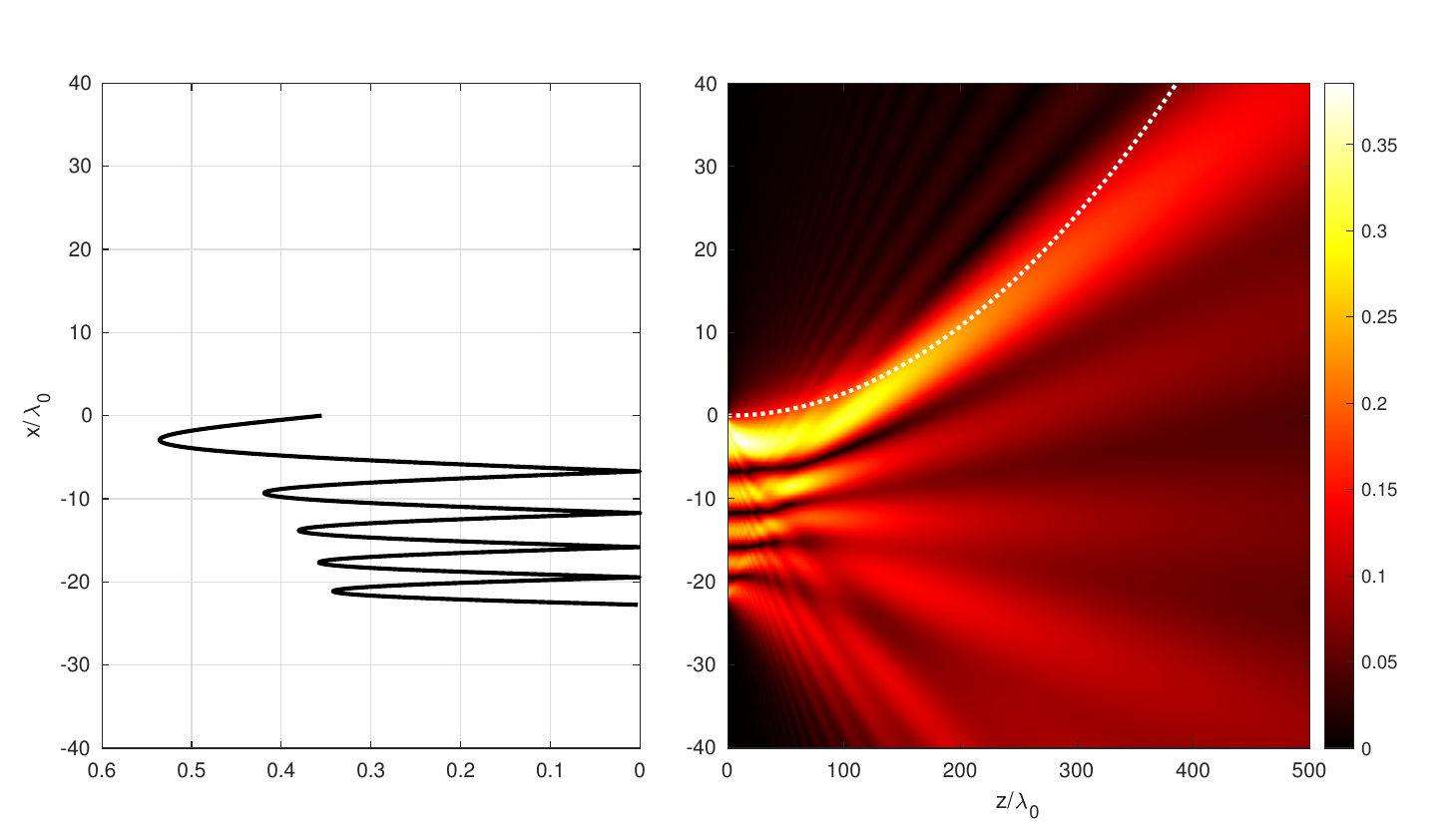} 
\caption{Left: Intensity of the aperture field in \eqref{eq_Ea} as a function of the transverse coordinate
with $\nu_\text{a}=\alpha_\text{a}=0$ and $\gamma_\text{a}=\tfrac{k_0}{18}$. Right:
Intensity distribution of the corresponding Airy beam as a function of $z$ and $x$,
when $x_\text{a}^{(1)}=-22.7 \, \lambda_0$ and $x_\text{a}^{(2)}=0$. 
The radiated beam is computed numerically by evaluating the Rayleigh-Sommerfeld diffraction integral in \eqref{eq:Ray-2}. 
The white dotted curve in the right-side panel represents the caustic trajectory in \eqref{eq:caustic-airy-gen}. 
All  spatial coordinates are normalized with respect to  the wavelength $\lambda_0$.}
\label{fig:fig_9_rev}
\end{figure}

\begin{figure}[t]
\centering
\includegraphics[width=\linewidth]{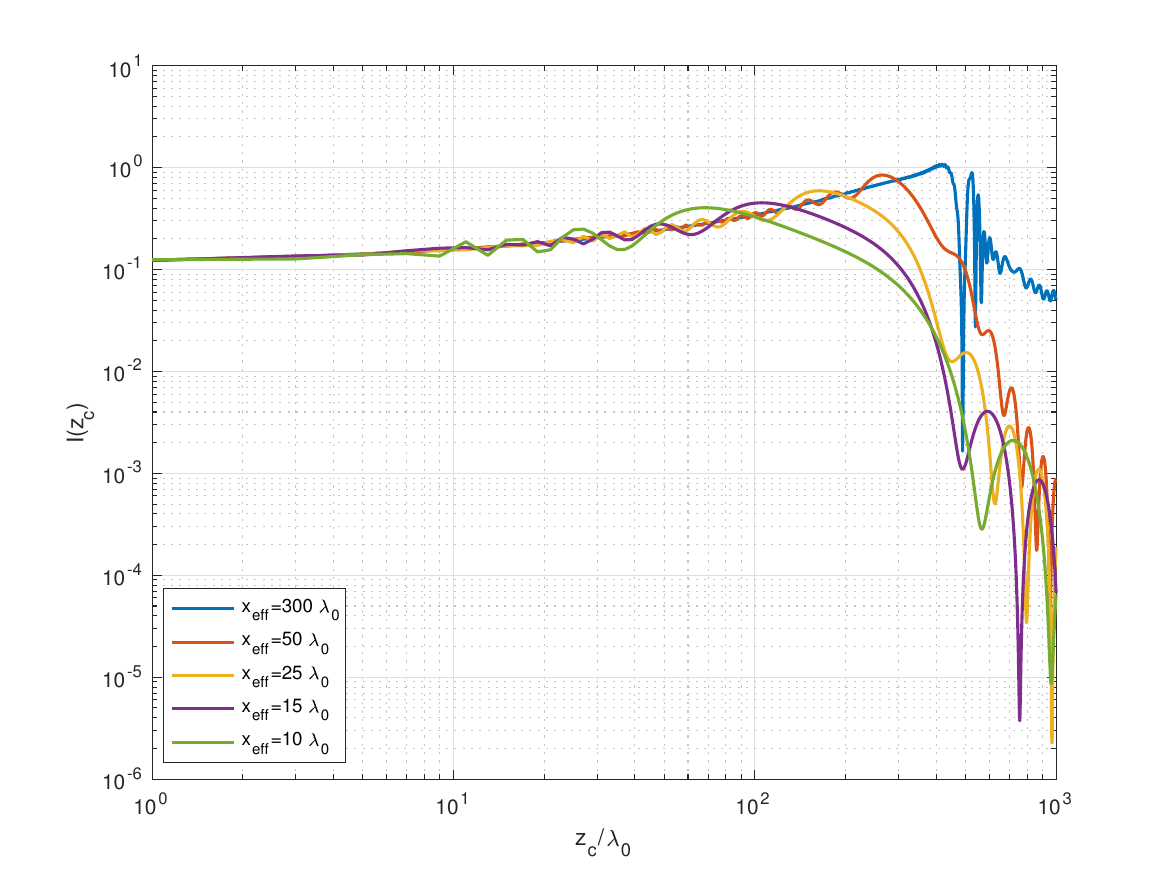} 
\caption{Intensity of the Airy beam along the caustic in \eqref{eq:caustic-airy-gen}
as a function of $z_\text{c}/\lambda_0$ for different values of the effective aperture size $x_\text{eff}$, 
with $\nu_\text{a}=\alpha_\text{a}=0$, $\gamma_\text{a}=\tfrac{k_0}{18}$, 
$x_\text{a}^{(1)}=-x_\text{eff}$, and $x_\text{a}^{(2)}=0$. 
The radiated beam is computed numerically by evaluating the Rayleigh-Sommerfeld diffraction integral in \eqref{eq:Ray-2}.}
\label{fig:fig_10_rev}
\end{figure}

Figure~\ref{fig:fig_9_rev} shows the intensity of 
the aperture field \eqref{eq_Ea} (left-side plot), with $\nu_\text{a}=\alpha_\text{a}=0$ and $\gamma_\text{a}=\tfrac{k_0}{18}$,
along with the corresponding intensity of the Airy beam (right-side plot), computed numerically from 
\eqref{eq:Ray-2} by setting $x_\text{a}^{(1)}=-22.7 \, \lambda_0$ and $x_\text{a}^{(2)}=0$.
According to \eqref{eq:frau}, the Fraunhofer distance is $z_\text{F} \approx 1600 \, \lambda_0$ in this setting.
The numerical results are in good agreement with the theoretical prediction from \eqref{eq:xeff}, which gives a maximum 
link distance of $z_\text{max} \approx 290 \, \lambda_0$ for an effective aperture length $x_\text{eff}=22.7 \, \lambda_0$. 
Beyond this distance, the  aperture truncation causes the beam to gradually lose its diffraction-resistant character and begin spreading.
This relationship between the maximum propagation distance and the effective aperture size is further illustrated in Fig.~\ref{fig:fig_10_rev}, where 
the intensity $I(z_\text{c})$ of the Airy beam along its parabolic caustic is plotted as a function of $z_\text{c}$, for different values
of $x_\text{eff}/\lambda_0 \in \{10, 15, 25, 50, 300\}$. 
The corresponding values of the Fraunhofer distance are 
$z_\text{F}/\lambda_0 \in \{314.1593, 706.8583, 1963.5, 7854, 282740\}$, respectively.
Each curve is obtained by evaluating
the Rayleigh-Sommerfeld diffraction integral in \eqref{eq:Ray-2}, substituting the caustic equation in \eqref{eq:caustic-airy-gen}, 
and computing the squared magnitude of the field  
for $x_\text{a}^{(1)}=-x_\text{eff}$ and $x_\text{a}^{(2)}=0$.
As predicted by \eqref{eq:xeff},
the field intensity along the caustic
remains nearly constant within the diffraction-resisting range,
validating the diffraction-resistant behavior of the beam up to a distance governed by
the effective aperture size $x_\text{eff}$. Beyond this range, the intensity starts to decrease, indicating the onset of diffraction effects due to aperture truncation.

It is worth noting that the maximum distance $z_{\max}$ in \eqref{eq:xeff}
achievable with a given effective aperture size $x_{\mathrm{eff}}$ is
substantially shorter than the Fraunhofer distance $z_{\mathrm{F}}$
associated with the same aperture (see also Fig.~\ref{fig:fig_10_rev}). The
diffraction-resisting and self-accelerating advantages of
finite-aperture Airy beams thus hold over a performance-defined
range, in full analogy with the effective Rayleigh distance
\cite{CuiDai2024} and the effective near-field range
\cite{Kludze2026} characterized in the communications literature for
focused near-field beams (see the discussion in Section~\ref{sec:Diff-theory}).

\begin{figure}[t]
\centering
\includegraphics[width=\linewidth]{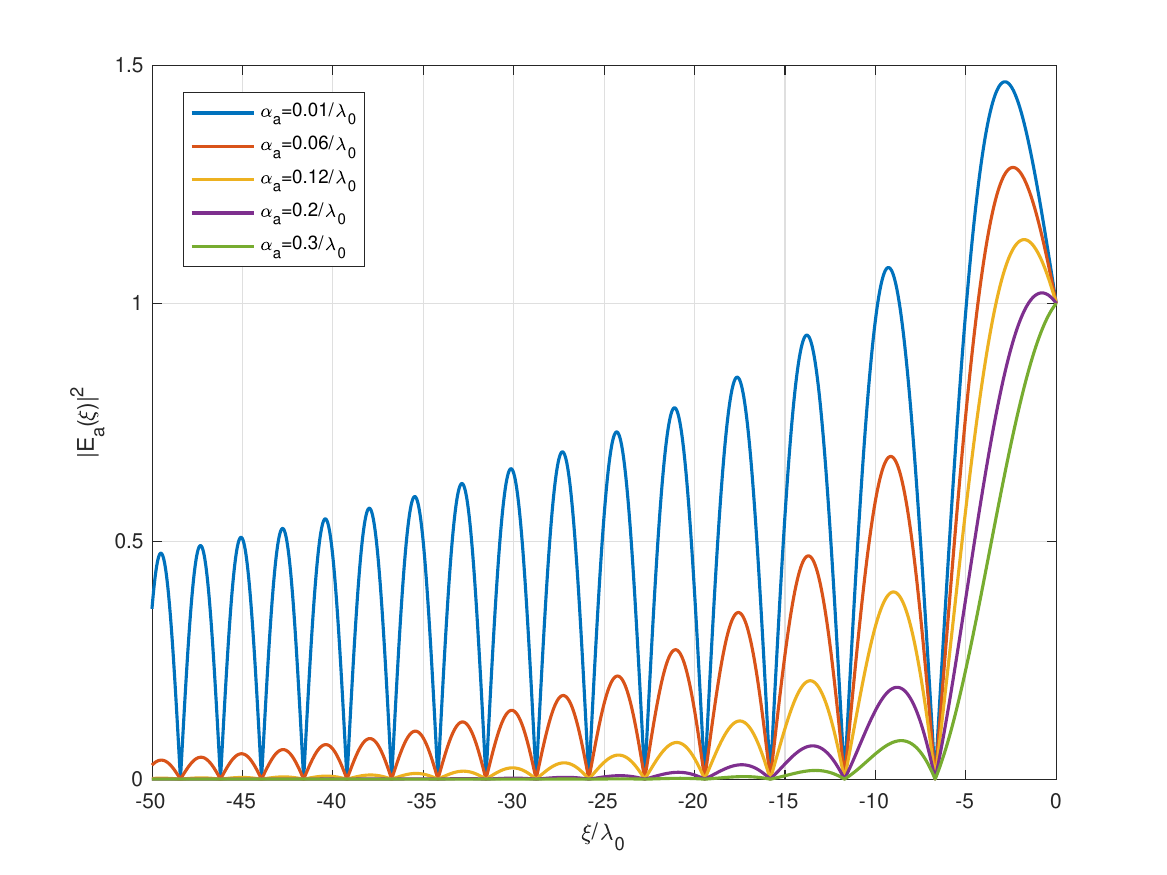} 
\caption{Intensity of the aperture Airy field in \eqref{eq_Ea} as a function of $\xi/{\lambda_0}$ for various values
of $\alpha_\text{a}$, with $\nu_\text{a}=0$ and 
$\gamma_\text{a} = \tfrac{k_0}{18}$.} 
\label{fig:fig_11_rev}
\end{figure}

\begin{figure}[t]
\centering
\includegraphics[width=\linewidth]{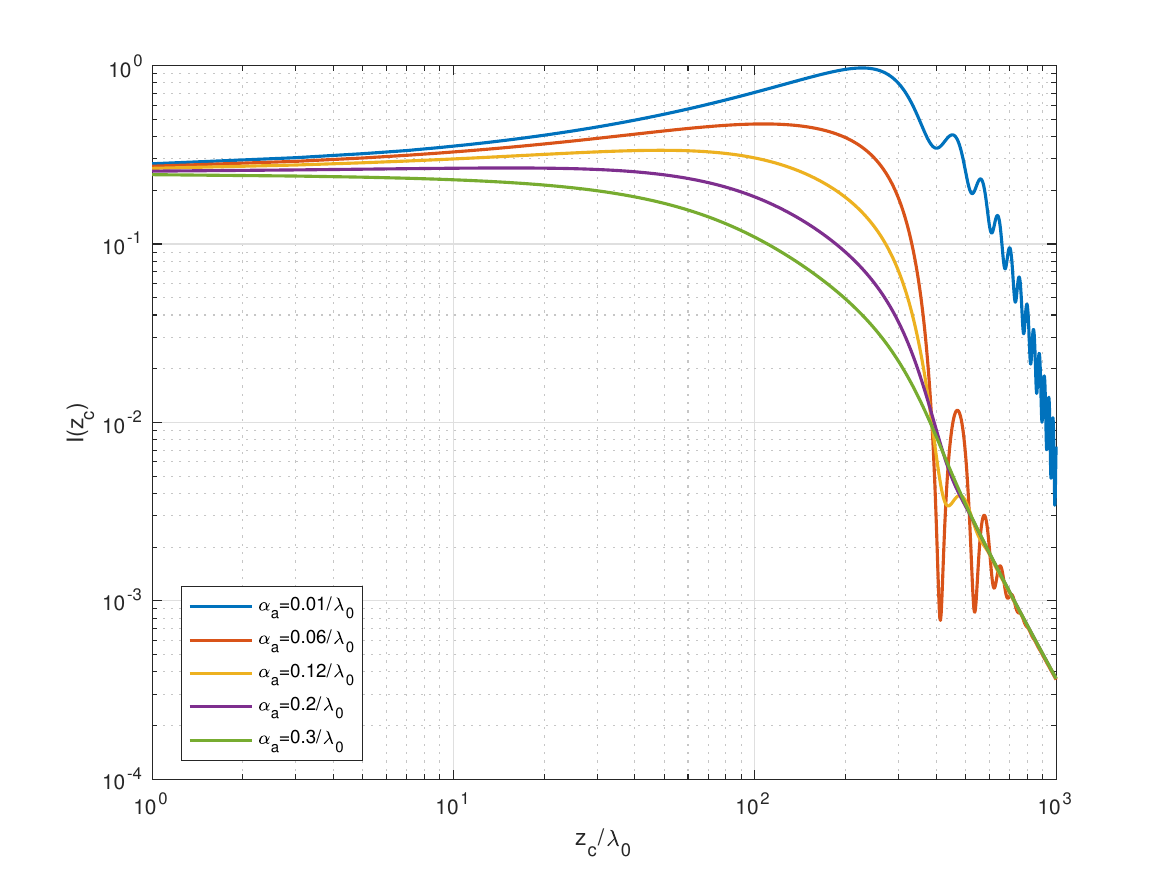} 
\caption{Intensity of the exponentially modulated Airy beam in \eqref{eq:airy-def} along the caustic 
in \eqref{eq:caustic-airy-gen}
as a function of $z_\text{c}/\lambda_0$ for various values of the apodization
parameter $\alpha_\text{a}$, with $\nu_\text{a}=0$ and $\gamma_\text{a}=\tfrac{k_0}{18}$.}
\label{fig:fig_12_rev}
\end{figure}

Physical truncation of the aperture and exponential modu\-lation of the Airy aperture field  are two equivalent strategies to limit the aperture size. In fact, the apodization parameter 
$\alpha_\text{a}$ in \eqref{eq_Ea} effectively controls the tapering of the field's amplitude profile, thereby allowing the designer to confine the beam spatially without  completely sacrificing its key propagation characteristics.
This relationship is illustrated in Fig.~\ref{fig:fig_11_rev}, which shows the intensity 
$|\E_\text a(\xi)|^2$ of the aperture Airy field for various
values of the apodization parameter $\alpha_\text{a}$, with $\nu_\text{a}=0$ and 
$\gamma_\text{a} = \tfrac{k_0}{18}$ in \eqref{eq_Ea}. 
The reciprocal of the apodization parameter $1/\alpha_\text{a}$ effectively quantifies the spatial range over which the aperture field
$\E_\text a(\xi)$ maintains a significant portion of its maximum amplitude. Specifically, at
$\xi=-1/\alpha_\text{a}$,  the aperture field decays to approximately 
$36.8 \%$ of its peak value $\E_\text a(0)$, and at $\xi=-3/\alpha_\text{a}$ it drops to
about $5\%$. Therefore, exponential modulation through the apodization parameter can be interpreted as 
a form of {\em soft} spatial truncation, in contrast to the {\em hard}  truncation achieved by directly limiting the effective aperture length.
To further support this interpretation,  Fig.~\ref{fig:fig_12_rev} illustrates
the intensity $I(z_\text{c})$ of the exponentially modulated Airy beam 
in \eqref{eq:airy-def} along its parabolic caustic for different values of
the apodization parameter $\alpha_\text{a}$, assuming $\nu_\text{a}=0$ and $\gamma_\text{a}=\tfrac{k_0}{18}$.
For exponentially modulated Airy beams radiated by infinite-size aperture, 
the Fraunhofer distance in \eqref{eq:frau} is still valid 
by replacing $x_\text{a}^\text{max}$ by a multiple integer of
$1/\alpha_\text{a}$, i.e., 
\be
\overline{z}_\text{F} \eqdef \frac{\pi \, n^2}{\lambda_0 \, \alpha_\text{a}^2}\:, 
\label{eq:frau-apo}
\ee
with the integer $n$ determining the effective extent of the infinite-size transmitting aperture
such that $\E_\text a(n/\alpha_\text{a})$ is equal to 
a given per cent of the peak value $\E_\text a(0)$. 
By choosing $n=3$, the values of the Fraunhofer distance 
corresponding to $\alpha_\text{a} \in \{0.01/\lambda_0,0.06/\lambda_0,0.12/\lambda_0,0.2/\lambda_0,0.3/\lambda_0\}$
are $\overline{z}_\text{F}/\lambda_0 \in \{282740, 7854, 1963.5, 706.8583, 314.1593\}$, respectively.
The results of Fig.~\ref{fig:fig_12_rev} closely mirror the trends observed in 
Fig.~\ref{fig:fig_10_rev} under hard truncation, reinforcing the notion 
that apodization operates as a smooth, continuous spatial cut-off.
Naturally, both techniques can be combined  to tailor diffraction-resistant Airy 
beams in practice.

\subsection{Airy beam generation: impact of spatial 
discretization and uniform quantization}
\label{sec:sampling}

Airy beams can be synthesized either by modulating 
amplitude and phase simultaneously or phase only along metasurfaces 
or antenna arrays~\cite{Liu_ArXiV2025,Ye_ArXiV2025,Inserra_2022,
Feng.2020,Wang.2023}.
As anticipated in Section~IV-\ref{sec:apert}, the spatial scaling factor 
$\gamma_\text{a}$ controls not only the propagation range of the 
Airy beam but also the spatial frequency content of the aperture 
field, which in turn determines the minimum sampling step required 
for accurate discrete implementation. In this case, the desired 
continuous aperture excitation must be spatially discretized. To 
assess the impact of spatial sampling, it is convenient to examine 
the spectrum of the (apodized) Airy aperture field 
$E_\text{a}(\xi) = U_\text{a}\,\Ai(\gamma_\text{a}\,\xi)\, 
e^{\alpha_\text{a}\xi}\,e^{-j\nu_\text{a}\xi}$, 
whose Fourier transform, defined as
$\tilde{E}_\text{a}(k_x) \triangleq \int_{-\infty}^{+\infty}
\E_\text{a}(\xi)\, e^{-j k_x \xi}\, \mathrm{d}\xi$, is given by
\begin{equation}
\tilde{E}_\text{a}(k_x)
= \frac{U_\text{a}}{\gamma_\text{a}} \, 
e^{\,j \frac{(k_x+\nu_\text{a}+j\alpha_\text{a})^3}{3\gamma_\text{a}^3}}, 
\label{eq:FT_airy_apod}
\end{equation}
where $k_x$ denotes the spatial angular frequency along the 
$x$-axis.
We may quantify the effective bandwidth from the normalized amplitude
spectrum: by taking the magnitude of~\eqref{eq:FT_airy_apod} and normalizing
it with respect to its peak value $|\tilde{E}_\text{a}(-\nu_\text{a})|$,
we obtain
\begin{equation}
\frac{|\tilde{E}_\text{a}(k_x)|}{|\tilde{E}_\text{a}(-\nu_\text{a})|}
= \exp\!\left(-\frac{(k_x+\nu_\text{a})^2}{2\,\sigma_\text{a}^2}\right),
\label{eq:gauss-spectrum}
\end{equation}
where $\sigma_\text{a} \triangleq \sqrt{\gamma_\text{a}^3/(2\alpha_\text{a})}$
represents the standard deviation of the Gaussian-shaped amplitude spectrum.
The linear phase $\nu_\text{a}$ merely shifts the spectrum along the
$k_x$-axis, leaving both its Gaussian shape and its width $\sigma_\text{a}$
unchanged.
From this expression, 
let $\banda_\text{sp}$ (in rad/m) denote the 
effective one-sided spatial bandwidth of the 
apodized Airy aperture field,
we can enforce a Nyquist-type condition 
on the spatial sampling rate $\Delta_\text{sp}$, i.e., $2\pi/\Delta_\text{sp} \ge 2 \banda_\text{sp}$ 
(or, equivalently, $\Delta_\text{sp} \le \pi/\banda_\text{sp}$), ensuring that the 
sampled aperture can reproduce the intended Airy-beam evolution 
without aliasing. In particular, a convenient rule-of-thumb is 
to retain the spectral interval $k_x \in [-3\sigma_\text{a},\,
3\sigma_\text{a}]$, which contains more than $99\%$ of the total 
signal energy. Under this choice, the effective one-sided bandwidth 
can be set to
\begin{equation}
\banda_\text{sp} = 3\sigma_\text{a} = 
3\sqrt{\frac{\gamma_\text{a}^{3}}{2\alpha_\text{a}}} \: ,
\label{eq:B_3sigma}
\end{equation}
and the corresponding spatial sampling step becomes
\begin{equation}
\Delta_\text{sp} \le \frac{\pi}{\banda_\text{sp}}
= \frac{\pi}{3\sigma_\text{a}}
= \frac{\pi}{3}\sqrt{\frac{2\alpha_\text{a}}{\gamma_\text{a}^{3}}}\:,
\label{eq:delta_3sigma}
\end{equation}
which shows that larger values of $\gamma_\text{a}$ require a 
finer spatial resolution, i.e., smaller values of $\Delta_\text{sp}$, 
whereas larger values of the apodization factor $\alpha_\text{a}$ 
provide a tighter spectral confinement and therefore allow a 
\emph{larger} sampling step, i.e., a coarser spatial resolution. 
This result is consistent with the physical interpretation of 
$\alpha_\text{a}$ as a soft spatial truncation of the aperture 
field: a larger apodization confines the field to a smaller 
spatial region, reducing the effective bandwidth and relaxing 
the sampling requirement.

In practice, the digital implementation of an 
Airy aperture entails not only spatial sampling but also 
quantization of the aperture excitation, which is typically 
the more critical impairment once the sampling condition on 
$\Delta_\text{sp}$ is satisfied. Although architectures enabling joint 
amplitude and phase modulation have been proposed for Airy-beam 
synthesis, \emph{phase-only} implementations are of particular 
practical interest due to the widespread adoption of phase-coding 
metasurfaces in communication-oriented platforms.
In~\cite{Feng.2020}, a phase-only aperture field is obtained by 
extracting the phase of the (real-valued) Airy beam excitation 
over a finite aperture, i.e.,
\begin{equation}
\Phi_\text{a}(\xi) =- \arg\!\big(E_\text{a}(\xi)\big),
\qquad 
\xi \in \big[x_\text{a}^{(1)},\,x_\text{a}^{(2)}\big],
\label{eq:phase_from_real_airy}
\end{equation}
and by imposing the corresponding phase-only field
\begin{equation}
E_\text{a}^{(\mathrm{ph})}(\xi) = 
\exp\!\big(-j\Phi_\text{a}(\xi)\big).
\label{eq:phase_only_aperture}
\end{equation}
Since $\E_\text{a}(\xi)$ is real-valued, $\Phi_\text{a}(\xi)$ takes only the
two values $0$ and $\pi$ (modulo $2\pi$), depending on the sign of
$\E_\text{a}(\xi)$; hence, \eqref{eq:phase_only_aperture} reduces to a binary
$\{+1,-1\}$ coding of the aperture. Note that, for such a binary profile,
the sign convention adopted in \eqref{eq:phase_from_real_airy} is immaterial,
since $e^{-j\pi} = e^{+j\pi} = -1$.

\begin{figure}[t]
\centering
\includegraphics[width=\linewidth]{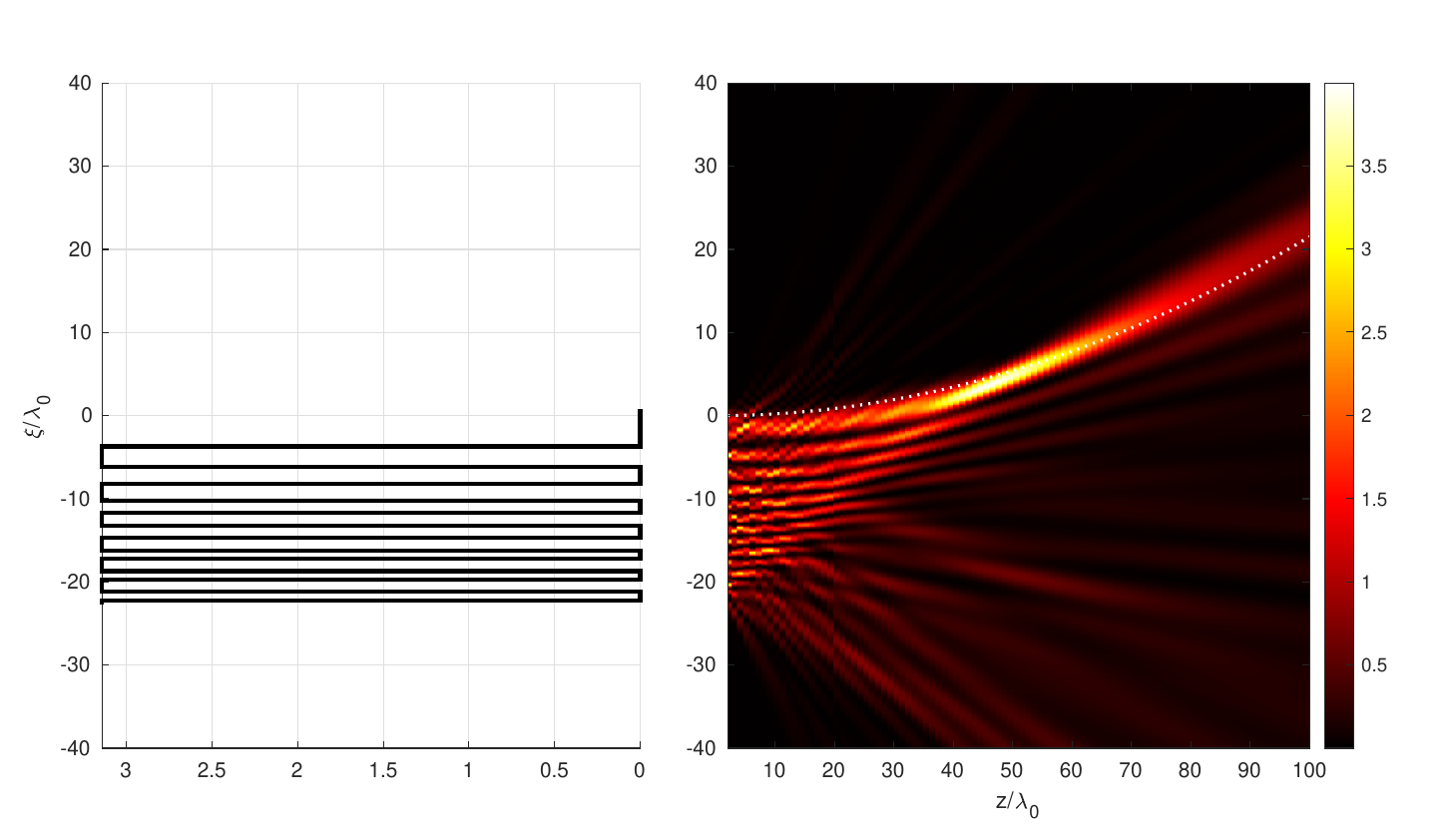} 
\caption{Left: $1$-bit coding phase profile as 
a function of the $\xi$-coordinate, with $\Delta_\text{sp} = \lambda_0/2$, 
$\alpha_\text{a} = 0.1/\lambda_0$, $\nu_\text{a} = 0$, and 
$\gamma_\text{a} = k_0/9$. Right: Intensity distribution of the 
corresponding Airy beam as a function of $z$, when 
$x^{(1)}_\text{a} = -22.7\,\lambda_0$ and $x^{(2)}_\text{a} = 0$. 
The radiated beam is computed numerically by evaluating the 
Rayleigh-Sommerfeld diffraction integral 
in~\eqref{eq:Ray-2}. The white dotted curve in the right-side 
panel represents the caustic trajectory 
in~\eqref{eq:caustic-airy-gen}. All spatial coordinates are 
normalized with respect to the wavelength $\lambda_0$.}
\label{fig:fig_12_1_rev}
\end{figure}

To provide an illustrative example of Airy-beam implementation 
via a $1$-bit metasurface, we consider an Airy beam with 
$\alpha_\text{a} = 0.1/\lambda_0$, $\nu_\text{a} = 0$, and 
$\gamma_\text{a} = k_0/9$, which yields a required spatial 
sampling step $\Delta_\text{sp} \le 0.8\,\lambda_0$ from 
eq.~\eqref{eq:delta_3sigma}. By selecting an inter-element 
spacing of $\lambda_0/2$ — well within the Nyquist limit — and 
adopting a phase-only implementation, the resulting $1$-bit 
coding profile is shown in Fig.~\ref{fig:fig_12_1_rev} 
(left-side plot), along with the corresponding intensity of 
the radiated Airy beam (right-side plot).
We observe that the radiated beam closely follows the nominal 
caustic trajectory, indicated by the white dotted curve, 
confirming that $1$-bit phase coding is both natural and 
sufficient for Airy beam synthesis on binary metasurfaces. 
Moreover, the field amplitude along the caustic is less 
uniform than in the case where both amplitude and phase are 
controlled at the aperture. The abrupt phase variations imposed 
on a finite aperture enhance edge-diffraction effects, leading 
to additional diffractive artifacts in the radiated field. 
These artifacts become more pronounced as the aperture size 
decreases, and can be mitigated by increasing the aperture 
extent or by adopting amplitude tapering in conjunction with 
phase coding.

\subsection{Link budget for exponentially modulated Airy beams}
\label{sec:link}

With reference to Fig.~\ref{fig:fig_1_rev}, 
we evaluate the free-space path loss associated with the communication link between 
the transmitting aperture and a receiver located near the caustic
at a longitudinal distance \( z_\text{c} > 0 \). In the far-field
regime, the Friis formula \cite{Rappaport}  predicts that the free-space path 
loss increases proportionally to the square of the distance 
between transmitter and receiver. 
However, in the radiative near-field region, the assumptions underlying the Friis formula no longer hold,  
and a more appropriate link budget must be formulated \cite{Guerb2024}.

We consider an infinite-size transmitting aperture
with an electric field distribution given by \eqref{eq_Ea}, assuming zero initial launch angle, 
i.e., $\nu_\text{a}=0$.
The 1-D receiving aperture extends infinitely along the $y$-axis and spans the transverse $x$-direction from $x_\text{r}^{(1)}=x_\text{c}-\Delta x_\text{r}$ to $x_\text{r}^{(2)}=x_\text{c}$.
For an exponentially modulated Airy field with unit normalized energy, ensured by choosing the scaling constant $U_\text{a}$ as 
in \eqref{eq:U0}, 
the free-space path loss in decibels is given by
\be
L_\text{dB} \eqdef - 10 \log_{10}(\en_r),
\ee
where
\be
\en_\text{r} \eqdef  \int_{x_\text{c}-\Delta x_\text{r}}^{x_\text{c}} |\E(z_\text{c},x)|^2 \, \mathrm{d}x \\
\label{eq:pow2}
\ee
represents the energy collected by the receiving aperture.
Substituting \eqref{eq:airy-def} into \eqref{eq:pow2} and using the caustic expression in \eqref{eq:caustic-airy-gen}, 
a change of variable yields
\be
\en_\text{r} = U_\text{a}^2 \, e^{-\frac{z_\text{c}^2 \, \gamma_\text{a}^3 \, \alpha_\text{a}}{2 \, k_0^2}} \int_{-\Delta x_\text{r}}^{0}
\left| \Ai \left(\gamma_\text{a} \, v - j \frac{z_\text{c} \, \gamma_\text{a} \, \alpha_\text{a}}{k_0} \right) \right|^2
e^{2 \, \alpha_\text{a} v} \mathrm{d}v \:.
\label{eq:pow2-2}
\ee
The integral in \eqref{eq:pow2-2} depends weakly on $z_\text{c}$.
The dependence of $\en_\text{r}$ on $z_\text{c}$
is mainly governed by the exponential term 
\be
e^{-\frac{z_\text{c}^2 \, \gamma_\text{a}^3 \, \alpha_\text{a}}{2 \, k_0^2}}\:.
\label{eq:expterm}
\ee
To understand the dependence of the path loss on 
the transmission distance, we divide the radiative near-field zone
into two non-overlapping spatial regions: the former one is characterized
by the values of $z_\text{c}$ for which \eqref{eq:expterm} tends to unity and 
$\en_\text{r}$ is approximately constant ({\em diffraction-resisting region});
in the latter one, the exponential in \eqref{eq:pow2-2} dominates and  
$\en_\text{r}$ decays rapidly ({\em diffractive region}).

In the diffraction-resisting regime, we consider the spatial  region where 
condition \eqref{eq:cond-df-apo} holds. In this case, the
imaginary component of the argument of the Airy function in \eqref{eq:pow2-2}
is negligible, and the exponential term is approximately unity. 
As a result, \eqref{eq:pow2-2} simplifies to 
\be
\en_\text{r} \approx \frac{U_\text{a}^2}{\gamma_\text{a}} \int_{-\gamma_\text{a} \, \Delta x_\text{r}}^{0}
\Ai^2 \left(v \right) \, e^{\frac{2 \, \alpha_\text{a}}{\gamma_\text{a}} v} \, \mathrm{d}v \:.
\label{eq:pow2-2-approx}
\ee
This integral can be approximated in closed form by expanding the exponential term via its 
first-order Maclaurin series, 
$e^{\frac{2 \, \alpha_\text{a}}{\gamma_\text{a}} v} \approx 1+\tfrac{2 \, \alpha_\text{a}}{\gamma_\text{a}} v$, which holds
under the condition $\alpha_\text{a} \, \Delta x_\text{r} \ll 1$. Substituting into \eqref{eq:pow2-2-approx},
one obtains \eqref{eq:Er-2}, 
\begin{figure*}[!t]
\barr
\en_\text{r} & \approx \frac{U_\text{a}^2}{\gamma_\text{a}} \left[ \int_{-\gamma_\text{a} \, \Delta x_\text{r}}^{0}
\Ai^2 \left(v \right) \, \mathrm{d}v + \frac{2 \, \alpha_\text{a}}{\gamma_\text{a}}
\int_{-\gamma_\text{a} \, \Delta x_\text{r}}^{0}
v \, \Ai^2 \left(v \right) \, \mathrm{d}v
\right] \nonumber \\ & = 
\frac{U_\text{a}^2}{\gamma_\text{a}} \left\{ 
\left[ v \, \Ai^2 \left(v \right) - \dAi^2 \left(v \right)\right]_{v=-\gamma_\text{a} \, \Delta x_\text{r}}^{v=0}
+ \frac{2}{3}\frac{\alpha_\text{a}}{\gamma_\text{a}} 
\left[ 
\Ai(v) \, \dAi(v) -v \, \dAi^2 \left(v \right) + v^2 \, \Ai^2 \left(v \right)
\right]_{v=-\gamma_\text{a} \, \Delta x_\text{r}}^{v=0}
\right\}
\nonumber \\ & = 
\frac{U_\text{a}^2}{\gamma_\text{a}} \left[
\frac{2}{3}\frac{\alpha_\text{a}}{\gamma_\text{a}} \, \Ai(0) \, \dAi(0) - \dAi^2 \left(0\right) \right]
+  \frac{U_\text{a}^2}{\gamma_\text{a}} \Big \{ 
\gamma_\text{a} \, \Delta x_\text{r} \, \Ai^2 \left(-\gamma_\text{a} \, \Delta x_\text{r} \right) +
\dAi^2 \left(-\gamma_\text{a} \, \Delta x_\text{r} \right)
\nonumber \\ & - \frac{2}{3}\frac{\alpha_\text{a}}{\gamma_\text{a}} 
\left[ 
\Ai(-\gamma_\text{a} \, \Delta x_\text{r}) \, \dAi(-\gamma_\text{a} \, \Delta x_\text{r}) + \gamma_\text{a} \, \Delta x_\text{r} 
\, \dAi^2 \left(-\gamma_\text{a} \, \Delta x_\text{r} \right) + \gamma_\text{a}^2 \, \Delta x_\text{r}^2 \, \Ai^2 \left(-\gamma_\text{a} \, \Delta x_\text{r} \right)
\right]
\Big\}.
\label{eq:Er-2}
\earr
\hrule
\end{figure*}
which enables an analytical prediction of the constant path loss in the diffraction-resisting region.
In such a zone, the feeble dependence of $\en_\text{r}$ on $z_\text{c}$ arises from the fact that, under condition \eqref{eq:cond-df-apo},
the exponentially modulated Airy beam exhibits only weak diffraction. 

In the diffractive regime, the beam begins to diffract more significantly, resulting in a progressive decrease of the received energy and, consequently, an 
increase in path loss. The transition from the diffraction-resisting regime to the diffraction one
occurs when the exponent of the exponential term in \eqref{eq:expterm} becomes on the order of one, i.e., 
\be
\frac{z_\text{c}^2 \, \gamma_\text{a}^3 \, \alpha_\text{a}}{2 \, k_0^2} \approx 1 
\quad \Rightarrow \quad
z_\text{c}^\text{corner} = \sqrt{\frac{2 \, k_0^2}{\gamma_\text{a}^3 \, \alpha_\text{a}}} \: .
\label{eq:Zcorner}
\ee
The received energy $\en_\text{r}$ decays exponentially with $z_\text{c}^2$ beyond the corner point $z_\text{c}^\text{corner}$.
Equivalently, the free-space path loss $L_\text{dB}$ exhibits a parabolic trend as a function of distance $z_\text{c}$
in the diffractive regime. Comparing the corner point $z_\text{c}^\text{corner}$ with the 
Fraunhofer distance $\overline{z}_\text{F}$ in \eqref{eq:frau-apo} gives 
\be
\frac{z_\text{c}^\text{corner}}{\overline{z}_\text{F}} = \frac{2 \sqrt{2}}{n^2} \, \left(\frac{\alpha_\text{a}}{\gamma_\text{a}}\right)^{3/2},
\ee
which depends on the ratio between the apodization parameter $\alpha_\text{a}$ and the
spatial scaling factor $\gamma_\text{a}$, but not on the carrier wavelength $\lambda_0$. 

\begin{figure}[t]
\centering
\includegraphics[width=\linewidth]{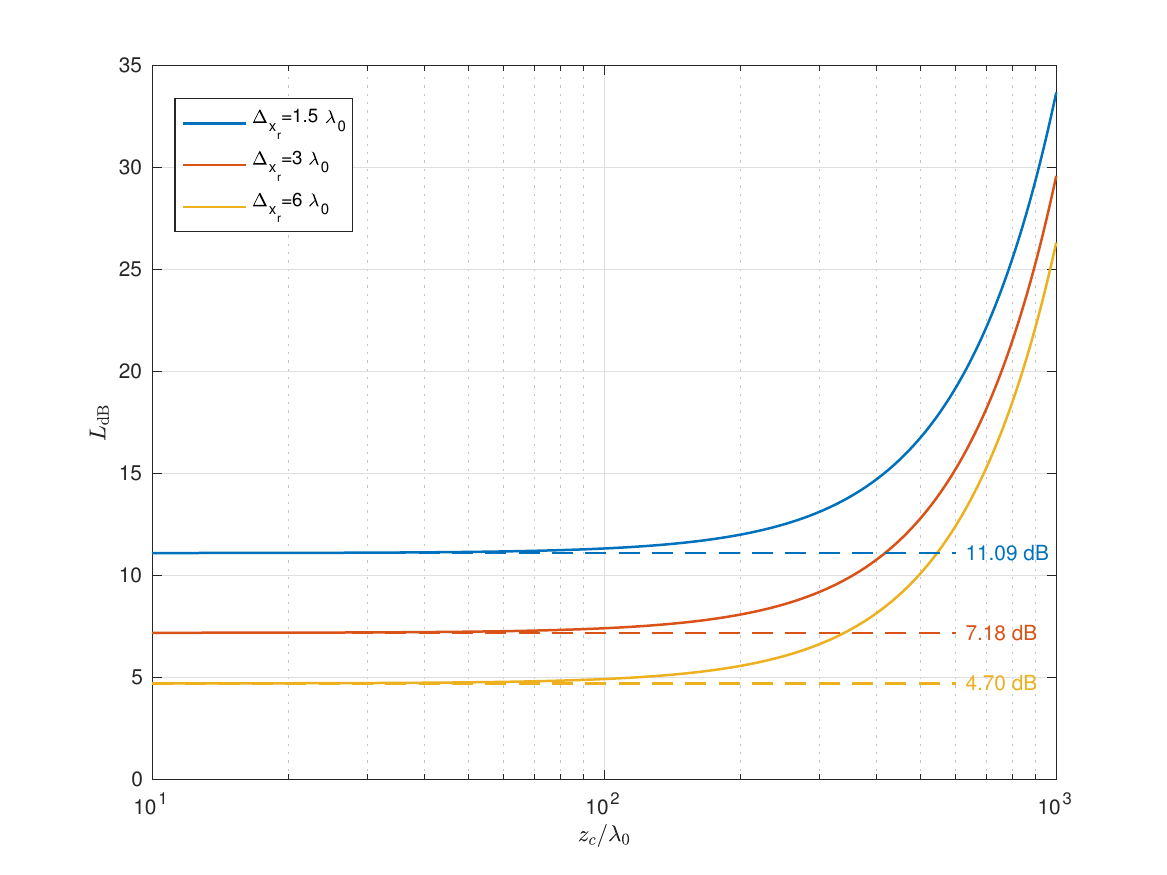} 
\caption{Path loss $L_\text{dB}$ as a function of the receiver's distance $z_\text{c}$ from an infinite-size transmit aperture
along the caustic in \eqref{eq:caustic-airy-gen}, for various aperture sizes $\Delta x_\text{r}$
of the receiver, with $\nu_\text{a}=0$, $\alpha_\text{a}=0.01/\lambda_0$, and $\gamma_\text{a}=\tfrac{k_0}{18}$.
All  spatial coordinates are normalized with respect to  the wavelength $\lambda_0$. The dashed lines indicate the values predicted 
by the analytical approximation in \eqref{eq:Er-2}
in the diffraction-resisting region.
}
\label{fig:fig_13_rev}
\end{figure}

\begin{figure}[t]
\centering
\includegraphics[width=\linewidth]{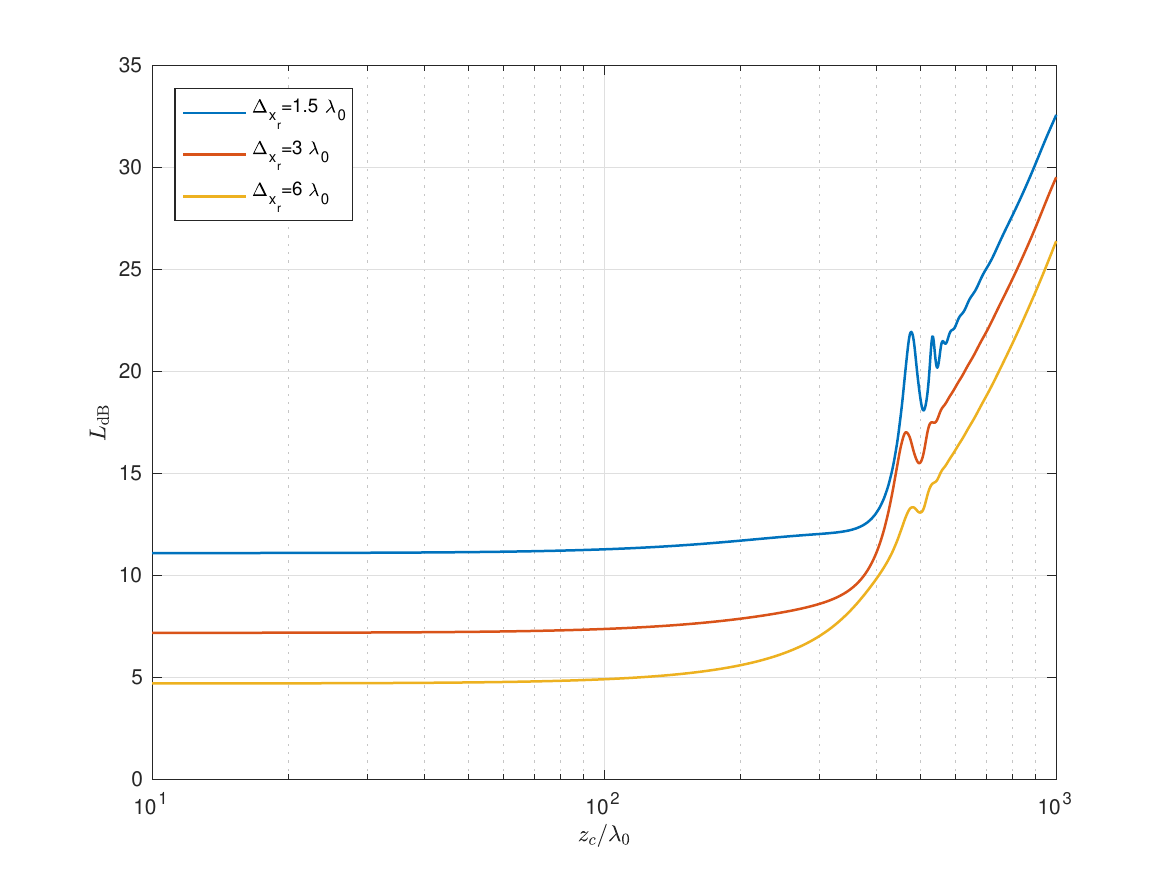} 
\caption{Path loss $L_\text{dB}$ as a function of the receiver's distance $z_\text{c}$ from a finite-size transmit aperture
along the caustic in \eqref{eq:caustic-airy-gen}, for various aperture sizes $\Delta x_\text{r}$
of the receiver, with $x_\text{a}^{(1)}=-300 \, \lambda_0$, $x_\text{a}^{(2)}=0$, 
$\nu_\text{a}=0$, $\alpha_\text{a}=0.01/\lambda_0$, and $\gamma_\text{a}=\tfrac{k_0}{18}$.
The path loss is computed by numerically solving \eqref{eq:Ray-2} and \eqref{eq:pow2},
with electric field distribution given by \eqref{eq_Ea}. 
All  spatial coordinates are normalized with respect to  the wavelength $\lambda_0$. 
}
\label{fig:fig_14_rev}
\end{figure}

Fig.~\ref{fig:fig_13_rev}  illustrates the path loss $L_\text{dB}$ as a function of the propagation distance 
$z_\text{c}$ along the caustic. The received energy $\en_\text{r}$ is computed from  \eqref{eq:pow2-2}, 
for various values of the receiver aperture width $\Delta x_\text{r}$, assuming $\nu_\text{a}=0$, $\alpha_\text{a}=0.01/\lambda_0$, and $\gamma_\text{a}=\tfrac{k_0}{18}$.
As predicted by the theoretical analysis, the path loss remains nearly constant within the diffraction-resisting
region, which extends to the corner point $z_\text{c}^\text{corner}=430.86 \, \lambda_0$.
Beyond this range, once significant diffraction sets in, the path loss in dB increases rapidly, following a square-law dependence on distance, as the Fraunhofer distance 
$\overline{z}_\text{F} = 282740 \, \lambda_0$ is approached.
As expected, enlarging the receiver aperture size  improves performance by capturing 
a greater portion of the beam's energy.
Remarkably, using the analytical approximation in \eqref{eq:Er-2}, the predicted constant path loss values
in the diffraction-resisting region for the three cases $\Delta x_\text{r} 
\in \{1.5 \, \lambda_0, 3 \, \lambda_0, 6 \, \lambda_0\}$
are $11.09$ dB, $7.18$ dB, and $4.70$ dB, respectively, which are in excellent agreement with
the numerical results of Fig.~\ref{fig:fig_13_rev}.
Consistent with the discussion in Sections~\ref{sec:Diff-theory} and 
IV-\ref{sec:apert}, the corner
distance $z_c^{\mathrm{corner}}$ in \eqref{eq:Zcorner} provides a further
performance-defined near-field range in the sense of
\cite{CuiDai2024,Kludze2026}. Here, this range is dictated by the link budget
rather than by beam-shape preservation, and it remains well below the Fraunhofer boundary $z_{\mathrm{F}}$.

The previous analysis of the path loss assumes a transmit aperture of infinite transverse extent. 
In Fig.~\ref{fig:fig_14_rev}, we additionally report the path loss $L_\text{dB}$ as a function 
of the propagation distance  $z_\text{c}$ along the caustic in the case of a finite aperture of 
size $x_\text{eff}=300 \, \lambda_0$. 
The electric field distribution is given by \eqref{eq_Ea},
with $\nu_\text{a}=0$, $\alpha_\text{a}=0.01/\lambda_0$, and $\gamma_\text{a}=\tfrac{k_0}{18}$.
We numerically calculate the integral \eqref{eq:pow2}, by evaluating
the Rayleigh-Sommerfeld diffraction integral in \eqref{eq:Ray-2}, substituting the caustic equation in \eqref{eq:caustic-airy-gen}, 
and computing the squared magnitude of the field  
for $x_\text{a}^{(1)}=-x_\text{eff}$ and $x_\text{a}^{(2)}=0$.
Results show a path loss profile
closely resembling
that  in Fig.~\ref{fig:fig_13_rev}. 
Interestingly, also in the case of a finite-size aperture, 
the formula in \eqref{eq:Er-2} provides a reliable estimate of
the constant path loss value in the diffraction-resisting regime.
This  further supports the interpretation
of spatial modulation as a soft 
truncation of the aperture.

\subsection{Polychromatic exponentially modulated Airy beams}
\label{sec:polyf}

Up to this point, we have considered the ideal 
case of a monochromatic field, which is both analytically 
tractable and can be closely approximated in practice.

We now extend the analysis to the more general 
case of a polychromatic wave, which is relevant for practical 
communication systems where the transmitted signal occupies a 
finite bandwidth around the carrier frequency $f_0$.
In this context, the bandpass representation 
in~\eqref{eq:rf}, with carrier frequency $f_0$, generalizes to
\be
\widetilde{\E}(z,x;t) = \Re 
\left\{\E(z,x;t)\,e^{j2\pi f_0 t}\right\},
\ee
where the complex envelope $\E(z,x;t)$ depends on both space 
and time, and its Fourier transform with respect to $t$ is 
given by
\be
\overline{\E}(z,x;f) = \int_{-\infty}^{+\infty} 
\E(z,x;t)\,e^{-j2\pi f t}\,{\rm d}t \: .
\ee
Each spectral component of the beam is associated with a 
frequency-dependent parabolic caustic and a corresponding 
diffraction-resisting region. As shown in 
Appendix~\ref{app:Poly}, these features can be derived 
from~\eqref{eq:caustic-airy-gen} 
and~\eqref{eq:cond-df-apo}, respectively, by replacing the 
carrier wavenumber $k_0$ with the frequency-dependent one
\be
k(f) = k_0\left(1+\frac{f}{f_0}\right),
\ee
which represents the wavenumber associated with the 
monochromatic component of the field at frequency $f+f_0$.
Both the curvature of the parabolic trajectory and the 
initial launch angle of the beam from the aperture vary with 
frequency. Moreover, relative to the carrier frequency $f_0$, 
the longitudinal range over which the beam maintains its 
diffraction-resistant character scales by a factor of 
$1+\tfrac{f}{f_0}$.
It is worth noting that the spatial sampling condition derived 
in eq.~\eqref{eq:delta_3sigma} for the 
monochromatic case extends naturally to the 
polychromatic setting by replacing $k_0$ with 
$k(f)$ in the spatial bandwidth 
expression~\eqref{eq:B_3sigma}. Note that this 
spatial bandwidth $B_\text{sp}$ is a function of 
the beam parameters $\gamma_\text{a}$ and 
$\alpha_\text{a}$, and is independent of the 
temporal signal bandwidth $\banda_\text{a}$ (in Hz).
Since $k(f) = k_0(1+f/f_0)$, the most stringent sampling 
requirement is imposed by the highest frequency component 
$f_0 + \banda_\text{a}/2$, which determines the minimum 
element spacing for alias-free implementation.

\begin{figure}[t]
\centering
\includegraphics[width=\linewidth]{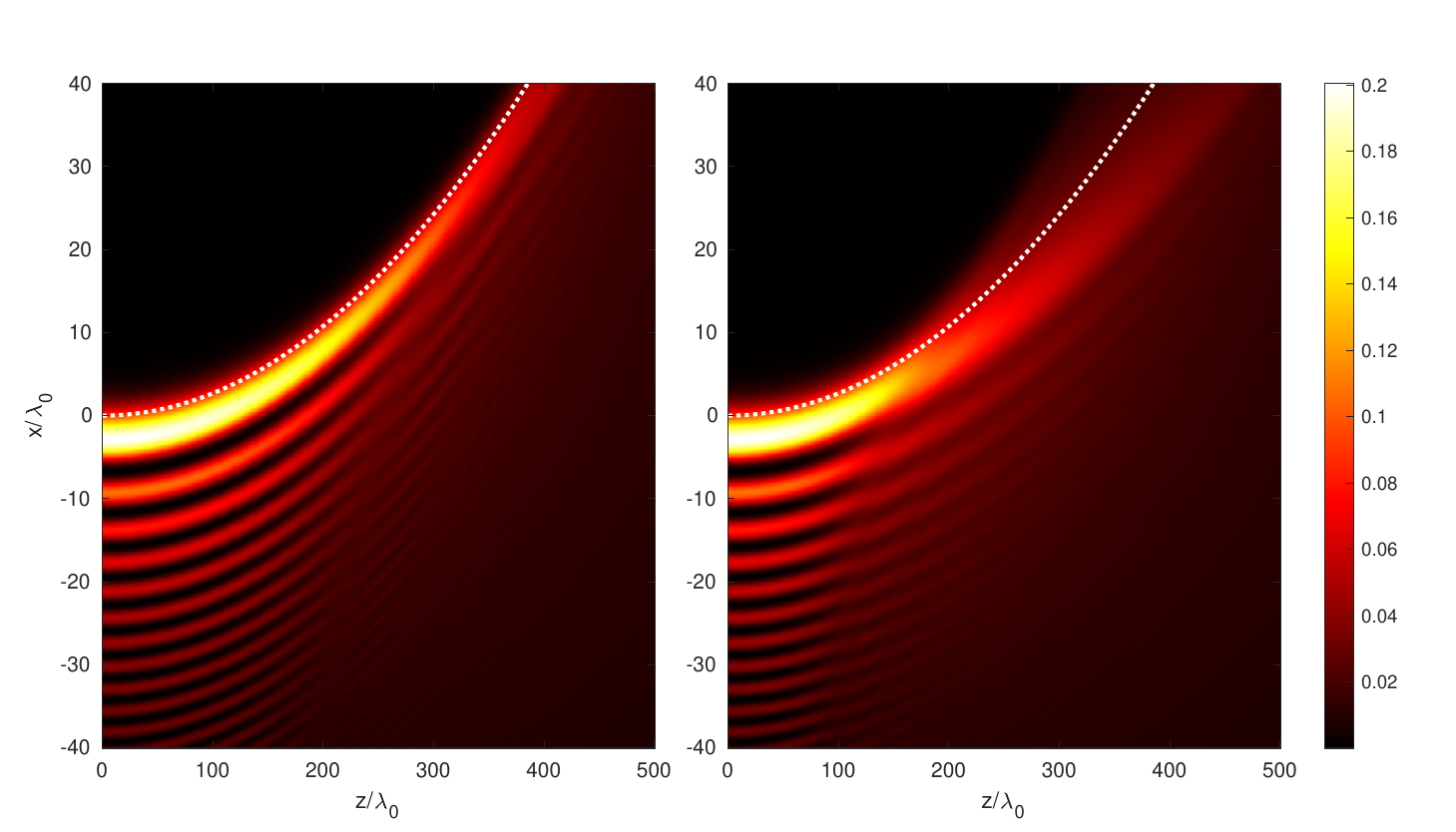} 
\caption{Intensity distribution of a pulsed 
Airy beam as a function of $z$ and $x$ for 
$\tfrac{\banda_\text{a}}{f_0}=0.1$ (left plot) and 
$\tfrac{\banda_\text{a}}{f_0}=0.4$ (right plot), when the 
aperture field is given 
by~\eqref{eq:pulsed-beam}--\eqref{eq_Ea-f}, with 
$\nu_\text{a}=0$, $\alpha_\text{a}=0.01/\lambda_0$, and 
$\gamma_\text{a}=\tfrac{k_0}{18}$. The white dotted curve 
represents the caustic trajectory 
in~\eqref{eq:caustic-airy-gen}, evaluated at the carrier 
frequency $f_0$. All spatial coordinates are normalized with 
respect to the carrier wavelength $\lambda_0$.}
\label{fig:fig_15_rev}
\end{figure}

For a finite-energy wave that is not strictly 
monochromatic, the intensity is given 
by~\eqref{eq:int-f-Nyquist-pap}, as derived in 
Appendix~\ref{app:Poly}.
\begin{figure*}[!t]
\be
I(z,x) = \frac{\overline{U}_\text{a}^2}{\banda_\text{a}^2}
\int_{-\frac{\banda_\text{a}}{2}}^{\frac{\banda_\text{a}}{2}} 
\left|\Ai\!\left(\gamma_\text{a}\,x
-\frac{z^2\,\gamma_\text{a}^4}{4\,k^2(f)}
-\frac{z\,\gamma_\text{a}\,\nu_\text{a}}{k(f)}
-j\frac{z\,\gamma_\text{a}\,\alpha_\text{a}}{k(f)}
\right)\right|^2 
e^{2\,\alpha_\text{a}\left(x
-\frac{z^2\,\gamma_\text{a}^3}{2\,k^2(f)}
-\frac{z\,\nu_\text{a}}{k(f)}\right)}
{\rm d}f \: .
\label{eq:int-f-Nyquist-pap}
\ee
\hrule
\end{figure*}
We now consider a digital communication system 
with symbol period $T_\text{s}$. For distortionless baseband 
data transmission using the ideal Nyquist pulse, the signal bandwidth 
is $\banda_\text{a} = \tfrac{1}{T_\text{s}}$. The 
corresponding value of $\overline{U}_\text{a}$, derived in 
Appendix~\ref{app:Poly}, is
\be
\overline{U}_\text{a} = \sqrt{\banda_\text{a}}\,
(8\,\pi\,\alpha_\text{a}\gamma_\text{a})^\frac{1}{4}\,
e^{-\frac{1}{3}\left(\frac{\alpha_\text{a}}{\gamma_\text{a}}
\right)^3}.
\ee
To illustrate the impact of temporal pulse shaping, 
Fig.~\ref{fig:fig_15_rev} shows the intensity distribution 
in~\eqref{eq:int-f-Nyquist-pap} of a pulsed Airy beam for 
different values of the signal bandwidth $\banda_\text{a}$ 
normalized to the carrier frequency $f_0$, assuming 
$\nu_\text{a}=0$, $\alpha_\text{a}=0.01/\lambda_0$, and 
$\gamma_\text{a}=\tfrac{k_0}{18}$.
Compared to the monochromatic case of Fig.~\ref{fig:fig_7_rev}, 
it is evident that a polychromatic exponentially modulated Airy 
beam maintains a similar propagation profile when 
$\banda_\text{a} \ll f_0$, corresponding to the {\em narrowband 
regime}, which is common in many practical communication 
scenarios. Based on this observation, we will henceforth focus 
on the case of a purely monochromatic wave. The key results 
obtained in the monochromatic regime are expected to hold, 
{\em mutatis mutandis}, for polychromatic beams under the 
narrowband assumption $\banda_\text{a} \ll f_0$.

\section{Self-accelerating paraxial Airy beams in partially obstructed communication links}
\label{sec:self-healing}

So far, we have assumed that the transmitting aperture operates in free space, i.e., it is positioned
far from the Earth and free of any obstruction. Although this assumption is valid
in a few simplified scenarios, it has served mainly to shed light on the fundamental propagation
mechanisms of self-accelerating beams and to establish meaningful performance bounds.
We now turn our attention to more realistic settings involving  partial obstruction along the propagation path  between the 
transmitting aperture and the receiver. 
This is the critical scenario in wireless communications as self-accelerating 
beams have been proposed to make wireless transmission more reliable in 
the presence of objects obstructing the LoS \cite{Guerb2024,Lee2025,Drou2025,GueEUCAP2024,Liu_ArXiV2025,Canals_ArXiV2025}.

Specifically, we consider the presence of an obstacle that partially blocks the link.
Two different models are analyzed: 
\begin{itemize}
\item{
In 
Subsection~V-\ref{sec:knife-diffraction}, we adopt the classical
{\em knife-edge} diffraction model.}
\item{
In Subsection~V-\ref{sec:soft-obstacle},
the obstacle is represented
as an infinite barrier with a bell-shaped absorption profile, modeling a {\em soft}   attenuation effect.}

\end{itemize}

\begin{figure}[t]
\centering
\includegraphics[width=\linewidth]{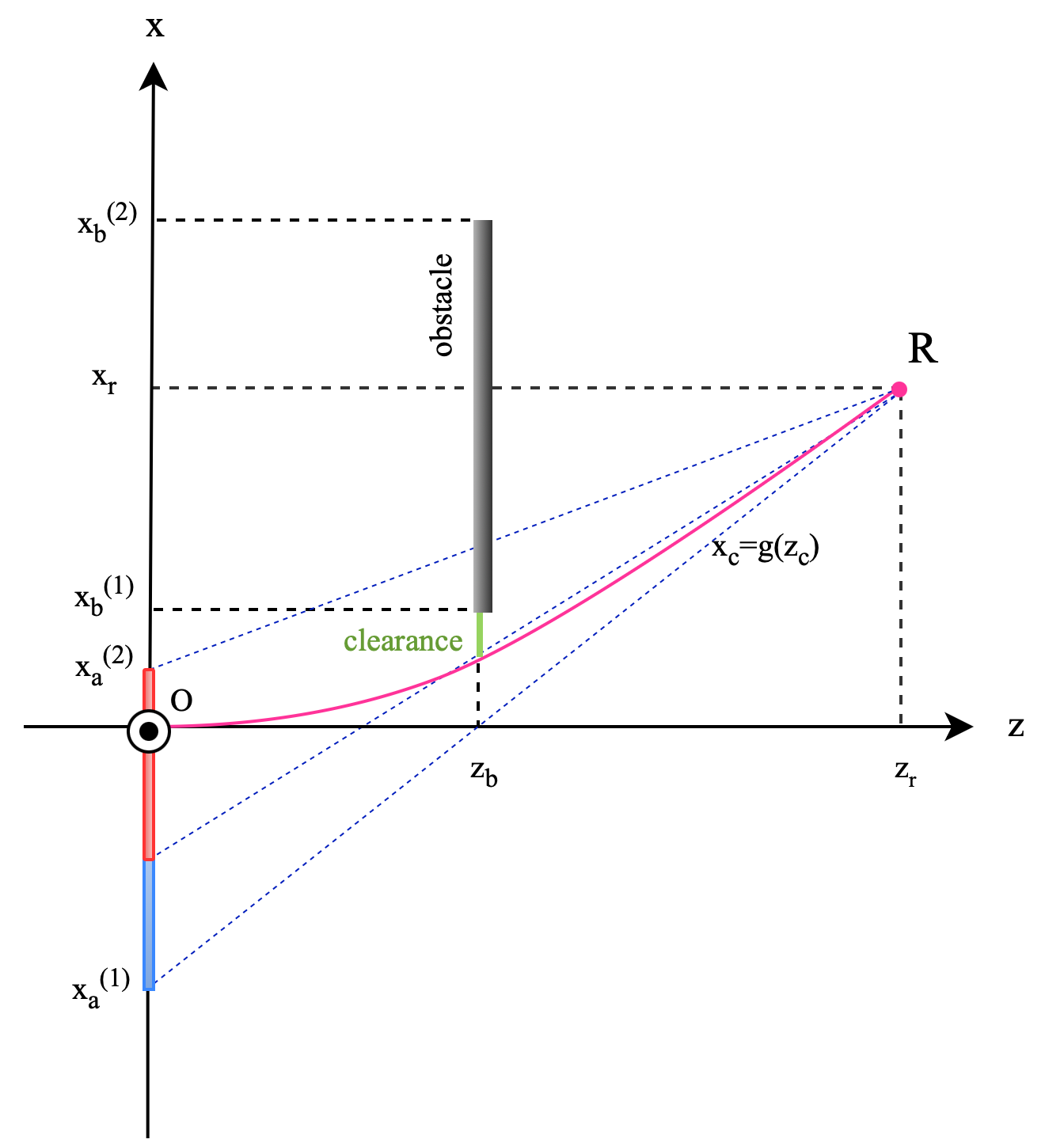} 
\caption{Geometry of knife-edge diffraction.}
\label{fig:fig_16_rev}
\end{figure}

\subsection{Knife-edge diffraction in the paraxial regime}
\label{sec:knife-diffraction}

To assess the impact of an obstruction on the propagating wavefront, we adopt the 
{\em knife-edge diffraction} model, in which
the obstacle is idealized as a  sharp, impenetrable (perfectly electric conducting) edge. Specifically, the knife-edge 
is positioned at a distance $z_\text{b}>0$ from the aperture (see Fig.~\ref{fig:fig_16_rev}).
The obstructive surface is perpendicular to the direct path between the transmitting aperture and the receiver,  
extending infinitely along the $y$-direction. Along the transverse $x$-axis,  it spans the interval from $x_\text{b}^{(1)}$ to $x_\text{b}^{(2)}$.
This model assumes an idealized geometry in which the diffracting edge is infinitesimally thin, perfectly straight, 
acting as a sharp  boundary that generates secondary wavelets in accordance with the Huygens-Fresnel principle.

Let $x_\text{c}=g(z_\text{c})$ denote the trajectory of the beam, such as the parabolic curve of the ideal 
Airy beam depicted in Fig.~\ref{fig:fig_16_rev}.  We define the vertical position of the lower edge of the obstacle at the obstruction plane  $z_\text{b}$ as
\be
 x_\text{b}^{(1)} = g(z_\text{b}) + \Delta, 
\label{eq:xB}
\ee
where $\Delta$ represents the beam {\em clearance} quantifying the vertical offset between the beam's caustic and the bottom edge of the obstacle. 
The  clearance $\Delta$ can take either sign:  A positive clearance ($\Delta > 0$) indicates that the obstacle lies above  
the beam trajectory,  and thus the beam is not directly blocked;
a negative clearance ($\Delta < 0$) implies that the obstacle intersects the caustics, causing partial or full  obstruction.
Intuitively, when the beam clearance is sufficiently large and positive, the obstacle exerts negligible influence, and the field at the receiver approximates its free-space counterpart.
Conversely, as $\Delta$ becomes negative, an increasing portion of the beam's energy is intercepted by the obstacle, resulting in
significant loss relative to the unblocked case.\footnote{An analogous argument holds if
the top edge of the obstacle approaches or intersects  
the beam path from below.}
Before quantifying this behavior using classical diffraction theory, 
we now formally define the concepts of LoS and NLoS as they apply to the present setting.

The ray-optics interpretation of self-accelerating beams reveals that their apparent
spatial acceleration arises from the presence of a curved caustic that envelopes a family of straight
rays launched from the aperture, as illustrated in Fig.~\ref{fig:fig_3_rev}.
Any rays that are intercepted by an obstacle {\em en route} to the receiver cannot contribute to the formation of the caustic. 
As a consequence, the portion of the aperture that maintains
LoS to the receiver (referred to as {\em unobstructed aperture}) contributes effectively to the generation of the self-accelerating beam. 
In practical terms, if the 
unobstructed portion of the aperture is significantly reduced (or vanishes altogether)
the resulting beam {\em cannot}  maintain its characteristic self-accelerating trajectory 
or exhibit diffraction-resistant behavior.

Let $\E(z,x)=u(z,x) \, e^{-j k_0 z}$ denote the field propagating from the aperture plane $z=0$ 
up to the obstructive plane at $z=z_\text{b}$. The  field diffracted beyond the
obstacle, i.e., for $z>z_\text{b}$  is described by the Rayleigh-Sommerfeld diffraction integral as
\be
\E_\text{d}(z,x) = 
\frac{1}{2 j} \int_{\mathcal{O}_\text{d}} 
\frac{k_0 \, (z-z_\text{b})}{\rho_\text{d}(\xi)} \, \E(z_\text{b},\xi)\, \text{H}^{(2)}_1(k_0 \, \rho_\text{d}(\xi)) \, \mathrm{d}\xi,
\label{eq:diff-field}
\ee
where $\mathcal{O}_\text{d} \eqdef (-\infty,x_\text{b}^{(1)}) \cup (x_\text{b}^{(2)},+\infty)$ defines the
integration domain and 
$\rho_\text{d}(\xi) \eqdef \sqrt{(x-\xi)^2+(z-z_\text{b})^2}$.
It is interesting to observe that, when $z_\text{b}=0$, eq.~\eqref{eq:diff-field} yields the 
field radiated by an infinite-size aperture that is blocked by
an opaque strip occupying the spatial region $(x_\text{b}^{(1)},x_\text{b}^{(2)})$
on the aperture plane at $z = 0$.
In the paraxial regime, under the Fresnel approximation $|x-\xi | \ll z-z_\text{b}$, and using the 
asymptotic form in \eqref{eq:approx-hank} of the Hankel function for large arguments,
the diffracted field in \eqref{eq:diff-field} 
for $z>z_\text{b}$
can be approximated as
\be
\E_\text{d}(z,x)=u_\text{d}(z,x) \, e^{-j k_0 z},
\label{eq:field-decomp-diff}
\ee
where
\be
u_\text{d}(z,x) = \sqrt{\frac{j}{\lambda_0 (z-z_\text{b})}}
\int_{\mathcal{O}_\text{d}} u(z_\text{b},\xi) \,  e^{-j \frac{k_0}{2 (z-z_\text{b})} (x-\xi)^2} \, \mathrm{d}\xi  \: .
\label{eq:pert-ap}
\ee
A similar expression holds for the diffracted field in the region $z < z_b$, where the integration is 
carried out over the complementary region (i.e., the obstacle), and a minus sign is included to account for the reflection coefficient.

In general, the integral in~\eqref{eq:pert-ap} does not admit 
a closed-form expression and must be evaluated numerically to 
assess the impact of the knife-edge obstacle on the beam's 
propagation.
In the forthcoming numerical study, we compare three 
different beamforming strategies. The first employs the Airy beam, 
which has been extensively discussed in Section~\ref{sec:airy-free}. 
In this case, the field $u(z_\text{b},\xi)$ in~\eqref{eq:pert-ap} 
is computed by using the aperture field given in~\eqref{eq_Ea}, 
with parameters $\nu_\text{a}=0$, $\alpha_\text{a}=0.1/\lambda_0$, 
and $\gamma_\text{a}=k_0/9$. The corresponding beam at the 
obstruction plane is obtained by numerically 
evaluating~\eqref{eq:Ray-2}, considering a finite-size aperture 
defined by $x_\text{a}^{(1)}=-13\,\lambda_0$ and 
$x_\text{a}^{(2)}=13\,\lambda_0$, which yields an overall aperture 
width of $\Delta x_\text{a}=26\,\lambda_0$ (full aperture).
The second strategy utilizes a Gaussian beam, 
referred to as ``Gaussian (collimated; full aperture)'' in the 
plots, which follows a straight-line propagation path, and 
therefore does not exhibit self-acceleration (see 
Appendix~\ref{app:Gauss} for details). The Gaussian beam is also 
launched from the same finite aperture of width 
$\Delta x_\text{a}=26\,\lambda_0$.
The receiver is positioned at coordinates
$(z_\text{r}, x_\text{r}) = (40\,\lambda_0,\, 3.27\,\lambda_0)$, i.e., at an
intermediate location between the nominal caustic of the Airy beam -- which,
according to~\eqref{eq:caustic-airy-gen}, lies at $x = 3.45\,\lambda_0$ for
$z_\text{r} = 40\,\lambda_0$ -- and the straight line~\eqref{eq:gauss-caustic},
along which the field intensity of the collimated Gaussian beam is maximized,
located at $x = 3.20\,\lambda_0$. Such an intermediate placement avoids
biasing the comparison in favor of either beam.
The third strategy uses a Gaussian beam focused 
at the receiver, i.e., with its minimum waist at $z_\text{r}$ 
(see Appendix~\ref{app:Gauss}), defined over the portion of the 
transmitting aperture that remains in line-of-sight with the 
receiver, i.e., the subregion from which rays are not blocked by 
the obstacle (``residual'' aperture). It is worth emphasizing 
that this comparison is not entirely fair with respect to the 
other two strategies: unlike the Airy and collimated Gaussian 
beams, synthesizing a Gaussian beam focused at the receiver and 
defined over the residual aperture requires additional side 
information, namely the positions of both the receiver and the 
obstacle. It is therefore included as a performance upper bound 
rather than a practically equivalent baseline.
As for the Airy beam, the diffracted field in~\eqref{eq:pert-ap} 
for both the Gaussian cases is numerically 
computed through~\eqref{eq:Ray-2}, using the aperture field 
in~\eqref{eq:Ea-Gauss}, with beam waist radius 
$\omega_\text{a}=\sqrt{6}\,\lambda_0$, and launch-angle parameter 
$\mu_\text{a}=0.08\,k_0$.

\begin{figure*}[!t]
\centering
\includegraphics[width=\textwidth]
{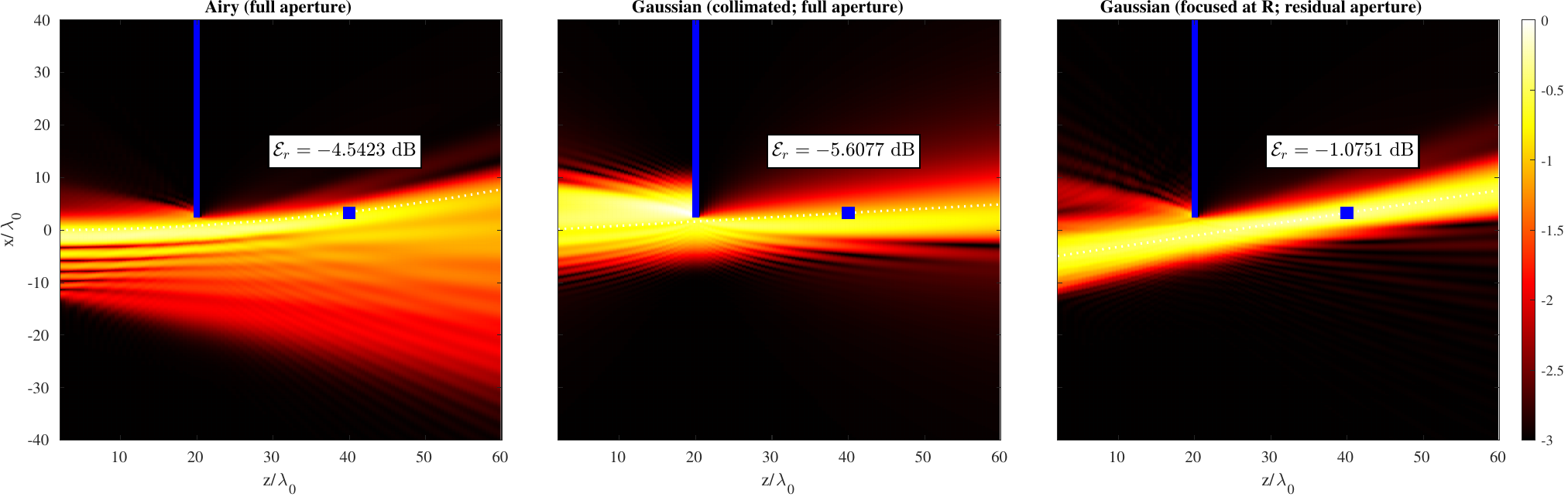}\par\vspace{0.6em}
\includegraphics[width=\textwidth]
{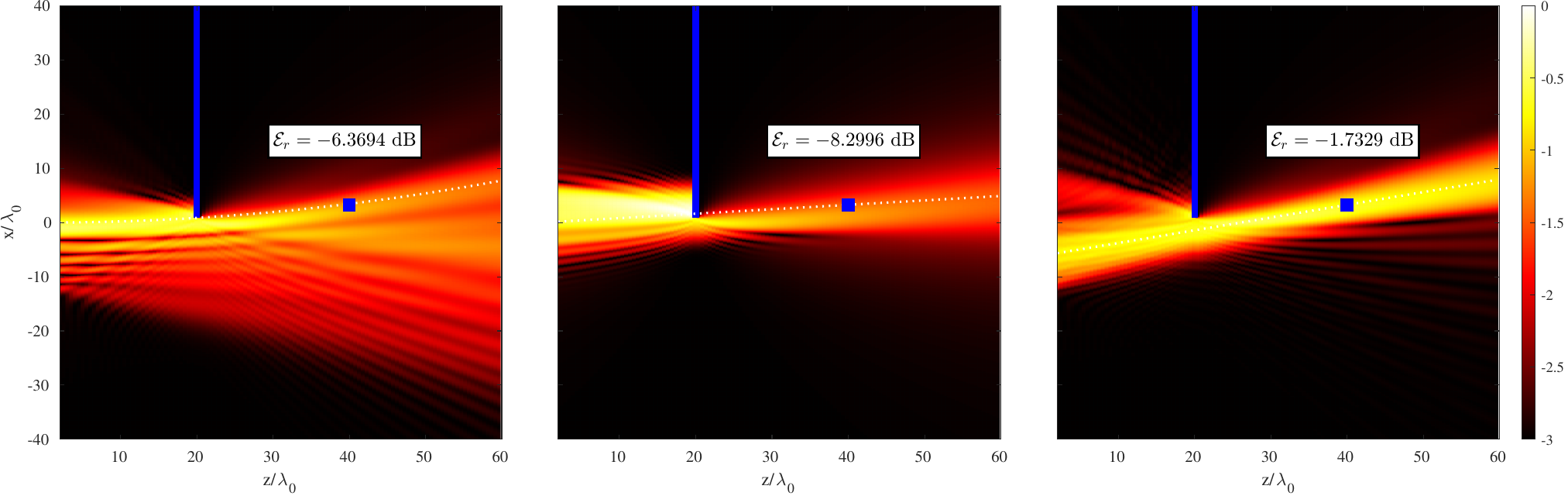}\par\vspace{0.6em}
\includegraphics[width=\textwidth]
{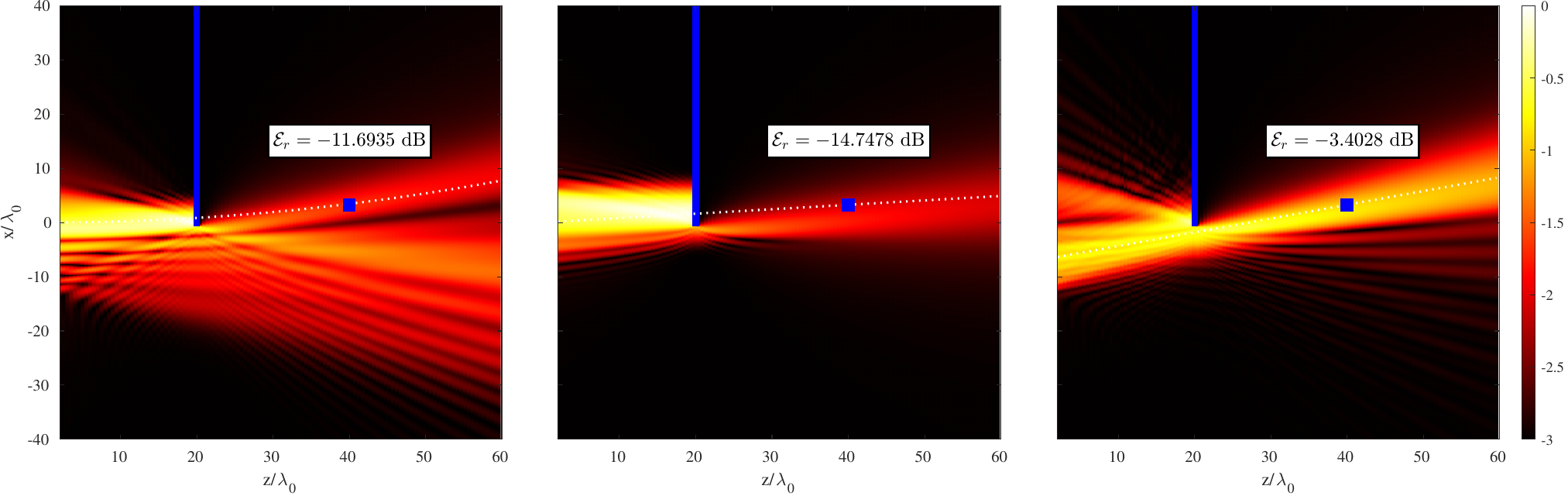}
\caption{Intensity distributions (in 
$\log_{10}$ scale) versus propagation distance $z$ and 
transverse coordinate $x$ for: (i)~Airy beam 
($\nu_\text{a}=0$, $\alpha_\text{a}=0.1/\lambda_0$, and 
$\gamma_\text{a}=k_0/9$); (ii)~collimated Gaussian beam 
($\omega_\text{a}=\sqrt{6}\,\lambda_0$ and 
$\mu_\text{a}=0.08\,k_0$); and (iii)~Gaussian beam focused 
at the receiver and defined over the residual aperture (same 
$\omega_\text{a}$ and $\mu_\text{a}$). The white dotted curve 
denotes the caustic trajectory. The finite-width transmit 
aperture extends from $x_\text{a}^{(1)}=-13\,\lambda_0$ to 
$x_\text{a}^{(2)}=13\,\lambda_0$ and is partially obstructed 
by a semi-infinite knife-edge obstacle located at distance 
$z_\text{b}=20\,\lambda_0$ from the aperture, where 
$x_\text{b}^{(2)}=+\infty$ and $x_\text{b}^{(1)}$ is given 
by~\eqref{eq:xB}. The obstacle is shown as a thick blue line, 
and the receiver is indicated by a blue square at 
$(40\,\lambda_0,\,3.27\,\lambda_0)$. All spatial coordinates 
are normalized by the wavelength $\lambda_0$. \textbf{Top 
row:} clearance $\Delta=1.5\,\lambda_0$. \textbf{Middle row:} 
clearance $\Delta=0$. \textbf{Bottom row:} clearance 
$\Delta=-1.5\,\lambda_0$.}
\label{fig:fieldmaps_3rows}
\end{figure*}

Figure~\ref{fig:fieldmaps_3rows} shows the 
intensity distributions of the three beams under comparison for 
a semi-infinite knife-edge obstacle located at distance 
$z_\text{b}=20\,\lambda_0$ from the aperture, for three values 
of the beam clearance $\Delta \in \{1.5\,\lambda_0, 0, 
-1.5\,\lambda_0\}$ defined in~\eqref{eq:xB}. It can be observed 
that, relative to the collimated Gaussian beam, the Airy beam 
yields higher values of received energy $\mathcal{E}_\text{r}$ 
at the receiver in the considered NLoS settings, owing to its 
parabolic propagation trajectory. This advantage directly 
translates into higher SNR at the receiver. However, its 
performance degrades as the size of the unobstructed aperture 
decreases, confirming that the self-healing capability of Airy 
beams is critically dependent on the portion of the transmit 
aperture that maintains LoS with the receiver. With reference 
to the focused Gaussian beam defined over the residual aperture, 
we observe that in all the scenarios considered in 
Fig.~\ref{fig:fieldmaps_3rows}, it provides significant 
performance gains over both the Airy and the collimated Gaussian 
beams, consistent with the findings of~\cite{Inserra_2022}. 
However, as noted above, this comparison is not entirely fair, 
since it requires side information on both the receiver and 
obstacle positions that is not needed by the other two 
strategies. We expect that, by leveraging such information, one can tailor 
the Airy-beam trajectory to bypass the blocker and maximize the 
delivered power at the destination. 

Along these lines, 
\cite{Ghasempour} shows that an Airy profile can be synthesized 
via a cubic phase mask augmented with quadratic (focusing) and 
linear (steering) terms, enabling the beam to be formed and 
directed toward the receiver with improved performance. 
Similarly, the optimization-based frameworks 
of~\cite{Liu_ArXiV2025} and~\cite{Yao_2026} provide systematic 
methods for jointly optimizing beam steering, focusing, and 
curving parameters in blockage-prone environments, and represent 
natural extensions of the wave-optics analysis developed here 
toward practical system design. The connection between the 
aperture phase profile derived in eq.~\eqref{eq_Ea-airy-gen} and the 
cubic phase parametrization adopted in these works also 
highlights the complementarity between the continuous-aperture 
framework of the present paper and discrete-array or 
metasurface-based implementations. However, a full 
optimization of the Airy-beam trajectory in partially obstructed 
scenarios lies beyond the scope of the present paper and will be 
investigated in future work.

\begin{figure}[t]
\centering
\includegraphics[width=\linewidth]
{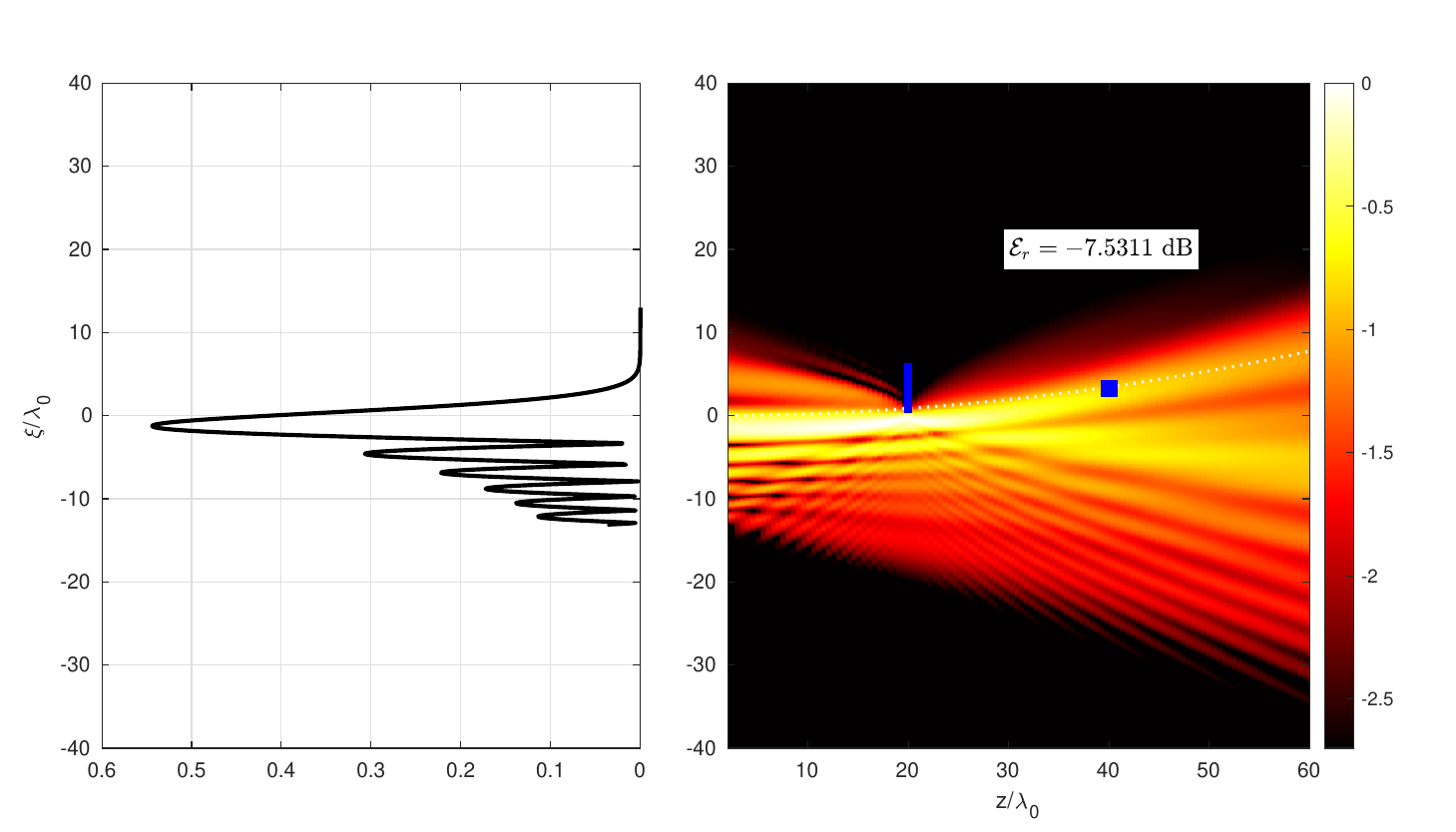}
\vspace{-3mm}
\includegraphics[width=\linewidth]
{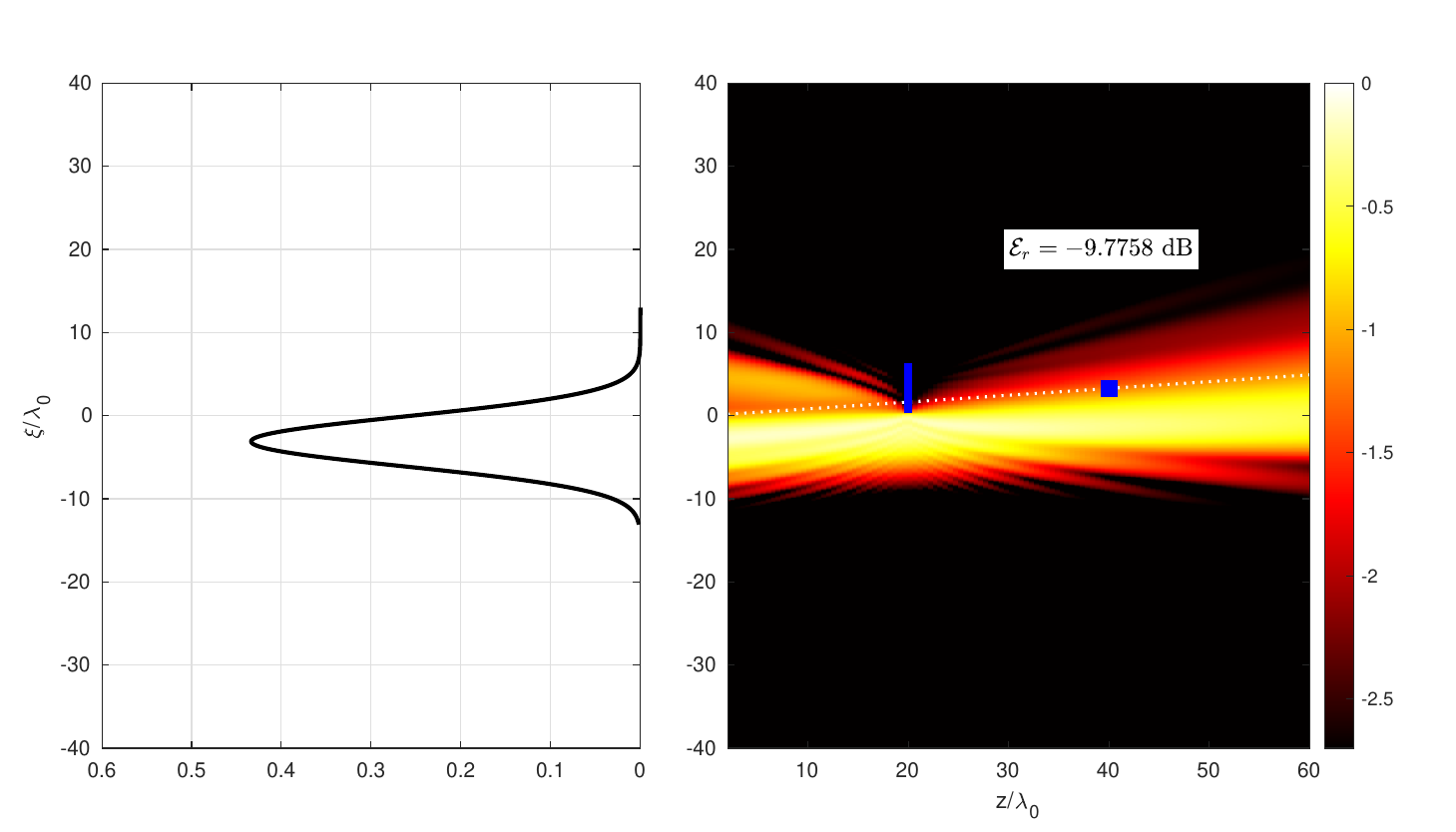}
\vspace{-3mm}
\caption{Finite-width transmit aperture extending from 
$x_\text{a}^{(1)}=-13\,\lambda_0$ to 
$x_\text{a}^{(2)}=13\,\lambda_0$, partially obstructed by a 
finite-height knife-edge obstacle located at distance 
$z_\text{b}=20\,\lambda_0$ from the aperture, where 
$x_\text{b}^{(2)}=6.27\,\lambda_0$ and 
$x_\text{b}^{(1)}=0.27\,\lambda_0$. The blue thick line 
represents the obstacle, and the blue square indicates the 
receiver located at $(40\,\lambda_0,\,3.27\,\lambda_0)$. All 
spatial coordinates are normalized with respect to the 
wavelength $\lambda_0$. Top-left: Intensity of the aperture 
field~\eqref{eq_Ea} as a function of the $\xi$-coordinate, 
with $\nu_\text{a}=0$, $\alpha_\text{a}=0.1/\lambda_0$, and 
$\gamma_\text{a}=k_0/9$. Top-right: Intensity distribution 
(in $\log_{10}$ scale) of the corresponding Airy beam as a 
function of $z$. The white dotted curve represents 
the caustic trajectory in~\eqref{eq:caustic-airy-gen}. 
Bottom panels: Corresponding results for a spatially shifted 
Gaussian aperture field~\eqref{eq:Ea-Gauss}, with 
$\omega_\text{a}=\sqrt{6}\,\lambda_0$ and 
$\mu_\text{a}=0.08\,k_0$. The white dotted line represents 
the linear trajectory in~\eqref{eq:gauss-caustic}.}
\label{fig:fig_21_rev}
\end{figure}

\begin{figure}[t]
\centering
\includegraphics[width=\linewidth]
{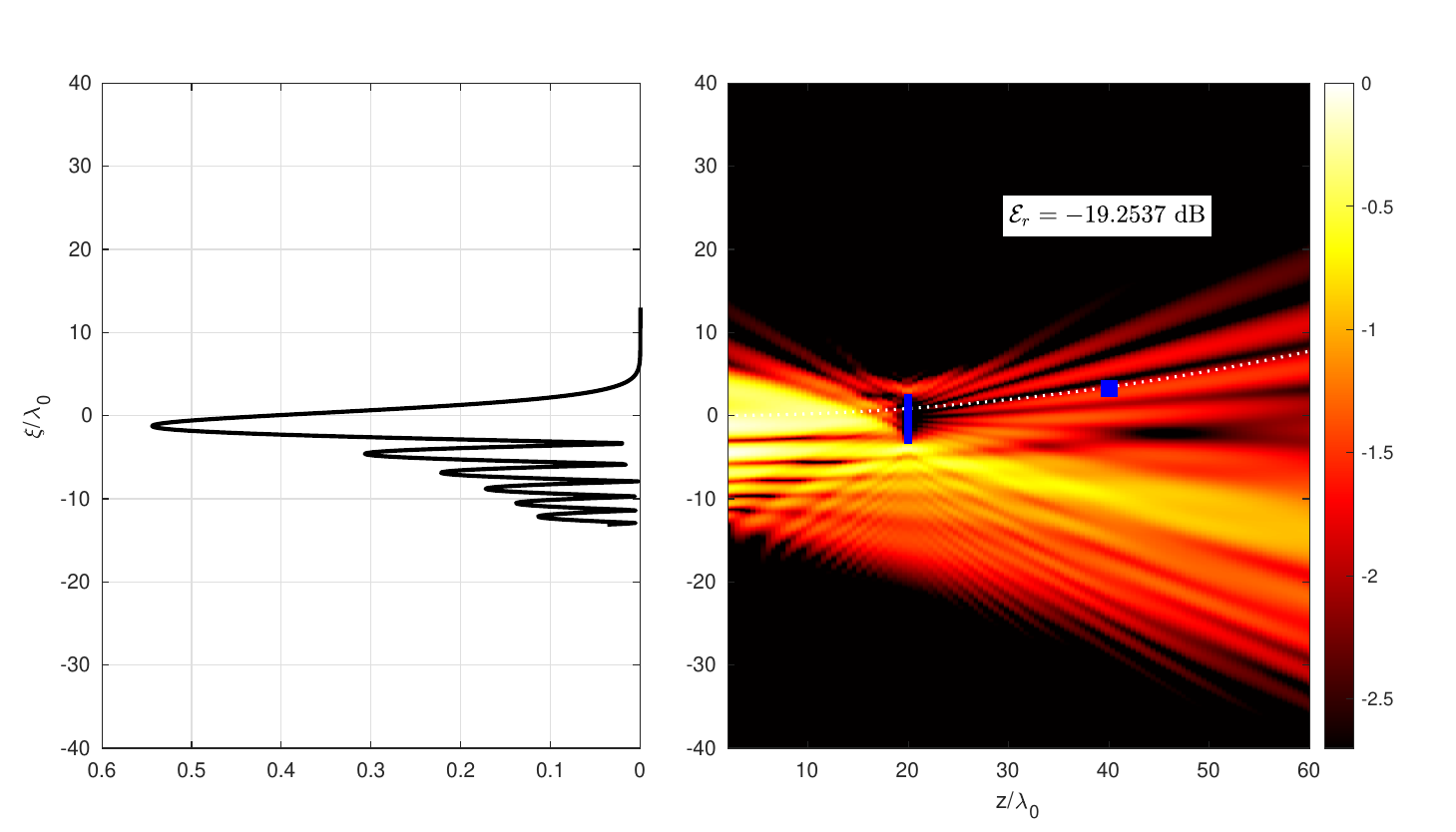}
\vspace{-3mm}
\includegraphics[width=\linewidth]
{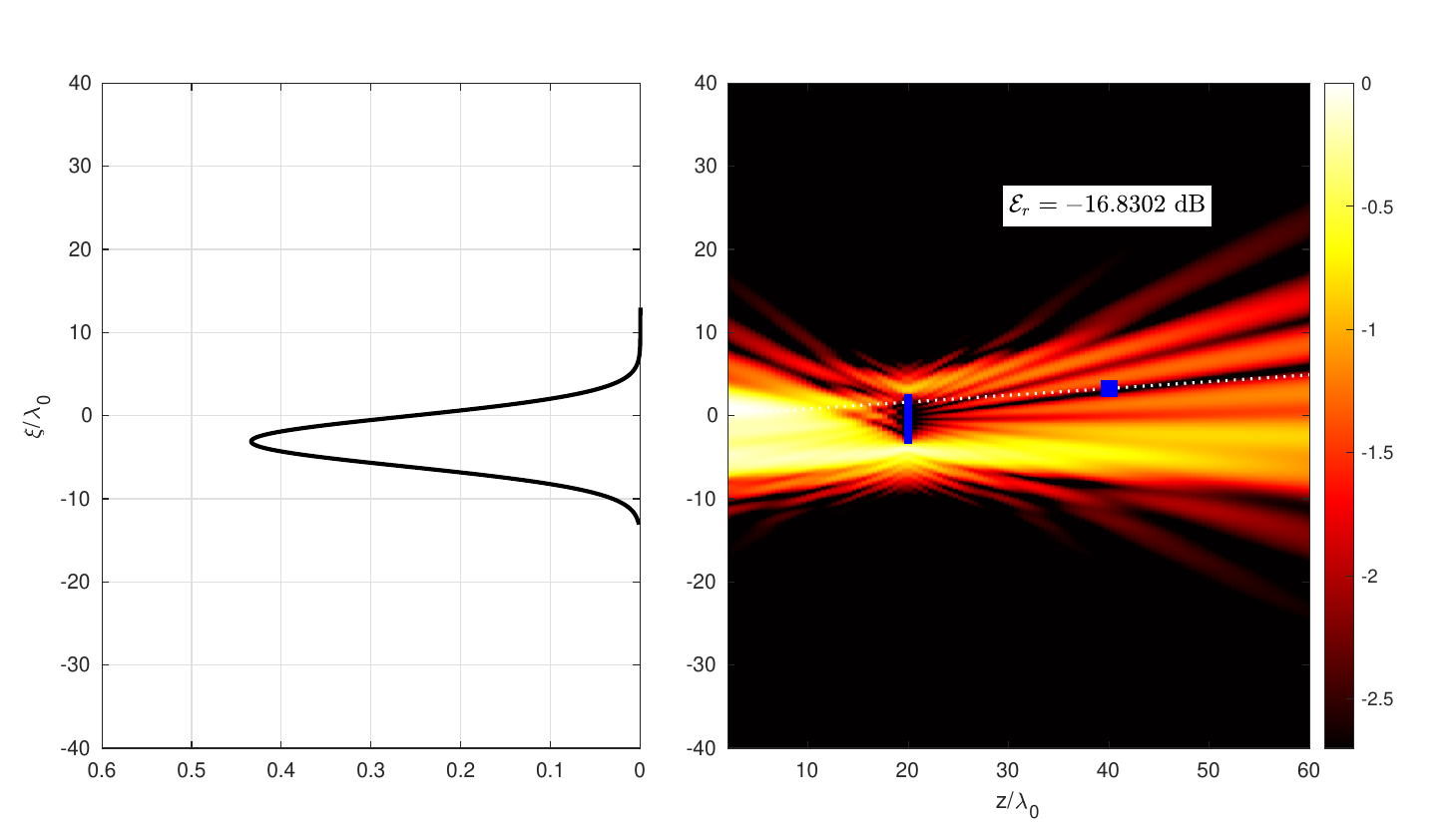}
\vspace{-3mm}
\caption{Finite-width transmit aperture extending from 
$x_\text{a}^{(1)}=-13\,\lambda_0$ to 
$x_\text{a}^{(2)}=13\,\lambda_0$, partially obstructed by a 
finite-height knife-edge obstacle located at distance 
$z_\text{b}=20\,\lambda_0$ from the aperture, where 
$x_\text{b}^{(2)}=2.6\,\lambda_0$ and 
$x_\text{b}^{(1)}=-3.4\,\lambda_0$. The blue thick line 
represents the obstacle, and the blue square indicates the 
receiver located at $(40\,\lambda_0,\,3.27\,\lambda_0)$. All 
spatial coordinates are normalized with respect to the 
wavelength $\lambda_0$. Top-left: Intensity of the aperture 
field~\eqref{eq_Ea} as a function of the $\xi$-coordinate, 
with $\nu_\text{a}=0$, $\alpha_\text{a}=0.1/\lambda_0$, and 
$\gamma_\text{a}=k_0/9$. Top-right: Intensity distribution 
(in $\log_{10}$ scale) of the corresponding Airy beam as a 
function of $z$. The white dotted curve represents 
the caustic trajectory in~\eqref{eq:caustic-airy-gen}. 
Bottom panels: Corresponding results for a spatially shifted 
Gaussian aperture field~\eqref{eq:Ea-Gauss}, with 
$\omega_\text{a}=\sqrt{6}\,\lambda_0$ and 
$\mu_\text{a}=0.08\,k_0$. The white dotted line represents 
the linear trajectory in~\eqref{eq:gauss-caustic}.}
\label{fig:fig_22_rev}
\end{figure}

\begin{figure}[t]
\centering
\includegraphics[width=\linewidth]
{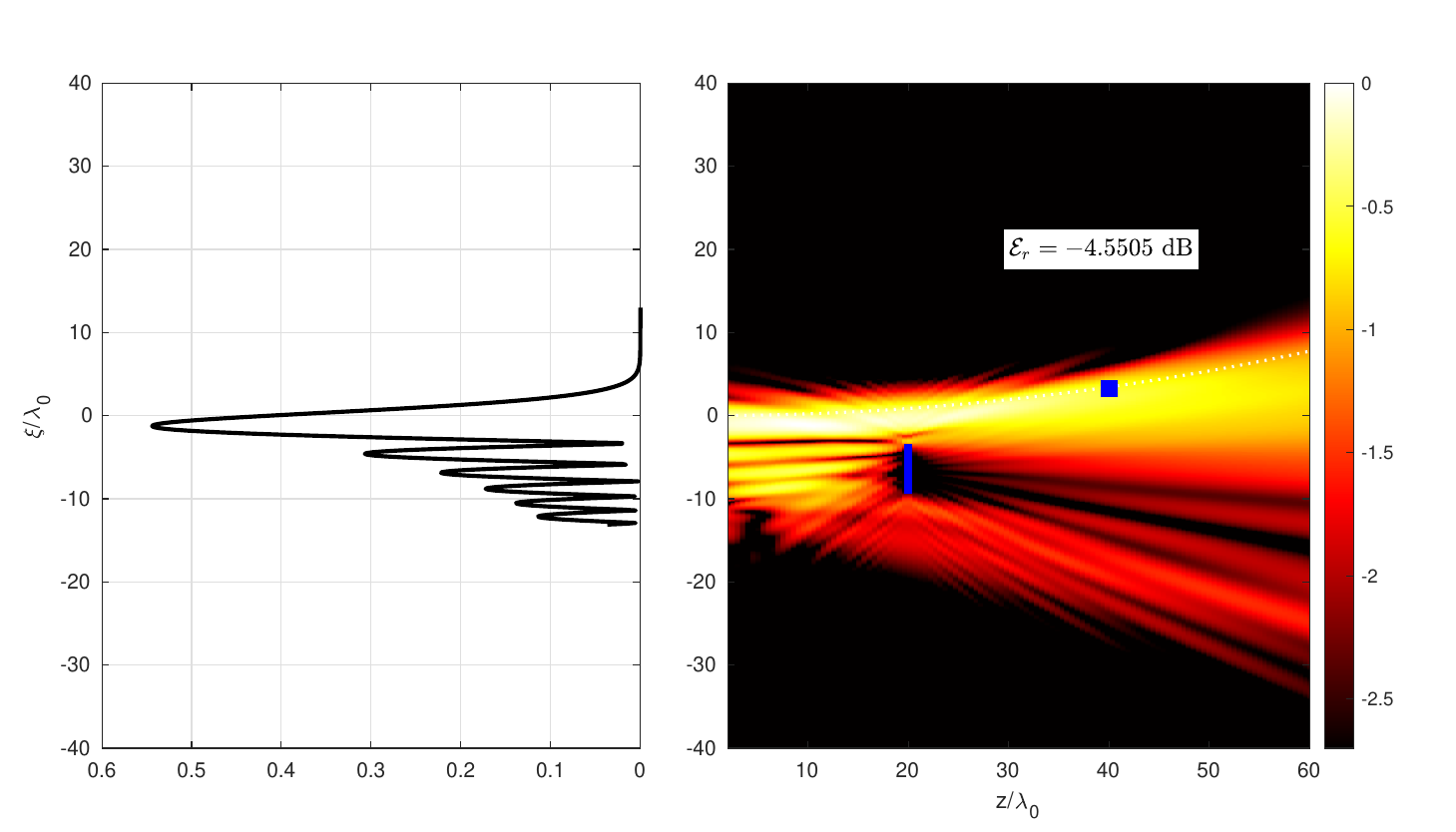}
\vspace{-3mm}
\includegraphics[width=\linewidth]
{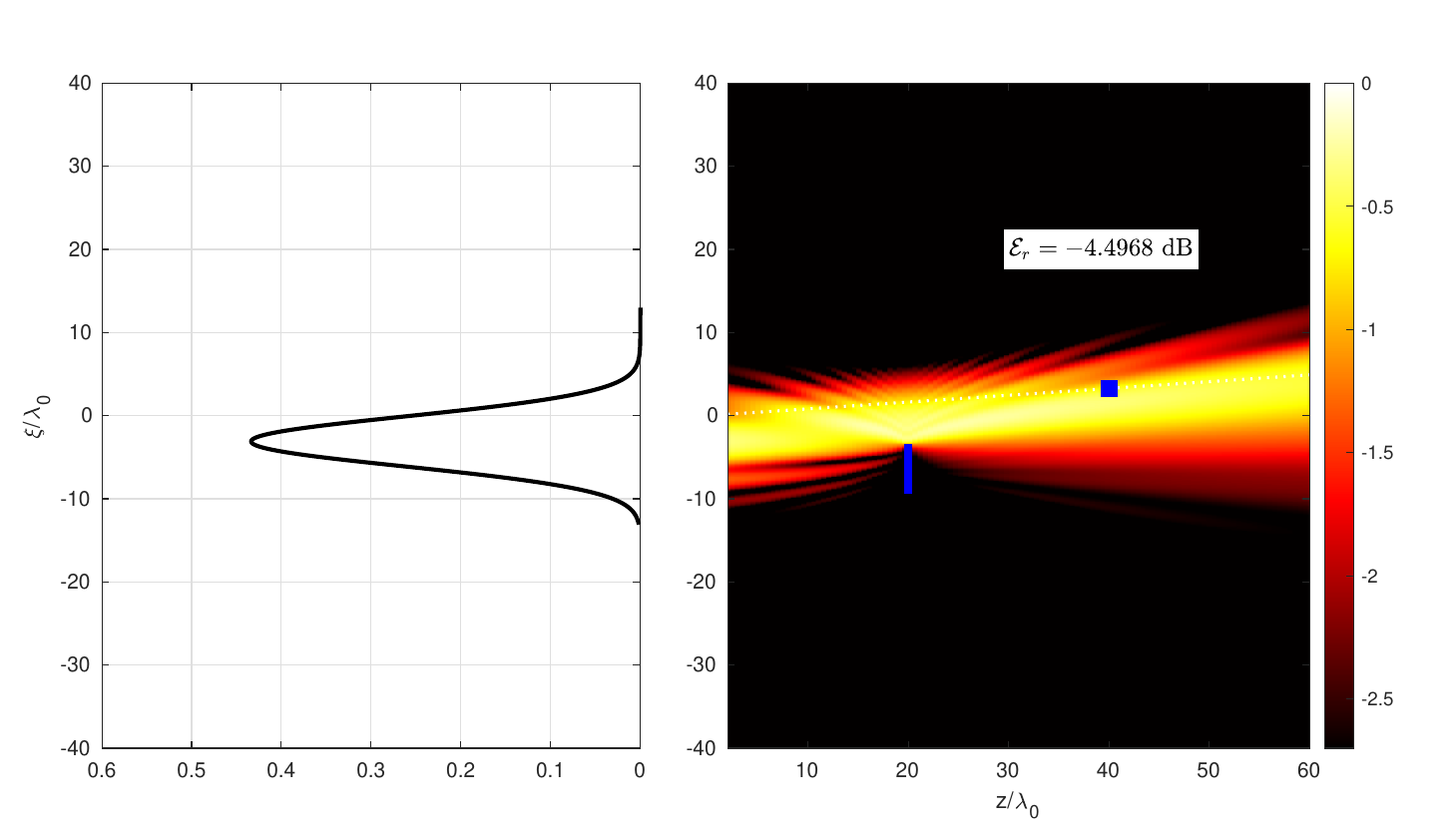}
\vspace{-3mm}
\caption{Finite-width transmit aperture extending from 
$x_\text{a}^{(1)}=-13\,\lambda_0$ to 
$x_\text{a}^{(2)}=13\,\lambda_0$, partially obstructed by a 
finite-height knife-edge obstacle located at distance 
$z_\text{b}=20\,\lambda_0$ from the aperture, where 
$x_\text{b}^{(2)}=-3.4\,\lambda_0$ and 
$x_\text{b}^{(1)}=-9.4\,\lambda_0$. The blue thick line 
represents the obstacle, and the blue square indicates the 
receiver located at $(40\,\lambda_0,\,3.27\,\lambda_0)$. All 
spatial coordinates are normalized with respect to the 
wavelength $\lambda_0$. Top-left: Intensity of the aperture 
field~\eqref{eq_Ea} as a function of the $\xi$-coordinate, 
with $\nu_\text{a}=0$, $\alpha_\text{a}=0.1/\lambda_0$, and 
$\gamma_\text{a}=k_0/9$. Top-right: Intensity distribution 
(in $\log_{10}$ scale) of the corresponding Airy beam as a 
function of $z$. The white dotted curve represents 
the caustic trajectory in~\eqref{eq:caustic-airy-gen}. 
Bottom panels: Corresponding results for a spatially shifted 
Gaussian aperture field~\eqref{eq:Ea-Gauss}, with 
$\omega_\text{a}=\sqrt{6}\,\lambda_0$ and 
$\mu_\text{a}=0.08\,k_0$. The white dotted line represents 
the linear trajectory in~\eqref{eq:gauss-caustic}.}
\label{fig:fig_23_rev}
\end{figure}

Figures~\ref{fig:fig_21_rev}, \ref{fig:fig_22_rev}, 
and~\ref{fig:fig_23_rev} examine the case of a knife-edge 
obstacle with finite height.
In this case, we compare only the Airy and 
the collimated Gaussian beams, since the residual-aperture 
definition may not be applicable when the obstacle obstructs 
central regions of the transmit aperture.
In such an NLoS scenario, the obstacle height equals $6\,\lambda_0$, 
which is smaller than the aperture width 
$\Delta x_\text{a}=26\,\lambda_0$.
Once again, the receiver is positioned at coordinates 
$(z_\text{r},x_\text{r})=(40\,\lambda_0,\,3.27\,\lambda_0)$.
In Fig.~\ref{fig:fig_21_rev}, the obstacle is symmetrically 
positioned along the $x$-axis with respect to the receiver, 
with its top edge at 
$x_\text{b}^{(2)}=x_\text{r}+3\,\lambda_0$ and its bottom 
edge at $x_\text{b}^{(1)}=x_\text{r}-3\,\lambda_0$.

To quantify the portion of the transmitting 
aperture that is obstructed by the obstacle, we define the 
energy metric
\be
\en_\text{obs} \eqdef \int_{(x_\text{a}^{(1)},x_\text{a}^{(2)}) 
\cap (x_\text{b}^{(1)},x_\text{b}^{(2)})} 
|\E_\text{a}(\xi)|^2\,\mathrm{d}\xi,
\ee
where the integration domain 
$(x_\text{a}^{(1)},x_\text{a}^{(2)}) \cap 
(x_\text{b}^{(1)},x_\text{b}^{(2)})$ identifies the subregion 
of the aperture from which the rays are blocked by the obstacle.
Quite simply, the metric $\en_\text{obs}$ quantifies the 
portion of the total transmit energy $\en_\text{a}$ that is 
intercepted by the obstacle and therefore does not contribute 
to the field strength at the receiver.
The corresponding value of the obstructed energy fraction is 
$\frac{\en_\text{obs}}{\en_\text{a}}=20\%$ for both the Airy 
and the collimated Gaussian beams. Roughly, the obstacle dims 
the main lobe of the aperture field in~\eqref{eq_Ea} for 
$\xi>0$.
In Fig.~\ref{fig:fig_22_rev}, the bottom edge of the obstacle 
is positioned at the first zero of the Airy beam, i.e., 
$x_\text{b}^{(1)}=-3.4\,\lambda_0$, while the top edge is 
located at 
$x_\text{b}^{(2)}=-3.4\,\lambda_0+6\,\lambda_0=2.6\,\lambda_0$. 
In this configuration, the obstructed energy fraction amounts 
to approximately $\frac{\en_\text{obs}}{\en_\text{a}}\approx 
75\%$ for both beams. About $64\%$ of the main lobe of the 
aperture field in~\eqref{eq_Ea} is dimmed.
In Fig.~\ref{fig:fig_23_rev}, the top edge of the obstacle is 
positioned at the first zero of the Airy beam, i.e., 
$x_\text{b}^{(2)}=-3.4\,\lambda_0$, while the bottom edge is 
located at 
$x_\text{b}^{(1)}=-3.4\,\lambda_0-6\,\lambda_0=-9.4\,\lambda_0$. 
In such a case, we have 
$\frac{\en_\text{obs}}{\en_\text{a}}\approx 18\%$ for both 
beams. The obstacle dims the first three secondary lobes of 
the aperture field in~\eqref{eq_Ea}, while its main lobe 
remains in LoS with the receiver.
The remaining simulation parameters used to generate the 
results in Figs.~\ref{fig:fig_21_rev}--\ref{fig:fig_23_rev} 
are identical to those specified for 
Fig.~\ref{fig:fieldmaps_3rows}.

It can be argued from Fig.~\ref{fig:fig_21_rev} that the 
self-accelerating Airy beam delivers higher field strength 
at the receiver when its main lobe on the negative $\xi$-axis 
falls within the unobstructed aperture, outperforming the 
collimated Gaussian beam with a gain of approximately 
$2.2$~dB. Airy and Gaussian beams perform comparably for 
an obstruction of the secondary lobes of the aperture field 
in~\eqref{eq_Ea} as in Fig.~\ref{fig:fig_23_rev}. However, 
in the case of a $64\%$ obstruction of the main lobe 
in~\eqref{eq_Ea} (see Fig.~\ref{fig:fig_22_rev}), the Airy 
beam exhibits an energy penalty at the receiver of about 
$2.4$~dB relative to the Gaussian beam. These results 
obtained in the case of a finite-height knife-edge obstacle 
are in line with those observed for a semi-infinite 
knife-edge obstruction.

These results, taken together, allow us to 
refine the assessment of~\cite{Inserra_2022} regarding the 
practical applicability of Airy beams. 
While~\cite{Inserra_2022} concludes, from an optimization 
perspective on discrete apertures, that near-field focusing 
on the residual LoS aperture generally outperforms 
self-accelerating beams, our wave-optics analysis reveals 
that this conclusion holds mainly when the main lobe of the 
aperture Airy field is heavily obstructed. When instead the 
obstruction affects predominantly the secondary lobes, 
preserving the main lobe in LoS, Airy beams exhibit robust 
self-healing and outperform collimated Gaussian beams, 
confirming that their potential advantage is not negligible 
but rather tightly linked to the specific obstruction 
geometry.

\emph{Computational methodology:} The 
Rayleigh-Sommerfeld diffraction integral was evaluated by 
direct numerical quadrature using the composite trapezoidal 
rule on a uniform aperture grid. No geometric theory of 
diffraction (GTD) or uniform theory of diffraction (UTD) 
approximations are employed: the obstacle is modeled 
directly at the propagation level as a hard truncation of 
the integration domain. Table~\ref{tab:rs_discretization} 
lists the discretization parameters used for the Airy, 
collimated Gaussian, and focused Gaussian beams in the 
scenario of Fig.~\ref{fig:fieldmaps_3rows}. For the 
Gaussian beam focused at the receiver and defined over the 
residual aperture, the parameters in 
Table~\ref{tab:rs_discretization} refer to the 
zero-clearance case ($\Delta=0$). Specifically, we report 
the grid sizes, the spatial sampling steps, and the total 
number of generated samples.
For each field sample $(x,z)$, the 
Rayleigh-Sommerfeld integral is approximated by a 
trapezoidal sum over $N_a$ aperture samples, requiring 
$\mathcal{O}(N_a)$ operations per sample. Over an 
observation grid with $N_x$ points along $x$ and $N_z$ 
points along $z$, the overall run-time scales as 
$\mathcal{O}(N_a N_x N_z)$. Equivalently, the FLOP count 
can be expressed as $\mathcal{F}\approx n_s\,N_a N_x N_z$, 
where $n_s$ denotes the average number of floating-point 
operations per summand in the trapezoidal evaluation, 
including the evaluation of the Hankel function and the 
complex exponential. In our simulations, this yields 
$\mathcal{F}\approx 2.446 \times 10^7\times n_s$ FLOPs 
for the Airy and collimated Gaussian beams, and 
$\mathcal{F}\approx 1.303 \times 10^7\times n_s$ FLOPs 
for the focused Gaussian beam.

\begin{table}[t]
\centering
\caption{Parameters for numerical Rayleigh-Sommerfeld 
integration.}
\label{tab:rs_discretization}
\setlength{\tabcolsep}{4pt}
\renewcommand{\arraystretch}{1.05}
\begin{tabular}{lcccc}
\hline
Grid & Min & Max & Step & Samples \\
\hline
Aperture (Airy/Gauss.\ collim.) 
& $-13\lambda_0$ & $13\lambda_0$ 
& $0.1\lambda_0$ & $261$ \\
Aperture (Gauss.\ focus.\ res.\ apert.) 
& $-13\lambda_0$ & $0.8 \lambda_0$ 
& $0.1\lambda_0$ & $139$ \\
Axis $x$ 
& $-40\lambda_0$ & $40\lambda_0$ 
& $0.1\lambda_0$ & $801$ \\
Axis $z$ 
& $2\lambda_0$ & $60\lambda_0$ 
& $0.5\lambda_0$ & $117$ \\
\hline
\end{tabular}
\end{table}

\subsection{Paraxial Airy beams in the presence of a soft obstacle}
\label{sec:soft-obstacle}

Fig.~\ref{fig:fieldmaps_3rows} clearly illustrates one of the most remarkable properties of 
diffraction-resistant beams: Their ability to self-reconstruct during propagation after encountering an
obstacle, a phenomenon known as {\em self-healing} \cite{Broky-2008}.
This property can be intuitively explained through
the ray-optics interpretation of self-accelerating beams.
As discussed in Section~\ref{sec:self-acc-beam}, 
such beams are composed of rays that emerge sideways from distant points 
across the transmit aperture. As a result, even when part of the aperture is blocked by an obstacle, these peripheral rays contribute to the reconstruction of the wavefront beyond the obstruction.
In principle, self-healing is a potential feature of any beam that admits an 
angular spectrum representation \cite{Aiello}. 

The self-healing properties of Airy beams can be studied analyti\-cally in specific scenarios.
A mathematically  tractable case arises when 
the beam is partially obstructed by an infinite opaque screen characterized by 
a Gaussian attenuation profile. 
This model was examined in  \cite{Chu-2012}, where the obstacle was assumed to lie directly on the aperture plane at $z=0$. 
In this setting, it was shown that the resulting diffracted field can be decomposed into two components: 
An Airy beam and a perturbative term that decreases with distance. Importantly, beyond a certain longitudinal distance, referred to as the {\em self-healing distance}, the perturbation becomes negligible, and the beam regains its original Airy-like shape. 
This distance depends on the size of the obstacle 
and the portion of the incident  beam that is blocked. 
Despite its elegant analytical formulation, the configuration analyzed in \cite{Chu-2012} is of limited relevance for wireless communications scenarios, where obstruction typically occurs in the radiative near-field propagation path rather than directly at the aperture plane. 

Herein,  we extend the approach proposed in \cite{Chu-2012} by relaxing the constraint that confines the obstacle to the aperture plane, allowing instead its placement  at an arbitrary position $z_\text{b}>0$. 
In particular, we model the obstruction as a \emph{soft} barrier, i.e., a spatially smooth, partially transparent screen, which extends 
infinitely along both the $x$- and $y$-axes. The {\em transmittance} of the obstacle varies continuously with the transverse 
coordinate $x$, and is defined as
\be
\tau_\text{obs}(x) \eqdef 1-a_\text{obs}(x), 
\ee
where the {\em absorption efficiency} $a_\text{obs}(x)$ follows a Gaussian profile given by 
\be
a_\text{obs}(x) = e^{-\frac{(x-\mu_\text{obs})^2}{2 \sigma_\text{obs}^2}} \quad \text{for $x \in \mathbb{R}$}. 
\ee
Here, $\mu_\text{obs}$ and  $\sigma_\text{obs}$ denote the center and effective width of the absorbing region, respectively.
This spatially varying attenuation captures the behavior of a soft obstruction, creating a gradual transition between fully transparent and fully opaque regions.
In this case, by resorting to the Huygens-Fresnel diffraction framework, 
the diffracted field in the paraxial regime can still be expressed  in the form of 
\eqref{eq:field-decomp-diff} for $z>z_\text{b}$, where now 
\begin{multline}
u_\text{d}(z,x) = \sqrt{\frac{j}{\lambda_0 (z-z_\text{b})}}
\\ \cdot 
\int_{-\infty}^{+\infty}
u(z_\text{b},\xi) \,  \tau_\text{obs}(\xi) \, e^{-j \frac{k_0}{2 (z-z_\text{b})} (x-\xi)^2} \, \mathrm{d}\xi,  
\label{eq:pert-ap-soft}
\end{multline}
with $u(z,x)$ denoting the beam that propagates from the aperture plane $z = 0$ up 
to the obstructive plane $z = z_\text{b}$.
The use of a soft obstacle model ensures the mathematical tractability of \eqref{eq:pert-ap-soft} and enables, as shown in the following proposition,  a closed-form expression for the  diffracted field  (see Appendix~\ref{app:prop-4} for the detailed derivation).

\vspace{2mm}
\begin{proposition}
\label{prop:pro}
For $z>z_\text{b}$, the field  $\Ep(z,x)$ diffracted by a soft Gaussian-shaped obstacle, as defined in \eqref{eq:pert-ap-soft}, can be expressed as  
\be
\Ep(z,x) = u(z,x)-p(z,x),
\label{eq:diff-field-prop}
\ee
where $u(z,x)$ denotes the Airy beam that would propagate in the absence of the obstacle, 
given by \eqref{eq:airy-def}, while
$p(z,x)$ is a perturbative term that characterizes the effect of the obstacle. This perturbative term takes the form:
\begin{multline}
p(z,x) = U_\text{b} \, \sqrt{\frac{\pi}{\lambda_0 (z-z_\text{b}) \, \eta_\text{b}(z)}} \, 
e^{-\frac{\gamma_\text{a}^3 \, \nu_\text{b}(z,x)}{8 \eta_\text{b}^2(z)}}
\\ 
\cdot \Ai \left(\delta_\text{b} - \frac{\gamma_\text{a}^4}{16 \, \eta_\text{b}^2(z)} + j \frac{\gamma_\text{a} \, \nu_\text{b}(z,x)}{2 \, \eta_\text{b}(z)} \right) 
\, e^{-j \psi_\text{b}(z,x)},
\label{eq:p-airy}
\end{multline}
where
\begin{multline}
\psi_\text{b}(z,x) \eqdef - \frac{\gamma_\text{a}^2}{4 \, \eta_\text{b}(z)} \left[ \delta_\text{b} - \frac{\gamma_\text{a}^4}{24 \, \eta_\text{b}^2(z)}
+\frac{\nu_\text{b}^2(z,x)}{\gamma_\text{a}^2}\right] \\
+ \frac{k_0 \, x^2}{2 \, (z-z_\text{b})} -\frac{\pi}{2} \:,
\label{eq:psib}
\end{multline}
with
\barr
\delta_\text{b} &\eqdef - \frac{z_\text{b}^2 \, \gamma_\text{a}^4}{4 \, k_0^2}- \frac{z_\text{b} \, \gamma_\text{a} \, \nu_\text{a}}{k_0}
- j \frac{z_\text{b} \, \gamma_\text{a} \, \alpha_\text{a}}{k_0}, 
\label{eq:deltab}
\\
U_\text{b} &\eqdef U_\text{a} \, e^{-\alpha_\text{a} \left(\frac{z_\text{b}^2 \, \gamma_\text{a}^3}{2 \, k_0^2}+\frac{z_\text{b} \, \nu_\text{a}}{k_0}\right)}
\, e^{-\frac{\mu_\text{obs}^2}{2 \sigma_\text{obs}^2}} 
\nonumber \\ & \hspace{10mm} \cdot 
e^{j \frac{z_\text{b} \, \gamma_\text{a}^2}{2 \, k_0} \left(\frac{z_\text{b}^2 \, \gamma_\text{a}^4}{6 \, k_0^2} 
+ \frac{z_\text{b} \, \gamma_\text{a} \, \nu_\text{a}}{k_0} 
- \frac{\alpha_\text{a}^2}{\gamma_\text{a}^2}+\frac{\nu_\text{a}^2}{\gamma_\text{a}^2} \right)}, 
\label{eq:Ub}
\\
\nu_\text{b}(z,x) &\eqdef  \alpha_\text{a} + \frac{\mu_\text{obs}}{\sigma_\text{obs}^2} -j \left[\nu_\text{a} + \frac{z_\text{b} \, \gamma_\text{a}^3}{2 \, k_0} - \frac{k_0 \, x}{z-z_\text{b}}\right], 
\label{eq:nub}
\\
\eta_\text{b}(z) & \eqdef  -\frac{k_0}{2 \, (z-z_\text{b})}+\frac{j}{2 \, \sigma_\text{obs}^2} \: .
\label{eq:etab}
\earr
\end{proposition}

\vspace{2mm}
It follows from Proposition~\ref{prop:pro}
that, for sufficiently large propagation distances beyond the obstacle, i.e., for 
$z -z_\text{b}\rightarrow + \infty$, 
and for fixed transverse coordinate $x$,
the auxiliary functions $\nu_\text{b}(z,x)$ and $\eta_\text{b}(z)$ become asymptotically independent of both $x$ and $z$, 
and the perturbative term $p(z,x)$  asymptotically vanishes. 
In this regime,  the diffracted field $u_\text d(z,x)$ tends to the unperturbed beam profile that would occur in the absence of the obstacle. This behavior confirms  the self-healing property of Airy beams after partial obstruction. Furthermore, $p(z,x)$ vanishes as the obstacle width tends to zero, i.e., $\sigma_\text{obs} \rightarrow 0$, in agreement with the intuition that a negligible perturbation induces a negligible diffraction effect. A similar behavior occurs when the obstacle position tends to $\pm\infty$, i.e., $\mu_\text{obs} \rightarrow \pm\infty$, corresponding to the case in which the obstacle is located in a region where the Airy beam amplitude is negligible.

\begin{figure}[t]
\centering
\includegraphics[width=\linewidth]{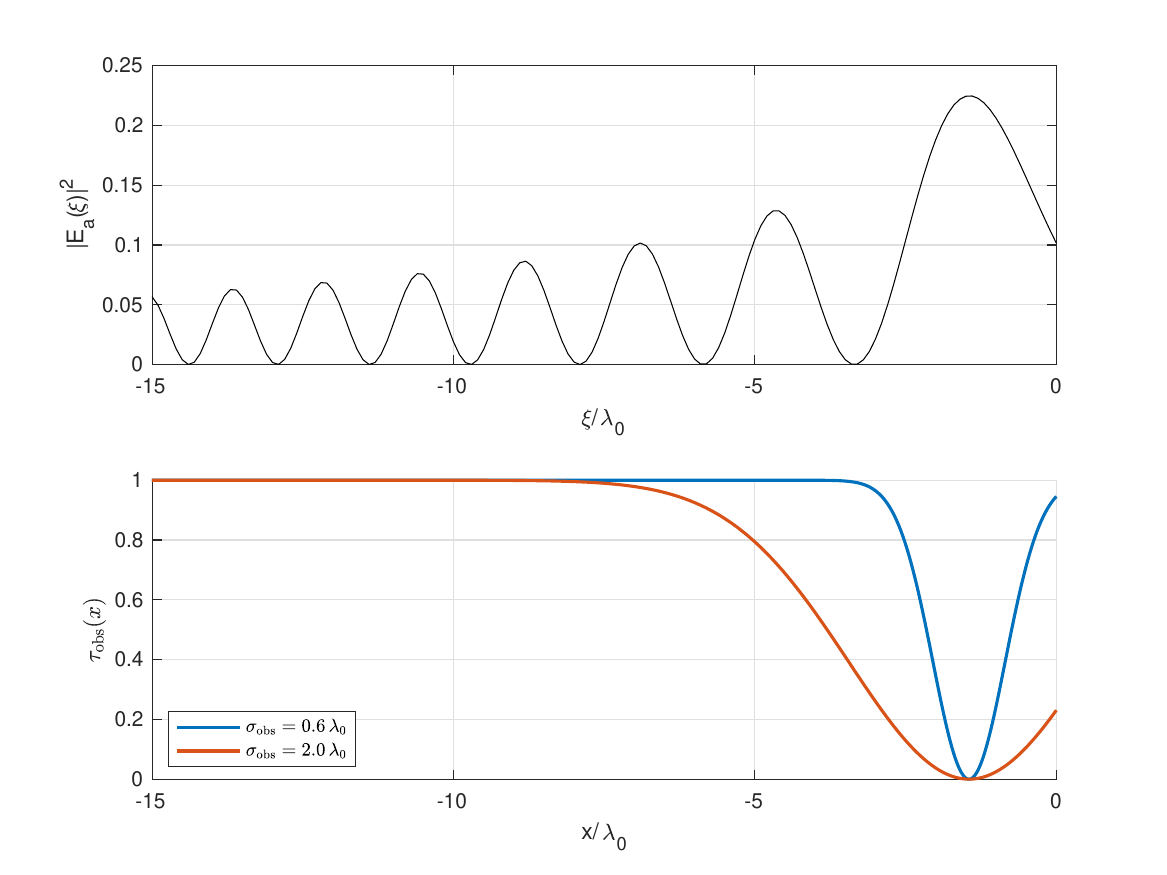} 
\caption{Top: Intensity of the aperture field $\E_\text a(\xi)$, with $\nu_\text{a}=0$, 
$\gamma_\text{a}=\tfrac{k_0}{9}$, and $\alpha_\text{a}=0.01/\lambda_0$. 
Bottom: Transmittance profile of a soft obstacle 
located at $z_\text{b} = 20 \, \lambda_0$ and centered at the main lobe of the Airy beam, i.e.,  
$\mu_\text{obs} = -1.45 \, \lambda_0$, for two different values of $\sigma_\text{obs}$. 
All  spatial coordinates are normalized with respect to the wavelength $\lambda_0$.}
\label{fig:fig_24_rev}
\end{figure}

\begin{figure}[t]
\centering
\includegraphics[width=\linewidth]{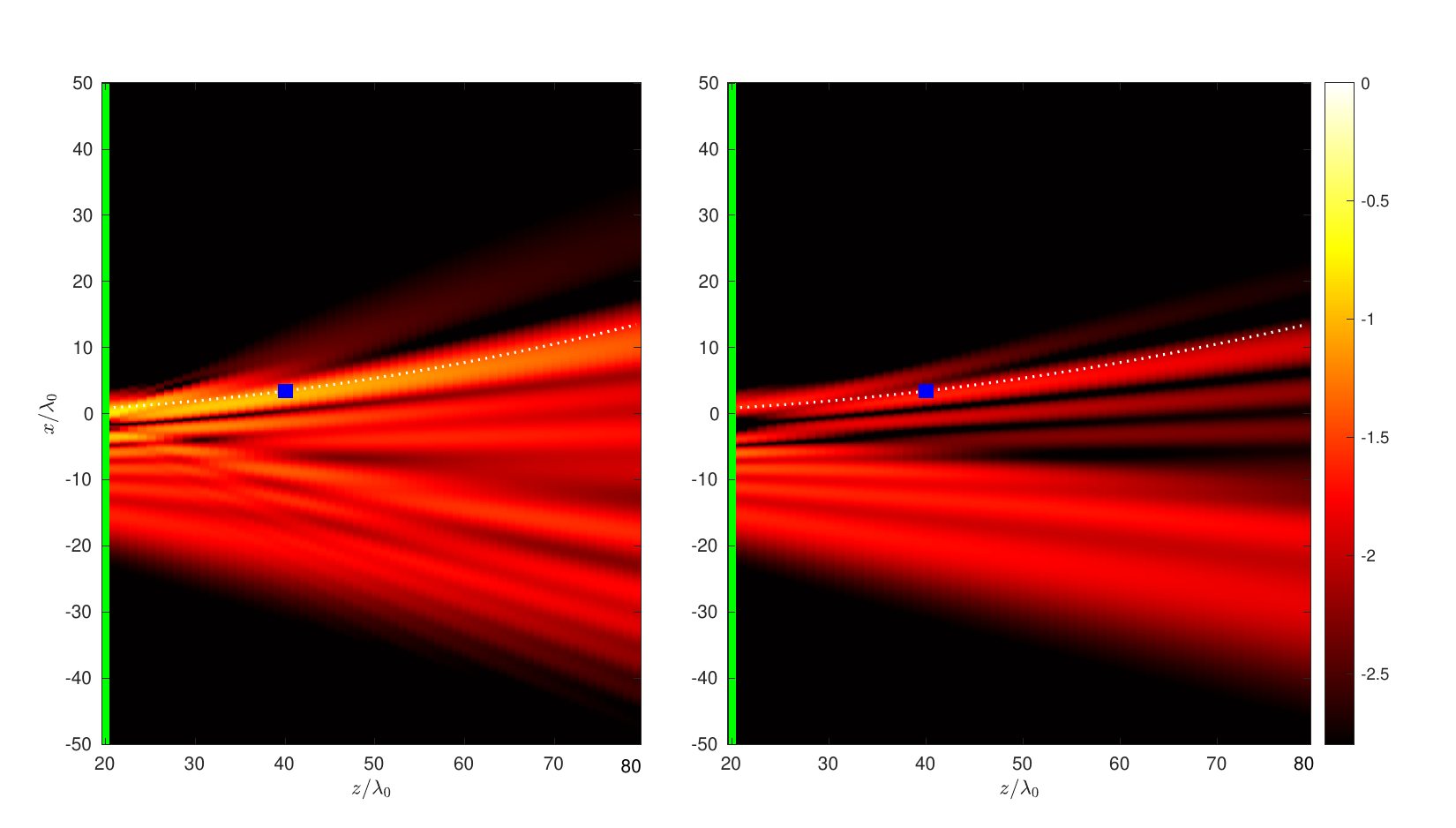}
\caption{
Intensity distribution of the diffracted field $\Ep(z,x)$ given by 
\eqref{eq:pert-ap-soft}, with 
$\nu_\text{a}=0$,  $\gamma_\text{a}=\tfrac{k_0}{9}$, $\alpha_\text{a}=0.01/\lambda_0$, 
$z_\text{b} = 20 \, \lambda_0$, and  $\mu_\text{obs} = -1.45 \, \lambda_0$,
for $\sigma_\text{obs}=0.6 \, \lambda_0$ (left plot),
and $\sigma_\text{obs}=2.0 \, \lambda_0$ (right plot).
The green thick line represents the soft obstacle. 
The white dotted curve represents the caustic trajectory in \eqref{eq:caustic-airy-gen}. 
All spatial coordinates are normalized with respect to the wavelength $\lambda_0$.
}
\label{fig:fig_25_rev}
\end{figure}

To numerically corroborate the self-healing capability of the Airy beam, we consider
an opaque screen with transmittance profile $\tau_\text{obs}(x)$ using different standard deviations
$\sigma_\text{obs}$ to emulate obstacles of varying widths. Specifically, Fig.~\ref{fig:fig_24_rev} 
illustrates two different transmittance profiles of the soft obstacle, positioned at $z_\text{b} = 20 \, \lambda_0$ and 
centered on the main lobe of the aperture Airy field $\E_\text{a}(\xi)$ defined by \eqref{eq_Ea}. The intensity of this field
is also shown in the top panel of Fig.~\ref{fig:fig_24_rev}, with parameters $\nu_\text{a}=0$, 
$\gamma_\text{a}=\tfrac{k_0}{9}$, and $\alpha_\text{a}=0.01/\lambda_0$. 
When $\sigma_\text{obs}=0.6 \, \lambda_0$, 
the obstacle attenuates only the main lobe of the aperture Airy field in \eqref{eq_Ea}.
On the other hand, for $\sigma_\text{obs} = 2.0 \, \lambda_0$, the attenuation profile broadens, affecting 
not only the main lobe but also the first two secondary lobes of $\E_\text{a}(\xi)$ in \eqref{eq_Ea}.
Based on this configuration, Fig.~\ref{fig:fig_25_rev} shows the resulting intensity of the diffracted field 
$\Ep(z,x)$  for the considered value of  
$\sigma_\text{obs}$, for $z>z_\text{b}$.
In accordance with the results of Subsection~V-\ref{sec:knife-diffraction} obtained 
by resorting to the knife-edge diffraction model, it can be argued that, 
when the main lobe of the aperture Airy field in \eqref{eq_Ea} is only slightly attenuated, i.e., $\sigma_\text{obs}=0.6 \, \lambda_0$, 
the diffracted beam follows quite closely the propagation 
dynamics  of the unperturbed Airy beam $u(z,x)$, thereby confirming the self-healing behavior. 
Physically, these results confirm the non-local nature of the Airy beam formation, as clearly illustrated in \cite{Kaganovsky-2010}. 
The unique propagation characteristics of the beam are not solely determined by the main lobe on the aperture 
but also result from rays 
originating in regions of the transmit aperture that are far from the maximum aperture field.
Strictly speaking, after an obstruction attenuates part of the main lobe, the remaining side lobes 
interfere constructively to regenerate the main lobe downstream.
As $\sigma_\text{obs}$ increases, the 
attenuation of the main lobe becomes more pronounced and 
extends beyond it, affecting also 
the secondary lobes of the aperture field in \eqref{eq_Ea}. 
This causes a gradual suppression of the beam's key features such as 
self-acceleration and resistance to diffraction.
This outcome is consistent with our theoretical analysis, which predicts that the amplitude of the perturbative term 
$p(z,x)$ grows rapidly as the obstacle size, characterized by $\sigma_\text{obs}$, increases. 
\begin{figure}[t]
\centering
\includegraphics[width=\linewidth]{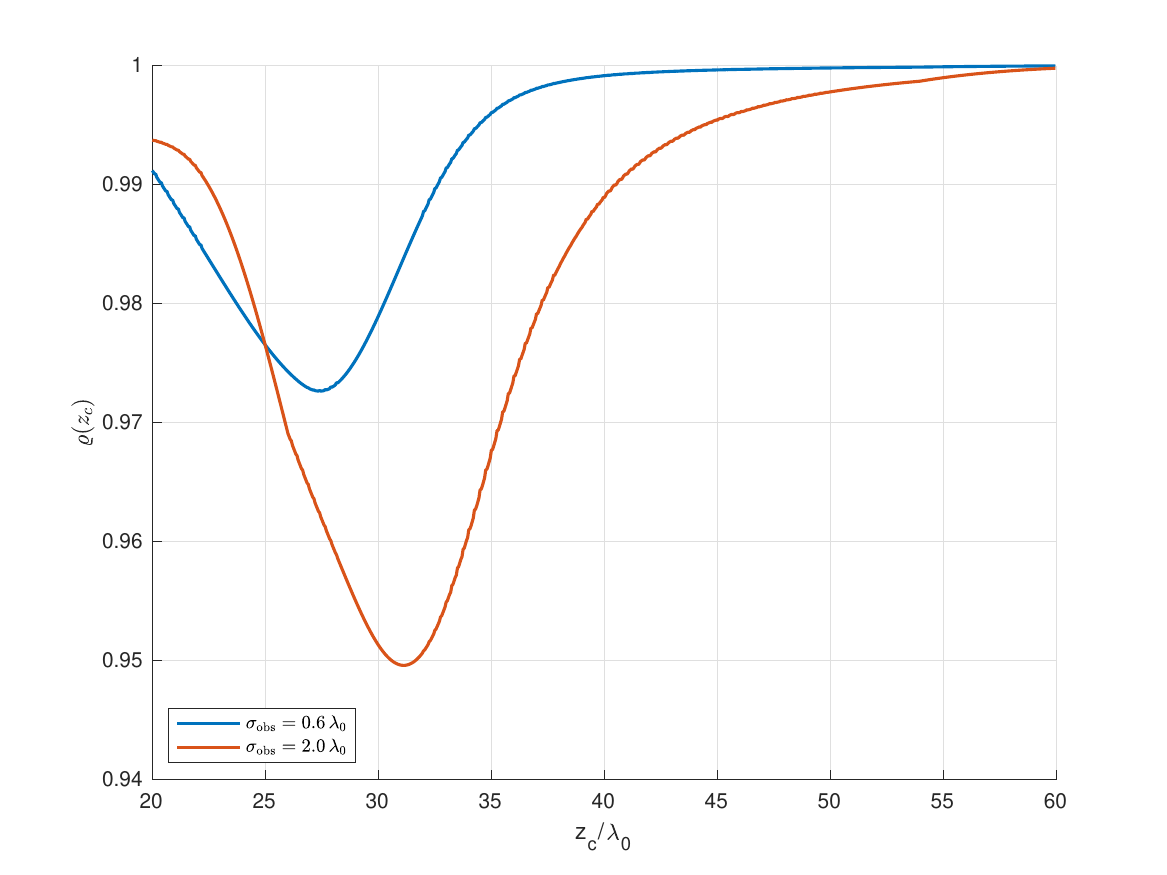} 
\caption{Similarity index between the diffracted field $\Ep(z_\text{c},x_\text{c})$ and the unperturbed beam $u(z_\text{c},x_\text{c})$
for two different values of $\sigma_\text{obs}$, with $\nu_\text{a}=0$,  $\gamma_\text{a}=\tfrac{k_0}{9}$, $\alpha_\text{a}=0.01/\lambda_0$, 
$z_\text{b} = 20 \, \lambda_0$, $\mu_\text{obs} = -1.45 \, \lambda_0$, and 
$\varepsilon=12 \, \lambda_0$.
The spatial coordinate $z_\text{c}$ is normalized with respect to the wavelength $\lambda_0$.}
\label{fig:fig_26_rev}
\end{figure}
Along the same scenario considered in Figs.~\ref{fig:fig_24_rev} and \ref{fig:fig_25_rev},
we present in Fig.~\ref{fig:fig_26_rev} an alternative perspective of the self-healing property of the Airy beam, 
by showing the (normalized) similarity index
\be
\varrho(z_\text{c}) \eqdef \frac{\left|\int_{z_c-\frac{\varepsilon}{2}}^{z_c+\frac{\varepsilon}{2}} \Ep(v,x_\text{c}) \, u^*(v,x_\text{c}) 
\, {\rm d}v \right|}{\left[\int_{z_c-\frac{\varepsilon}{2}}^{z_c+\frac{\varepsilon}{2}} |\Ep(v,x_\text{c})|^2 \, {\rm d}v\right]^\frac{1}{2}
\left[\int_{z_c-\frac{\varepsilon}{2}}^{z_c+\frac{\varepsilon}{2}} |u(v,x_\text{c})|^2 \, {\rm d}v\right]^\frac{1}{2}}.
\label{eq:varrho}
\ee
This metric quantifies the degree of similarity between the diffracted field $\Ep(z_\text{c},x_\text{c})$ and the unperturbed beam $u(z_\text{c},x_\text{c})$,
both evaluated along the nominal parabolic trajectory \eqref{eq:caustic-airy-gen}, with an integration window of width $\varepsilon=12 \, \lambda_0$.
The index $\varrho(z_\text{c})$ ranges from zero to one, where $\varrho(z_\text{c})=0$ indicates complete dissimilarity, while
$\varrho(z_\text{c})=1$ denotes perfect  similarity.
When $\sigma_\text{obs} = 0.60 \, \lambda_0$, the Airy beam 
reconstructs itself after encountering the obstacle, with a minimum self-healing distance
of about $15 \, \lambda_0$. As $\sigma_\text{obs}$ increases, two effects 
become evident from Proposition~\ref{prop:pro}: The self-healing distance grows, and the beam fails to recover its original profile. 
Indeed, when  $\sigma_\text{obs}=2.0 \, \lambda_0$, the 
similarity index approaches one only at a distance of about $60 \, \lambda_0$, where however 
the intensity of the unperturbed Airy field in \eqref{eq:airy-def} becomes negligible  due to the 
finite-energy constraint.

\begin{figure}[t]
\centering
\includegraphics[width=\linewidth]{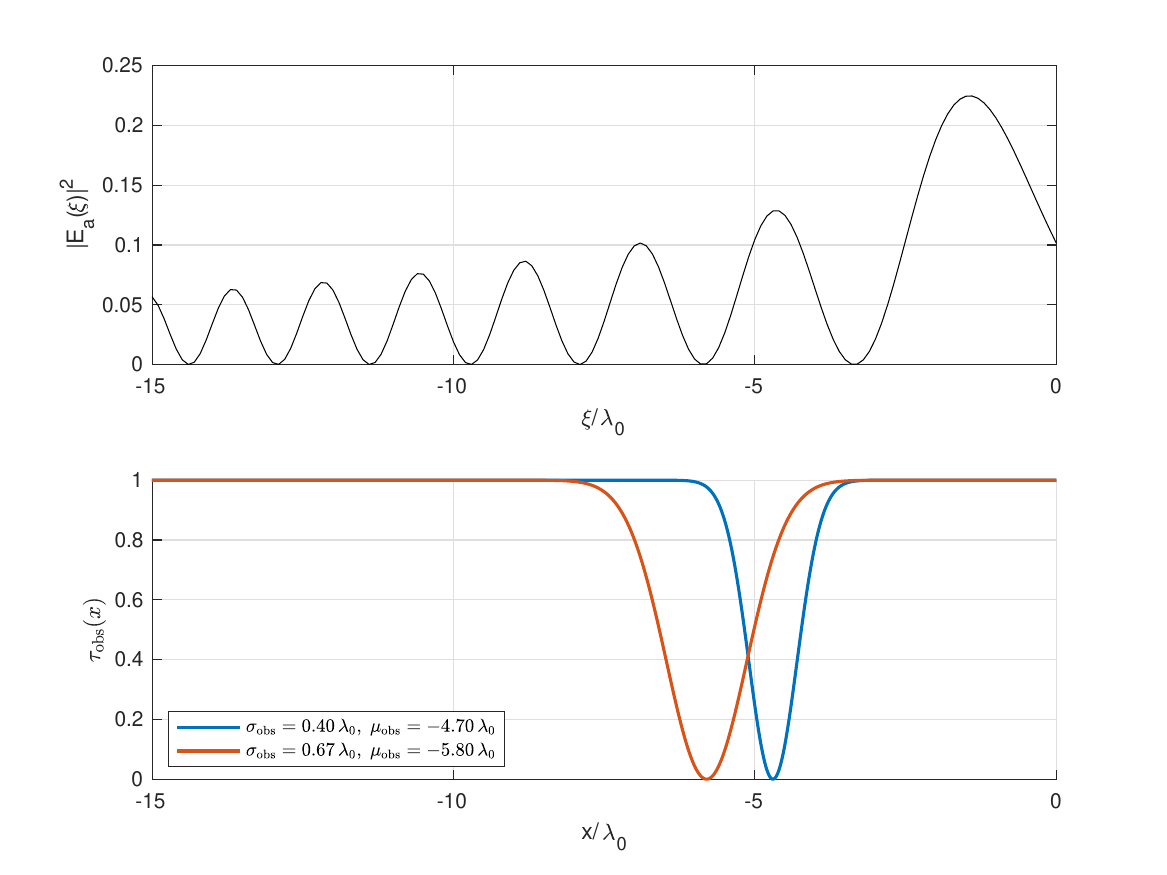} 
\caption{Top: Intensity of the aperture field $\E_\text a(\xi)$, with $\nu_\text{a}=0$, 
$\gamma_\text{a}=\tfrac{k_0}{9}$, and $\alpha_\text{a}=0.01/\lambda_0$. Bottom: Transmittance profile of a soft obstacle 
located at $z_\text{b} = 20 \, \lambda_0$, for two different values of
$(\mu_\text{obs},\sigma_\text{obs})$.
All  spatial coordinates are normalized with respect to the wavelength $\lambda_0$.}
\label{fig:fig_27_rev}
\end{figure}

\begin{figure}[t]
\centering
\includegraphics[width=\linewidth]{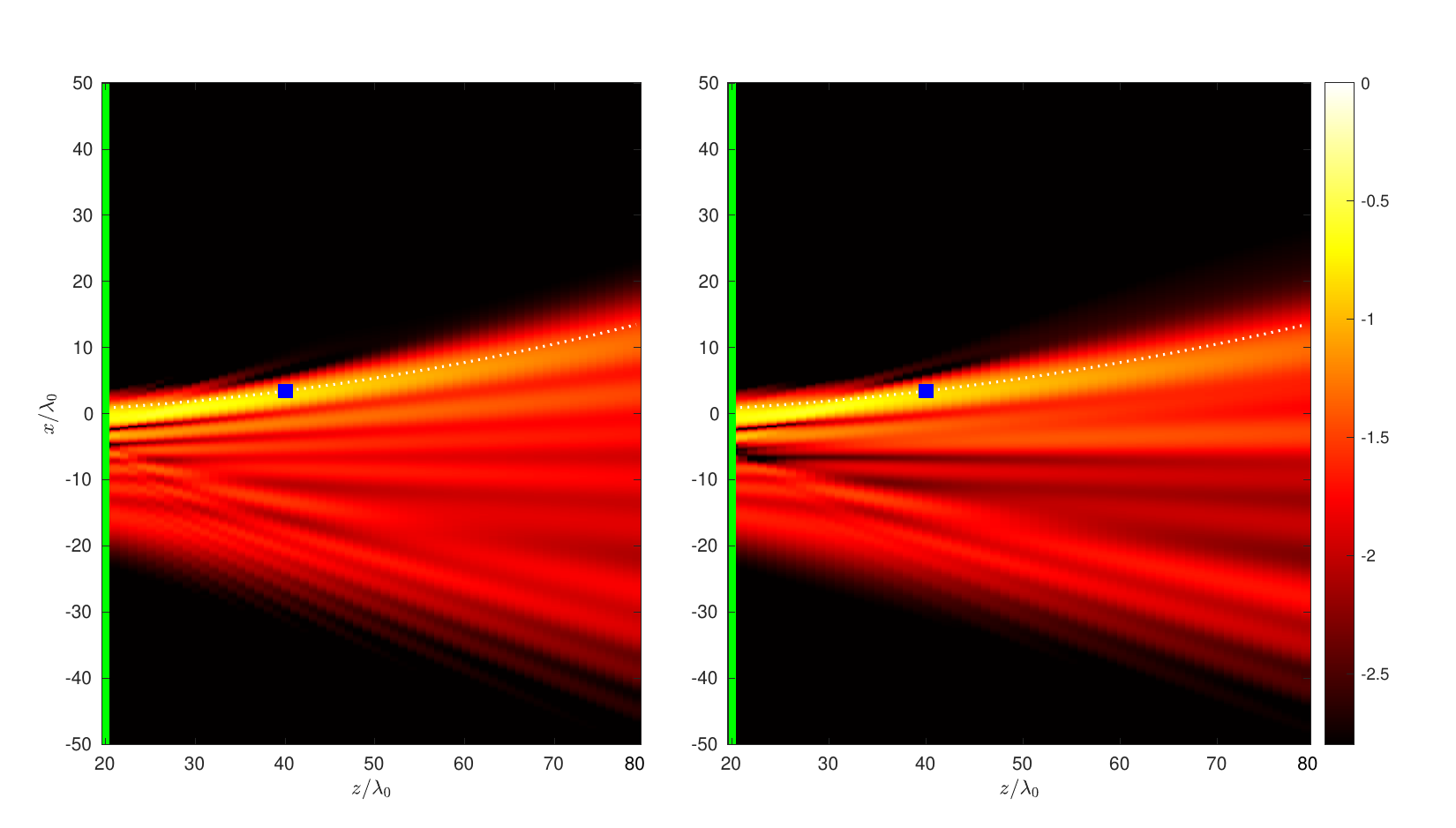} 
\caption{Intensity distribution of the diffracted field $\Ep(z,x)$ given by \eqref{eq:pert-ap-soft}, with 
$\nu_\text{a}=0$,  $\gamma_\text{a}=\tfrac{k_0}{9}$, $\alpha_\text{a}=0.01/\lambda_0$, 
$z_\text{b} = 20 \, \lambda_0$,
for $(\mu_\text{obs},\sigma_\text{obs})=(-4.70 \, \lambda_0, 0.40 \, \lambda_0)$ (left plot),
and $(\mu_\text{obs},\sigma_\text{obs})=(-5.80 \, \lambda_0, 0.67 \, \lambda_0)$ (right plot).
The green thick line represents the soft obstacle. 
The white dotted curve represents the caustic trajectory in \eqref{eq:caustic-airy-gen}. 
All spatial coordinates are normalized with respect to the wavelength $\lambda_0$.
}
\label{fig:fig_28_rev}
\end{figure}

\begin{figure}[t]
\centering
\includegraphics[width=\linewidth]{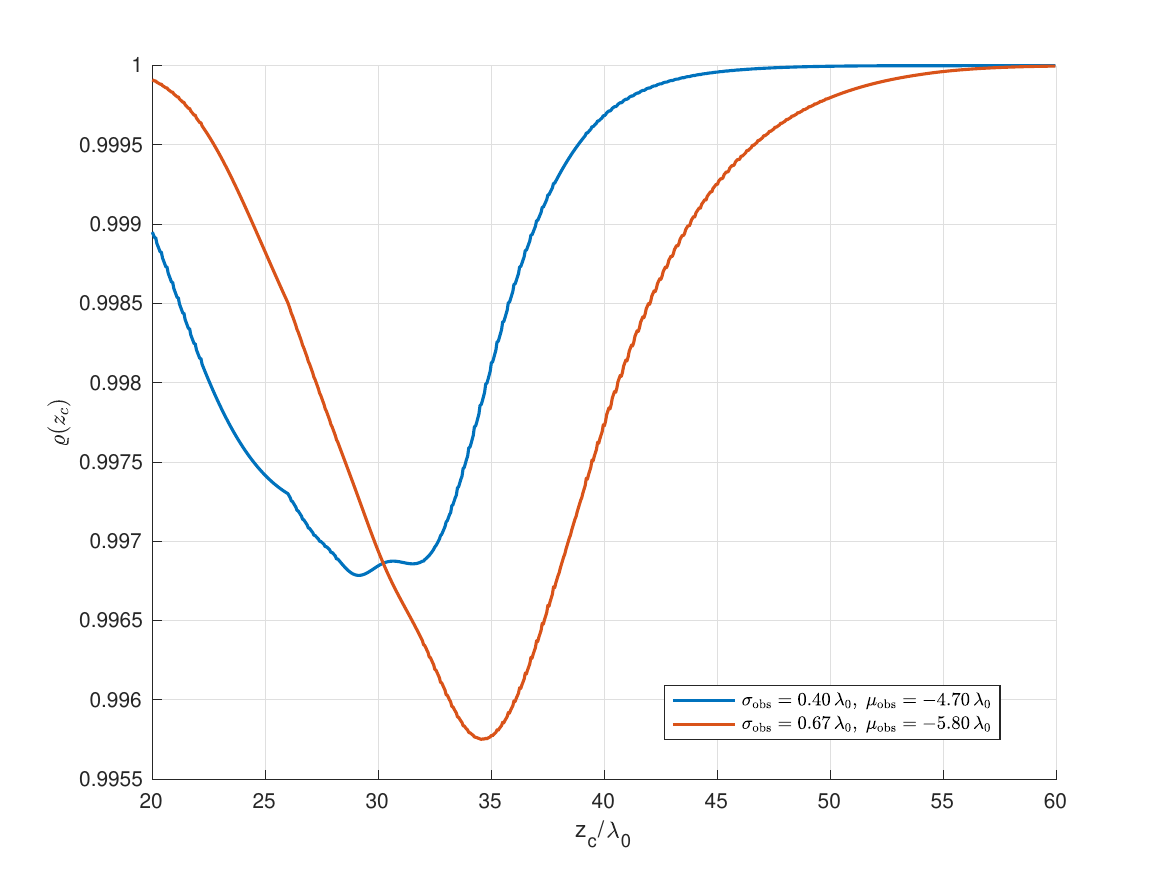} 
\caption{Similarity index between the diffracted field $\Ep(z_\text{c},x_\text{c})$ and the unperturbed beam $u(z_\text{c},x_\text{c})$
for two different values of $(\mu_\text{obs},\sigma_\text{obs})$, with $\nu_\text{a}=0$,  
$\gamma_\text{a}=\tfrac{k_0}{9}$, $\alpha_\text{a}=0.01/\lambda_0$, 
$z_\text{b} = 20 \, \lambda_0$, and  $\varepsilon=12 \, \lambda_0$.
The spatial coordinate $z_\text{c}$ is normalized with respect to the wavelength $\lambda_0$.}
\label{fig:fig_29_rev}
\end{figure}

We finally investigate the self-healing capability of the Airy beam when attenuation is applied exclusively to its secondary lobes at the aperture. To this end, 
we maintain the same simulation setting specified for 
Figs.~\ref{fig:fig_24_rev}--\ref{fig:fig_26_rev}, by considering 
now the transmittance profiles shown in Fig.~\ref{fig:fig_27_rev}. In the first case, 
the transmittance profile $\tau_\text{obs}(x)$ of the soft obstacle is centered at 
$\mu_\text{obs} = -4.70 \, \lambda_0$, with $\sigma_\text{obs} = 0.40 \, \lambda_0$,
thus attenuating only the first secondary lobe of
the aperture Airy field in \eqref{eq_Ea}. In the second case,  $\tau_\text{obs}(x)$ 
is centered at $\mu_\text{obs} = -5.80 \, \lambda_0$, with $\sigma_\text{obs} = 0.67 \, \lambda_0$, 
hence attenuating both the first and the second secondary lobes of $\E_\text{a}(\xi)$ in \eqref{eq_Ea}.
Fig.~\ref{fig:fig_28_rev} shows the corresponding intensity of the diffracted field $\Ep(z,x)$ for 
the considered values of $\sigma_\text{obs}$ and $\mu_\text{obs}$, for $z>z_\text{b}$. 
By comparison with Fig.~\ref{fig:fig_25_rev}, it can be inferred that the Airy beam exhibits 
a stronger self-healing capability when its main lobe on the aperture 
remains unattenuated, as this 
lobe carries the largest portion of the beam's energy.
This result is consistent with that of Subsection~V-\ref{sec:knife-diffraction} and
it is also corroborated by Fig.~\ref{fig:fig_29_rev}, which shows that attenuating the 
secondary lobes of the aperture Airy field in \eqref{eq_Ea} yields a higher similarity index $\varrho(z_\text{c})$, 
compared to the attenuation of the main lobe. Obviously, the similarity index decreases
and the self-healing distance grows as the number of attenuated 
secondary lobes increases.


\section{Conclusions and directions for future work}
\label{sec:concl}

This paper presented a comprehensive analysis of 
self-accelerating beams, with a focus on Airy beams radiated 
by continuous apertures in the radiative near-field. Through 
theoretical modeling and numerical validation, we examined 
their key characteristics, including propagation along 
parabolic caustics, diffraction-resistant behavior, and 
self-healing in the presence of obstructions. These effects 
were shown to depend on several interrelated parameters such 
as aperture size and energy, spectral shaping of the 
transmitted field, and the geometry and position of obstacles 
in NLoS scenarios.
Our results indicate that, in the presence of an obstacle 
positioned in the radiative near-field region at an 
intermediate distance along the communication link, Airy 
beams may outperform conventional Gaussian beams in 
controlled near-field settings by delivering higher received 
energy around their trajectory, 
provided that the main lobe of the Airy field on the aperture remains largely unobstructed, even when some of the secondary lobes are partially obstructed or attenuated. However, 
the performance advantage is scenario-dependent (distance 
of the obstacle from the aperture, portion of the main lobe 
on the aperture that is blocked/attenuated, receiver 
size/placement) and design-dependent (effective size of the 
aperture, apodization of the input Airy field, initial 
launch angle of the beam). This emphasizes the importance 
of transmitter design and environmental awareness in 
practical implementations of self-accelerating beams.

Based on the theoretical and numerical 
results of this paper, the following design guidelines can 
be drawn for the practical deployment of Airy beams in 
radiative near-field wireless links.
\begin{itemize}

\item \textit{Aperture sizing.} The 
effective aperture size must grow quadratically with the 
desired diffraction-resisting propagation range, as 
established by eq.~\eqref{eq:xeff}. This 
represents a fundamental constraint for practical system 
design: for a target link distance $z_\text{max}$, the 
minimum aperture width scales as $x_\text{eff} \propto 
\gamma_\text{a}^3 z_\text{max}^2/k_0^2$. At sub-THz 
frequencies, where $k_0$ is large, this constraint is 
relatively mild, making Airy beams practically realizable 
with apertures of moderate size.

\item \textit{Obstruction geometry.} 
Airy beams offer a performance advantage over collimated 
Gaussian beams when the main lobe of the aperture field 
remains in LoS with the receiver, even if the secondary 
lobes are partially obstructed. Conversely, when a large 
fraction of the main lobe is blocked, a focused Gaussian 
beam defined over the residual LoS aperture provides 
superior performance, consistent with the findings 
of~\cite{Inserra_2022}. These two regimes are 
quantitatively delineated by the obstructed energy 
fraction $\en_\text{obs}/\en_\text{a}$ introduced in 
Section~\ref{sec:self-healing}.

\item \textit{Apodization and 
diffraction-resisting range.} The apodization parameter 
$\alpha_\text{a}$ controls the trade-off between aperture 
localization and diffraction-resisting range: a smaller 
$\alpha_\text{a}$ extends the range over which the beam 
maintains its shape, as quantified by 
eq.~\eqref{eq:cond-df-apo-zc}, but requires a larger 
physical aperture to support the oscillatory tail of the 
Airy field.

\item \textit{Beam steering and adaptive 
control.} The initial launch angle of the parabolic 
caustic can be controlled independently via the linear 
phase parameter $\nu_\text{a}$, enabling conventional 
beam steering without affecting the self-acceleration 
property. In phased-array or metasurface implementations, 
time-varying $\nu_\text{a}(t)$ and $\gamma_\text{a}(t)$ 
can enable adaptive steering and curvature control, 
compatible with standard RIS control 
mechanisms~\cite{Malevich.2025} and with the 
trajectory-adaptive approaches 
of~\cite{Ye_ArXiV2025,Yao_2026}.

\item \textit{Hardware implementation.} 
The phase-only nature of the Airy aperture field, whose 
phase takes only two values ($0$ and $\pi$) since the 
aperture field is real-valued, makes it naturally suited 
for implementation on 1-bit binary metasurfaces, which 
are currently manufacturable at scale at sub-THz 
frequencies~\cite{Feng.2020,Wang.2023}. The minimum 
element spacing is determined by the Nyquist condition 
of eq.~\eqref{eq:delta_3sigma}, and $\lambda_0/2$ 
spacing is generally sufficient for the beam parameters 
considered in this paper.

\end{itemize}

The problem of optimizing the phase distribution of large-size 
apertures in near-field channels has recently received considerable 
attention~\cite{Uchimura.2026} and has strong ties with the emerging 
electromagnetic signal and information theory framework~\cite{Migliore,Renzo.2024}, 
which is especially tailored to characterize optimal designs and 
fundamental performance limits in near-field channels.
In this context, for example, it would be relevant 
to identify the optimal basis functions for transmission and 
reception by considering the presence of blocking objects at 
the design stage~\cite{Piestun}, and to compare such optimal 
designs with known accelerated beams, such as the Airy and 
Pearcey beams~\cite{Qin_Pearcey_2026}, analyzed in the 
present paper. Similarly, the connection between the 
wave-optics framework developed here and the system-level 
optimization approaches of~\cite{Liu_ArXiV2025,Yao_2026} 
represents a promising direction for future research, as 
it would enable the joint design of beam trajectory, 
aperture size, and hardware constraints within a unified 
communication-theoretic framework.

Despite their theoretical appeal, the deployment of Airy 
beams in real wireless systems remains challenging. Idealized 
assumptions, such as continuous aperture profiles and a 
wireless channel with a single obstacle, might not hold in 
practice. Implementation via holographic or programmable 
metasurfaces introduces discretization, quantization, and 
other hardware constraints that affect beam control, as 
analyzed in Section~IV-\ref{sec:sampling}. Furthermore, 
real-world systems have to operate within power, bandwidth, 
and fabrication limits that impose additional restrictions 
on beam design and steering. Extending the present 
continuous-aperture wave-optics framework to account for 
multiple simultaneous obstacles, three-dimensional beam 
geometries, and multiuser interference scenarios, along 
the lines of~\cite{Drou2025,Qin_Pearcey_2026,Yao_2026}, 
represents a natural and important direction for future 
research.

%
%

\appendices

\section{Approximation of \eqref{eq:int-2}
via the stationary-phase principle}
\label{app:PSP}

Consider a fixed point $(z,x)$ in the radiative near-field region. To analyze the local contribution of the aperture to the total field, we 
evaluate the integral in \eqref{eq:int-2} 
over a small neighborhood $[x_\text{a}-\epsilon,x_\text{a}+\epsilon]$
around an arbitrary point $x_\text{a} \in [x_\text{a}^{(1)},x_\text{a}^{(2)}]$, with $\epsilon>0$. Let us define: 
\be
I_\epsilon(x_\text{a}) \eqdef \int_{x_\text{a}-\epsilon}^{x_\text{a}+\epsilon} 
A_\text{a}(\xi) \, e^{-j Q_\xi(z,x)} \, \mathrm{d}\xi \:.
\ee
Assuming that the integrand $f(\xi)=A_\text{a}(\xi) \, e^{-j Q_\xi(z,x)}$
is smooth and can be locally approximated via its Taylor expansion about 
$x_\text{a}$, we obtain
\begin{multline}
I_\epsilon(x_\text{a}) \approx f(x_\text{a}) \int_{x_\text{a}-\epsilon}^{x_\text{a}+\epsilon} 
\left[1+\frac{\dot{f}(x_\text{a})}{f(x_\text{a})} \, (\xi-x_\text{a})
\right. \\ \left. 
+ \frac{1}{2} \, \frac{\ddot{f}(x_\text{a})}{f(x_\text{a})} \, (\xi-x_\text{a})^2\right] \mathrm{d}\xi. 
\end{multline}
The first and second derivatives of $f(\xi)$ at $x_\text{a}$  can be expressed as: 
\barr
\frac{\dot{f}(x_\text{a})}{f(x_\text{a})} & = \frac{\dot{A}_\text{a}(x_\text{a})}{A_\text{a}(x_\text{a})}
- j \, \dot{Q}_{x_\text{a}}(z,x),
\\
\frac{\ddot{f}(x_\text{a})}{f(x_\text{a})} & = \frac{\ddot{A}_\text{a}(x_\text{a})}{A_\text{a}(x_\text{a})}
- \left[\frac{\dot{A}_\text{a}(x_\text{a})}{A_\text{a}(x_\text{a})}\right]^2 
\nonumber \\
 & \hspace{10mm} - j \, \ddot{Q}_{x_\text{a}}(z,x) 
+ \left[\frac{\dot{f}(x_\text{a})}{f(x_\text{a})}\right]^2 \: .
\earr
Now, if the phase derivatives dominate over the amplitude variations, i.e.,
\be
\left| \dot{Q}_{x_\text{a}}(z,x) \right| \gg \left| \frac{\dot{A}_\text{a}(x_\text{a})}{A_\text{a}(x_\text{a})} \right|,
\label{eq:first-phasedom}
\ee
then we can approximate (see, e.g., \cite{Delprat-1992})
\be
\frac{\dot{f}(x_\text{a})}{f(x_\text{a})} \approx 
- j \, \dot{Q}_{x_\text{a}}(z,x) \: .
\ee
This condition is known as the 
{\em first-order phase-rate dominance}. 
Moreover, as shown in \cite{Delprat-1992}, the second derivative can be approximated as 
\be
\frac{\ddot{f}(x_\text{a})}{f(x_\text{a})} \approx - j \, \ddot{Q}_{x_\text{a}}(z,x) 
- \left[\dot{Q}_{x_\text{a}}(z,x)\right]^2,
\ee
under the additional condition that
\be
\left| \ddot{Q}_{x_\text{a}}(z,x) \right| \gg \left| \frac{\ddot{A}_\text{a}(x_\text{a})}{A_\text{a}(x_\text{a})}
- \left[\frac{\dot{A}_\text{a}(x_\text{a})}{A_\text{a}(x_\text{a})}\right]^2  \right|,
\label{eq:second-phasedom}
\ee
which is referred to as the
{\em second-order phase-rate dominance}. 

Under conditions \eqref{eq:first-phasedom} and \eqref{eq:second-phasedom}, the phase of the integrand dominates over the amplitude variations. In particular,
at a stationary point, where $\dot{Q}_{x_\text{a}}(z,x)=0$, 
the local behavior of the integral can be
accurately captured by the
second derivative of the phase, $\ddot{Q}_{x_\text{a}}(z,x)$, alone, since 
the contributions of the amplitude derivatives $\dot{A}_\text{a}(x_\text{a})$
and $\ddot{A}_\text{a}(x_\text{a})$ become negligible.
Indeed, let us suppose that $\dot{Q}_{x_\text{a}}(z,x)=0$ and 
$\dot{Q}_\xi(z,x) \neq 0$ elsewhere within the neighborhood $[x_\text{a}-\epsilon,x_\text{a}+\epsilon]$, ensuring the presence of a unique stationary point at $x_\text{a}$. 
Moreover, let 
$\mu \in \{\pm 1\}$ denote the sign of the nonzero second derivative, i.e., 
$\ddot{Q}_{x_\text{a}}(z,x)= \mu \, \big|\ddot{Q}_{x_\text{a}}(z,x)\big|$.
Under these conditions, we can approximate the integral as follows:
\begin{multline}
I_\epsilon(x_\text{a}) = e^{-j Q_{x_\text{a}}(z,x)} \int_{x_\text{a}-\epsilon}^{x_\text{a}+\epsilon} 
A_\text{a}(\xi) \, e^{-j [Q_\xi(z,x)-Q_{x_\text{a}}(z,x)]} \, \mathrm{d}\xi
\\ \approx f(x_\text{a})
\int_{x_\text{a}-\epsilon}^{x_\text{a}+\epsilon}
e^{-\frac{j}{2} \ddot{Q}_{x_\text{a}}(z,x) \, (\xi-x_\text{a})^2} \, \mathrm{d}\xi
\\ \approx f(x_\text{a}) \int_{-\infty}^{+\infty}
e^{-\frac{j}{2} \ddot{Q}_{x_\text{a}}(z,x) \, u^2} \, \mathrm{d}u
\\ = f(x_\text{a}) \sqrt{- \frac{2 \pi j}{\ddot{Q}_{x_\text{a}}(z,x)}}
\\ = f(x_\text{a}) \sqrt{\frac{2 \pi}{\big|\ddot{Q}_{x_\text{a}}(z,x)\big|}} \, \left(-j \, \mu\right)^{1/2}
\\ = f(x_\text{a}) \sqrt{\frac{2 \pi}{\big|\ddot{Q}_{x_\text{a}}(z,x)\big|}} \, e^{-j \frac{\pi}{4} \mu},
\label{app:fin}
\end{multline}
where we have approximated $Q_\xi(z,x)$ by its second-order Taylor expansion about  $\xi=x_\text{a}$,
\be
Q_\xi(z,x) \approx Q_{x_\text{a}}(z,x) + \frac{1}{2} \, \ddot{Q}_{x_\text{a}}(z,x) \, (\xi-x_\text{a})^2,
\ee
and have used the known Fresnel formula for evaluating the resulting Gaussian integral:
\be
\int_{-\infty}^{+\infty} e^{j a u^2} \, \mathrm{d}u = \sqrt{\frac{j \, \pi}{a}} \: .
\ee
This result provides a closed-form approximation for the integral near a stationary point. 
However, this approximation breaks down
when $\ddot{Q}_{x_\text{a}}(z,x)=0$, as the standard stationary-phase method no longer applies, and more refined uniform asymptotic expansions must be employed.

%
%

\section{Proof of Proposition~\ref{prop:1}}
\label{app:prop-1}

The caustic of a family of rays is defined as the envelope of those rays, or equivalently,
the locus  of intersections
of infinitesimally close rays. To derive the caustic associated with the ray family described by \eqref{eq:rays},
we consider two neighboring rays and determine the coordinates $(z_\text{c},x_\text{c})$ where they intersect. This leads to the system
\be
\begin{cases}
x_\text{c}= \xi + z_\text{c} \, \frac{\dot{\Phi}_\text a(\xi)}{k_0} \:,
\\
x_\text{c}= \xi + \Delta\xi + z_\text{c} \, \frac{\dot{\Phi}_\text a(\xi+\Delta\xi)}{k_0} \:,
\end{cases}
\label{eq:syst-app-1}
\ee
where $\Delta\xi>0$ is a small offset along the aperture.
Subtracting the first equation from the second and  
dividing  by $\Delta\xi$, we obtain 
\be
\begin{cases}
x_\text{c}= \xi + z_\text{c} \, \frac{\dot{\Phi}_\text a(\xi)}{k_0} \:;
\\
0= 1 + z_\text{c} \, \frac{\dot{\Phi}_\text a(\xi+\Delta\xi)-\dot{\Phi}_\text a(\xi)}{k_0 \, \Delta\xi} \:.
\end{cases}
\label{eq:syst-app-2}
\ee
As $\Delta\xi\rightarrow 0$, the difference quotient in the second equation approaches 
$\ddot{\Phi}_\text a(\xi)$. Therefore, in the limit,
we recover the system \eqref{eq:syst}, which defines the caustic curve. 

%
%
\section{Polychromatic exponentially modulated Airy beams}
\label{app:Poly}

The bandpass representation of a polychromatic wave, with
carrier frequency $f_0$, is given by
\be
\widetilde{\E}(z,x;t) = \Re 
\left\{\E(z,x; t) \, e^{ j 2 \pi f_0 t} \right\},
\label{eq:rf-poly}
\ee 
where the complex envelope $\E(z,x;t)$ depends on both space and time, and
its Fourier transform with respect to $t$ is given by 
\be
\overline{\E}(z,x; f) = \int_{-\infty}^{+\infty} \E(z,x; t) \, e^{-j 2 \pi f t} \, {\rm d}t \: .
\ee
The time-domain field $\E(z,x;t)$ can be interpreted as
a continuous linear superposition of time-harmonic fields
$\overline{\E}(z,x; f) \, e^{j 2 \pi f t}$ across all 
frequency components $f \in \mathbb{R}$: 
\be
\E(z,x; t) = \int_{-\infty}^{+\infty}  \overline{\E}(z,x; f) \, e^{j 2 \pi f t} \, {\rm d}f \: .
\label{eq:Four-synthesis}
\ee
Substituting \eqref{eq:Four-synthesis} into \eqref{eq:rf-poly} and then 
into \eqref{eq:max}, and exploiting the linearity of wave propagation, we find that $\overline{\E}(z,x; f)$ satisfies a 
{\em frequency-dependent} source-free Helmholtz equation:
\be
\frac{\partial^2 \, \overline{\E}(z,x; f)}{\partial x^2} + \frac{\partial^2 \, \overline{\E}(z,x; f)}{\partial z^2} + k^2(f) \, \overline{\E}(z,x; f) = 0,
\label{eq:Helm-f}
\ee
where 
\be
k(f) \eqdef \frac{2 \pi \, (f+f_0)}{c} = k_0 \left(1+\frac{f}{f_0}\right)
\ee
is the wave number associated with the 
monochromatic  component of the field at frequency $f+f_0$.
Following the same approach as in Section~\ref{sec:Diff-theory}, 
the solution to \eqref{eq:Helm-f} for $z>0$ can be expressed as 
\be
\overline{\E}(z,x; f) =  \frac{1}{2 j} \int_{x_\text{a}^{(1)}}^{x_\text{a}^{(2)}} 
\frac{k(f) \, z}{\rho(\xi)} \, \overline{\E}_\text a(\xi;f) \, \text{H}^{(2)}_1(k(f) \, \rho(\xi)) \, \mathrm{d}\xi,
\label{eq:Ray-2-f}
\ee
where $\overline{\E}_\text a(\xi;f)$ is the Fourier transform of the complex envelope 
of the aperture field,
and the distance $\rho(\xi)$ is defined as in \eqref{eq:Ray-2}.
The time-domain aperture field is then synthesized as  
\be
\E_\text a(\xi;t) = \int_{-\infty}^{+\infty} \overline{\E}_\text a(\xi;f) \, e^{j 2 \pi f t} \, {\rm d}f.
\ee
In digital communication applications, the relevant case involves 
a {\em pulsed} Airy source with finite spectral bandwidth, 
such that 
\be
\E_\text a(\xi;t)= \E_\text a(\xi) \, p(t), 
\label{eq:pulsed-beam}
\ee
where 
$\E_\text a(\xi)$ is typically an Airy-like function, and $p(t)$
is a {\em finite-energy} baseband pulse profile with Fourier transform $P(f)$.
The pulse $p(t)$ has significant energy only within a spectral support $\mathcal{W}_\text{a}$,
of bandwidth $\banda_\text{a}$ and centered around $f = 0$, i.e., 
$P(f) \approx 0$ for $f \not \in \mathcal{W}_\text{a}$.
Under the paraxial approximation and for an infinite-size aperture, let us consider the
field 
\be
\E_\text a(\xi) = \overline{U}_\text{a} \, \Ai(\gamma_\text{a} \, \xi) \, \, e^{\alpha_\text{a} \xi} \, e^{-j \nu_\text{a} \xi}.
\label{eq_Ea-f}
\ee
Substituting the Fourier transform of \eqref{eq:pulsed-beam} into 
\eqref{eq:Ray-2-f} yields 
\be
\overline{\E}(z,x; f) = P(f) \, \overline{u}(z,x;f) \, e^{-j k(f) z},
\ee
where
\begin{multline}
\overline{u}(z,x;f) = \overline{U}_\text{a} \, \Ai \left(\gamma_\text{a} \, x-\frac{z^2 \, \gamma_\text{a}^4}{4 \, k^2(f)}- \frac{z \, \gamma_\text{a} \, \nu_\text{a}}{k(f)}
- j \frac{z \, \gamma_\text{a} \, \alpha_\text{a}}{k(f)} \right)  \\
\cdot e^{\alpha_\text{a} \left( x-\frac{z^2 \, \gamma_\text{a}^3}{2 \, k^2(f)}-\frac{z \, \nu_\text{a}}{k(f)}\right)}  \, e^{-j \overline{\phi}(x,z;f)},
\label{eq:airy-def-f}
\end{multline}
and the accumulated phase is given by
\begin{multline}
\overline{\phi}(z,x;f) \eqdef \nu_\text{a} \, x +
\frac{z \, \gamma_\text{a}^2}{2 \, k(f)} \\ \cdot \left(\gamma_\text{a} \, x -\frac{z^2 \, \gamma_\text{a}^4}{6 \, k^2(f)} 
- 
 \frac{z \, \gamma_\text{a} \, \nu_\text{a}}{k(f)} 
  + \frac{\alpha_\text{a}^2}{\gamma_\text{a}^2}-\frac{\nu_\text{a}^2}{\gamma_\text{a}^2} \right) \: .
\end{multline}
The constant $\overline{U}_\text{a}$ is selected so that the total energy of the aperture field is normalized to unity:
\be
\overline{\en}_\text{a} \eqdef \int_{-\infty}^{+\infty} \int_{-\infty}^{+\infty} |\E_\text a(\xi;t)|^2 \, \mathrm{d}\xi \, \mathrm{d}t=1\:,
\label{eq:Ea-new-f}
\ee
where $\overline{\en}_\text{a}$ denotes the total 
spatio-temporal energy of the polychromatic aperture 
field, which generalizes the purely spatial energy 
$\en_\text{a}$ defined in~\eqref{eq:Ea-new} for the 
monochromatic case.
When the wave is finite in energy but not strictly monochromatic, a natural
generalization of the definition of intensity is given by
\barr
I(z,x) & \eqdef \int_{-\infty}^{+\infty} \left|\E(z,x; t)\right|^2 {\rm d}t 
\nonumber \\ &
= \int_{-\infty}^{+\infty} \left|\overline{\E}(z,x; f)\right|^2 {\rm d}f,
\label{eq:int-t}
\earr
where the second equality follows from Parseval's theorem.
Let us consider now a digital communication system with symbol period $T_\text{s}$. The ideal
Nyquist pulse for distortionless baseband data transmission is the sinc function
\be
p(t)=\frac{\sin(\pi \, \banda_\text{a} \, t)}{\pi \, \banda_\text{a} \, t}=
\text{sinc}(\banda_\text{a} \, t),
\ee
where $\banda_\text{a}=\tfrac{1}{T_\text{s}}$. The corresponding Fourier transform 
is the rectangular function 
\be
P(f) = \frac{1}{\banda_\text{a}} \, \Pi\left(\frac{f}{\banda_\text{a}}\right),
\ee
with $\Pi(v)=1$ for $|v|\le \tfrac{1}{2}$ and zero elsewhere. 
In this case, the intensity \eqref{eq:int-t} can be written 
as in \eqref{eq:int-f-Nyquist-pap}, with 
\be
\overline{U}_\text{a} = \sqrt{\banda_\text{a}} \, (8 \, \pi \, \alpha_\text{a} \gamma_\text{a})^\frac{1}{4} \, e^{-\frac{1}{3} \left(\frac{\alpha_\text{a}}{\gamma_\text{a}}\right)^3},
\label{eq:U0-f}
\ee
in compliance with the energy constraint in \eqref{eq:Ea-new-f}.

%
%

\section{Propagation of paraxial Gaussian beams in free space}
\label{app:Gauss}

A Gaussian beam is a fundamental solution to the paraxial Helmholtz equation, which describes the evolution of slowly varying wave envelopes under the assumption of small beam divergence. These beams are characterized by having an EM field amplitude that follows a Gaussian distribution in any transverse plane \cite{Siegman}.
To study their propagation, we 
consider a field launched from the aperture plane $z=0$ with 
a Gaussian distribution
\be
\E_\text a(\xi) = V_\text{a} \, e^{-\frac{\xi^2}{\omega_\text{a}^2}} \, e^{-j \mu_\text{a} \xi},
\label{eq:Ea-Gauss}
\ee
where $V_\text{a}>0$ and $\omega_\text{a} >0$ are real constants that define field amplitude and width, respectively, while
$\mu_\text{a}$ determines the 
launch angle.
The intensity of the aperture field 
in \eqref{eq:Ea-Gauss} is also Gaussian. Even for an unbounded aperture, the total energy 
is finite:
\be
\en_\text{a} = \int_{-\infty}^{+\infty} |\E_\text a(\xi)|^2 \, \mathrm{d}\xi = \sqrt{\frac{\pi}{2}} \,
V_\text{a}^2 \, \omega_\text{a} \:.
\label{eq:Ea-new-Gauss}
\ee
We normalize the energy by choosing 
$\en_\text{a} =1$, thus yielding 
\be
V_\text{a} =  \frac{1}{\sqrt{\omega_\text{a}}} \left(\frac{2}{\pi}\right)^\frac{1}{4} \:.
\label{eq:U0-Gauss}
\ee
The parameter $\omega_\text{a}$ is known as the {\em beam waist radius}, i.e., the point where the field amplitude falls to $1/e$ of its maximum.
Substituting \eqref{eq:Ea-Gauss} into the Huygens-Fresnel diffraction integral in \eqref{eq:int} 
and evaluating it over the entire real axis leads to an exact solution in the paraxial regime. By
using the Fourier transform of a Gaussian function,\footnote{\label{foot:Fourier}The
identity
$\int_{-\infty}^{+\infty}
e^{- \alpha \xi^2} \, e^{\pm j \beta \xi} \, {\rm d}\xi =\sqrt{\frac{\pi}{\alpha}}
\, e^{-\frac{\beta^2}{4 \alpha}}$ is employed.} 
the propagated field for $z>0$ can be expressed in the general form given by \eqref{eq:field-decomp}, where
(see, e.g., \cite{Siegman})
\begin{multline}
u(z,x) = V_\text{a} \, \sqrt{\frac{\omega_\text{a}}{w(z)}} \, e^{-\frac{\left(x-z \frac{\mu_\text{a}}{k_0}\right)^2}{w^2(z)}} \, e^{-j \frac{k_0}{2} \frac{\left(x-z \frac{\mu_\text{a}}{k_0}\right)^2}{R(z)}} 
\\  \cdot e^{-j \left( x- z \frac{\mu_\text{a}}{2 k_0}\right) \mu_\text{a}} \, e^{j \frac{1}{2} \arctan \left(\frac{z}{z_0}\right)}.
\label{eq:int-Gauss}
\end{multline}
Here, the functions
\barr
w(z) & \eqdef \omega_\text{a} \sqrt{1+\left(\frac{z}{z_0}\right)^2},
\\
R(z) & \eqdef z \left[ 1+ \left(\frac{z_0}{z}\right)^2\right]
\earr
represent the {\em beam radius} and {\em radius of curvature}, respectively, 
while
\be
z_0 \eqdef \frac{k_0 \, \omega_\text{a}^2}{2}
\ee
is known as the {\em Rayleigh range}. This quantity indicates the distance from the aperture at which the beam radius expands by a factor
$\sqrt{2}$, i.e.,  
$w(z_0)= \omega_\text{a} \sqrt{2}$. A larger beam waist leads to slower diffraction and broader propagation.
At $z=0$, the beam radius is minimized, $w(0)=\omega_\text{a}$, and the wavefront is planar, i.e., $R(0)=\infty$. 
As the beam propagates, the wavefront becomes curved, with the radius of curvature reaching a
minimum at $z=z_0$ and diverging again beyond that. 
Furthermore,  for any given $z$,  
the field intensity in \eqref{eq:int-Gauss} is maximized along the straight line 
\be
x = \frac{\mu_\text{a}}{k_0} \, z , 
\label{eq:gauss-caustic}
\ee
indicating that the beam follows a {\em linear} caustic, with slope determined by 
the parameter $\mu_\text{a}/k_0$. In contrast to Airy beams,  Gaussian beams 
do not satisfy the diffraction-resistant condition in 
\eqref{eq:diff-free}, and their transverse profile  depends on the propagation distance 
$z$.
To model a Gaussian beam with its minimum waist (or radius) at \(z=z'>0\) (and diverging for \(z>z'\)), it suffices to replace \(z\) with \(z-z'\) in \eqref{eq:int-Gauss}.

%
%

\section{Proof of Proposition~\ref{prop:pro}}
\label{app:prop-4}

Let $u(z_\text{b},x)$ denote the Airy beam incident on the obstacle plane at $z=z_\text{b}$, 
obtained by evaluating \eqref{eq:airy-def} at $z=z_\text{b}$.
Under the soft obstacle model, the diffraction integral in \eqref{eq:pert-ap} 
can be decomposed as 
\be
u_\text{d}(z,x) = u(z,x)-p(z,x) \:, 
\ee
where
\barr
u(z,x) & \eqdef \sqrt{\frac{j}{\lambda_0 (z-z_\text{b})}}
\int_{-\infty}^{+\infty}
u(z_\text{b},\xi) \, e^{-j \frac{k_0}{2 (z-z_\text{b})} (x-\xi)^2} \, \mathrm{d}\xi,
\\
p(z,x) & \eqdef \sqrt{\frac{j}{\lambda_0 (z-z_\text{b})}}
\nonumber \\ & \hspace{10mm} \cdot
\int_{-\infty}^{+\infty}
u(z_\text{b},\xi) \,  a_\text{obs}(\xi) \, e^{-j \frac{k_0}{2 (z-z_\text{b})} (x-\xi)^2} \, \mathrm{d}\xi.
\label{eq:pert-ap-soft-app}
\earr
Physically, $u(z,x)$ represents the free-space Airy beam in \eqref{eq:airy-def} propagating to the observation point $(z,x)$, 
while $p(z,x)$ captures the perturbation due to the obstacle.
Rewriting the perturbative term in a more tractable form yields:
\be
p(z,x) = U_\text{b} \, \sqrt{\frac{j}{\lambda_0 (z-z_\text{b})}} \, e^{-j \frac{k_0 x^2}{2(z-z_\text{b})}} \, I(z,x),
\label{eq:pippo}
\ee
where the integral $I(z,x)$ is defined as
\be
I(z,x) \eqdef  \int_{-\infty}^{+\infty} \Ai(\gamma_\text{a} \, \xi+\delta_\text{b}) \, e^{\nu_\text{b}(z,x) \, \xi} \, e^{j \eta_\text{b}(z) \, \xi^2} \, \mathrm{d}\xi
\label{eq:pluto},
\ee
with parameters $\delta_\text{b}$, $U_\text{b}$, $\nu_\text{b}(z,x)$, and $\eta_\text{b}(z)$ defined in 
\eqref{eq:deltab}, \eqref{eq:Ub}, \eqref{eq:nub}, and \eqref{eq:etab}, respectively.
Using the integral representation of the Airy function in \eqref{eq_fun-airy}, and 
exchanging the order of integration, we obtain after some manipulations: 
\begin{multline}
I(z,x) = \frac{1}{2 \pi} \int_{-\infty}^{+\infty} 
\left[ \int_{-\infty}^{+\infty} 
e^{j \eta_\text{b}(z) \, \xi^2} \, 
e^{j [\gamma_\text{a} v - j \nu_\text{b}(z,x)] \xi} \, \mathrm{d}\xi \right] 
\\ \cdot e^{j \frac{v^3}{3}} e^{j \delta_\text{b} v} \, \mathrm{d}v \: .
\label{eq:pluto-2}
\end{multline}
The inner integral is the Fourier transform
of a Gaussian, yielding:  
\be
I(z,x) = \sqrt{\frac{j}{4 \, \pi \, \eta_\text{b}(z)}}
\int_{-\infty}^{+\infty}
e^{j \frac{v^3}{3}} \, e^{j \delta_\text{b} v} 
\, e^{-j \frac{[\gamma_\text{a} v - j \nu_\text{b}(z,x)]^2}{4 \eta_\text{b}(z)}}
\, \mathrm{d}v \: .
\ee
This integral fits the canonical form (see, e.g., \cite{Vallee-2004})
\begin{multline}
\int_{-\infty}^{+\infty} e^{j\left( \frac{v^3}{3} + \alpha v^2 +\beta v\right)} \, \mathrm{d}v =
2 \, \pi \, e^{j \alpha \left (\frac{2}{3} \alpha^2-\beta \right)} \, \Ai(\beta-\alpha^2).
\end{multline}
Applying this, 
the expression for $I(z,x)$ becomes
\begin{multline}
I(z,x) = \sqrt{\frac{j \, \pi}{\eta_\text{b}(z)}} \, \Ai \left(\delta_\text{b} - \frac{\gamma_\text{a}^4}{16 \, \eta_\text{b}^2(z)} + j \frac{\gamma_\text{a} \, \nu_\text{b}(z,x)}{2 \, \eta_\text{b}(z)} \right) 
\\ 
\cdot e^{-\frac{\gamma_\text{a}^3 \, \nu_\text{b}(z,x)}{8 \eta_\text{b}^2(z)}} \, 
e^{j \frac{\gamma_\text{a}^2}{4 \eta_\text{b}(z)} \left[ \delta_\text{b} - \frac{\gamma_\text{a}^4}{24 \eta_\text{b}^2(z)}
+\frac{\nu_\text{b}^2(z,x)}{\gamma_\text{a}^2}\right]} \: .
\label{eq:paperino}
\end{multline}
Finally, substituting this into \eqref{eq:pippo}, the perturbative term $p(z,x)$ assumes the closed-form expression 
given by \eqref{eq:p-airy}, with the phase term $\psi_\text{b}(z,x)$ defined in \eqref{eq:psib}.


\begin{IEEEbiography}[{\includegraphics[width=1in,height=1.25in,clip,keepaspectratio]{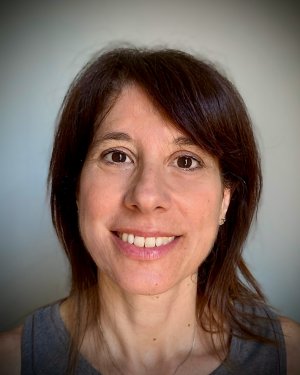}}]
{Donatella Darsena\,} (Senior Member, IEEE) received the Dr. Eng. degree {\em summa cum laude} in telecommunications engineering in 2001, and the Ph.D. degree in electronic and telecommunications engineering in 2005, both from the University of Napoli Federico II, Italy. From 2001 to 2002, she worked as embedded system designer in the Telecommunications, Peripherals and Automotive Group, STMicroelectronics, Milano, Italy. 
In 2005 she joined the Department of Engineering at Parthenope University of Napoli, Italy and worked first as an Assistant Professor and then as an Associate Professor from 2005 to 2022.
She is currently an Associate Professor in the Department of Electrical Engineering and Information Technology of the University of Napoli Federico II, Italy.
Her research interests are in the broad area of signal processing for communications, with current emphasis on reflected-power communications, orthogonal and nonorthogonal multiple access techniques, wireless system optimization, and physical-layer security.
Dr. Darsena has served as an Executive Editor for IEEE Communications Letters since 2023, Senior Editor for IEEE Access since 2024 and IEEE Signal Processing Letters since 2026. She was previously an Associate Editor of IEEE Access (from 2018 to 2023), IEEE Communications Letters (from 2016 to 2019), IEEE Signal Processing Letters (from 2020 to 2025), and Senior Area Editor of IEEE Communications Letters (from 2020 to 2023). 
\end{IEEEbiography}

\begin{IEEEbiography}[
{\includegraphics[width=1in,height=1.25in,clip,keepaspectratio]{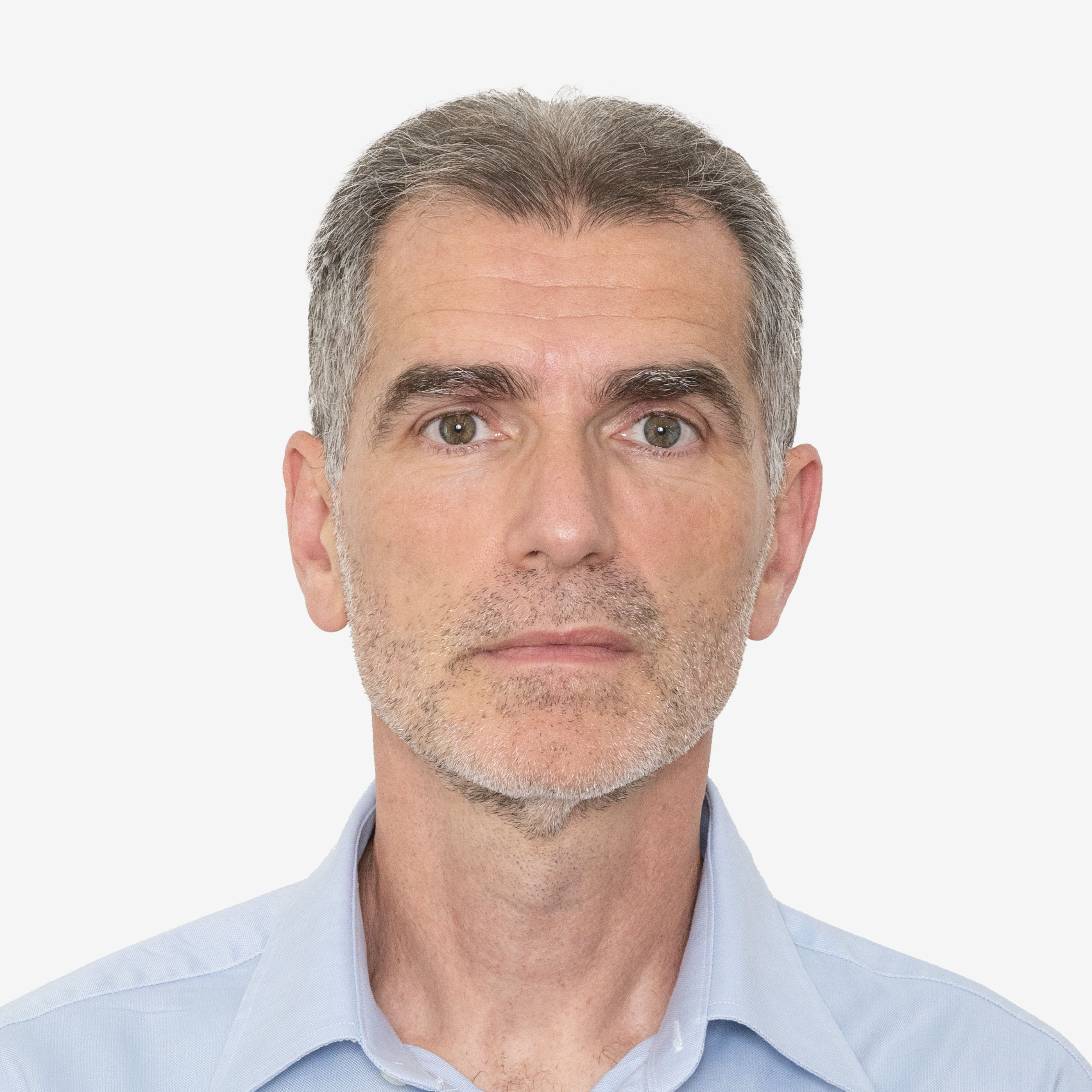}}]
{Francesco Verde\,} (Senior Member, IEEE) received the Dr. Eng. degree {\em summa cum laude} in electronic engineering from the Second University of Naples (SUN), Aversa, Italy, in 1998, and the Ph.D. degree in information engineering from the University of Naples Federico II, Italy, in 2002.
From December 2002 to August 2025, he was with the University of Naples Federico II, where he served as an Assistant Professor of signal theory and mobile communications and was promoted to Associate Professor of telecommunications in December 2011. Since September 2025, he has been an Associate Professor of telecommunications with the Department of Engineering, University of Campania Luigi Vanvitelli, Aversa, Italy.

His research interests include wireless communications via metamaterials, reflected-power communications, physical-layer security, unmanned aerial vehicle (UAV)-assisted communications, orthogonal and non-orthogonal multiple-access techniques, and space–time processing for cooperative and cognitive communication systems.
Prof. Verde has served on numerous technical program committees for major IEEE conferences in signal processing and wireless communications. He has been an Associate Editor of the IEEE Transactions on Vehicular Technology since 2022 and the IEEE Transactions on Wireless Communications since 2026. He was previously an Associate Editor of the IEEE Transactions on Signal Processing (2010–2014), IEEE Signal Processing Letters (2014–2018), and IEEE Transactions on Communications (2017–2022), as well as a Senior Area Editor of IEEE Signal Processing Letters (2018–2023). He also served as a Guest Editor for the EURASIP Journal on Advances in Signal Processing (2010) and Sensors (MDPI) (2018–2022).
\end{IEEEbiography}

\begin{IEEEbiography}[
{\includegraphics[width=1in,height=1.25in,clip,keepaspectratio]{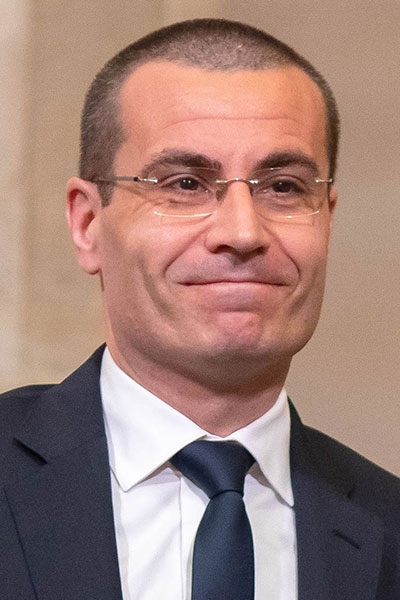}}]
{Marco Di Renzo\,} (Fellow, IEEE) received the Laurea (cum laude) and Ph.D. degrees in electrical engineering from the University of L’Aquila, Italy, in 2003 and 2007, respectively, and the Habilitation à Diriger des Recherches (Doctor of Science) degree from University Paris-Sud (currently Paris-Saclay University), France, in 2013. Currently, he is Chair Professor of Telecommunications Engineering, the Director of the Centre for Telecommunications Research, and the Head of the Telecommunications Group, Department of Engineering, King's College London, London, United Kingdom. He is also a CNRS Research Director (Professor) with the Institute of Electronics and Digital Technologies (IETR) at CNRS-CentraleSup\'elec, Rennes, France. He was a France-Nokia Chair of Excellence in ICT at the University of Oulu (Finland), a Tan Chin Tuan Exchange Fellow in Engineering at Nanyang Technological University (Singapore), a Fulbright Fellow at The City University of New York (USA), a Nokia Foundation Visiting Professor at Aalto University (Finland), and a Royal Academy of Engineering Distinguished Visiting Fellow at Queen's University Belfast (U.K.). He is a Fellow of the IEEE, IET, EURASIP, RSA, and AAIA; an Academician of AIIA; an Ordinary Member of the European Academy of Sciences and Arts, an Ordinary Member of the Academia Europaea, and an Ordinary Member of the Italian Academy of Technology and Engineering; an Ambassador of the European Association on Antennas and Propagation; and a Highly Cited Researcher. He has received several distinctions, including the Michel Monpetit Prize conferred by the French Academy of Sciences, the IEEE Communications Society Heinrich Hertz Award, and the IEEE Communications Society Marconi Prize Paper Award in Wireless Communications. Also, he is a principal investigator of an ERC Synergy grant on metasurface-based information processing. He served as the Editor-in-Chief of IEEE Communications Letters from 2019 to 2023, and as the Director of Journals and Chair of the Publications Misconduct Ad Hoc Committee of the IEEE Communications Society from 2024 to 2025. Currently, he sits on the IEEE-COMSOC Fellow Evaluation Standing Committee and on the Editorial Board of the Proceedings of the IEEE.
\end{IEEEbiography}

\begin{IEEEbiography}
[{\includegraphics[width=1in,height=1.25in,clip,keepaspectratio]{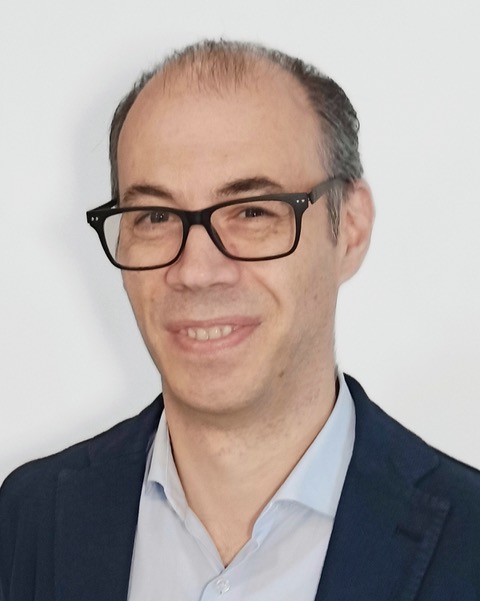}}]
{Vincenzo Galdi\,} (Fellow, IEEE) received the Laurea degree
\emph{summa cum laude} in electrical engineering and the Ph.D. degree
in applied electromagnetics from the University of Salerno, Salerno,
Italy, in 1995 and 1999, respectively.

He held research and visiting appointments at several international
institutions, including the European Space Research and Technology Centre
(ESTEC), Noordwijk, The Netherlands; Boston University, Boston, MA, USA;
the Massachusetts Institute of Technology, Cambridge, MA, USA; the
California Institute of Technology, Pasadena, CA, USA; and The University
of Texas at Austin, Austin, TX, USA. He is currently a Professor of
Electromagnetics with the Department of Engineering, University of
Sannio, Benevento, Italy, where he leads the Fields and Waves Laboratory.
He is also Co-Founder of the spinoff company MANTID, Benevento, Italy,
and of the startup company BioTag, Naples, Italy.

He has co-edited two books and coauthored more than 200 articles in
peer-reviewed international journals. He is a co-inventor of 13 patents.
His research interests include wave interactions with complex structures
and media, multiphysics metamaterials, smart propagation environments,
optical sensing, and gravitational interferometry.

Prof. Galdi is currently an Administrative Committee Member of the IEEE
Nanotechnology Council, representing the IEEE Antennas and Propagation
Society. He is a Fellow of Optica and of the American Physical Society, and
a Senior Member of the LIGO Scientific Collaboration. He received the
URSI Young Scientist Award in 2001 and the Outstanding Associate Editor
Award of the \emph{IEEE Transactions on Antennas and Propagation} in
2014.

He served as Chair of the Technical Program Committee of the
International Congress on Engineered Material Platforms for Novel Wave
Phenomena in 2018. He also served as a Topical/Track Chair for the IEEE
International Symposium on Antennas and Propagation and USNC-URSI Radio
Science Meeting from 2016 to 2017 and from 2020 to 2023, and as
Organizer/Chair of several topical workshops and special sessions.

He is currently an Associate Editor of \emph{Reviews of Optoelectronics}.
He served as an Associate Editor from 2013 to 2014, Senior Associate
Editor from 2015 to 2016, and Track Editor from 2016 to 2020 of the
\emph{IEEE Transactions on Antennas and Propagation}. He also served as
an Associate Editor of \emph{Optics Express} from 2019 to 2025 and as a
regular reviewer for numerous journals, conferences, and funding
agencies.
\end{IEEEbiography}

\end{document}